\documentclass[aps,prb,showpacs,twocolumn,superscriptaddress]{revtex4-1}
\usepackage{amsmath}
\usepackage{amssymb,amscd,bm,dsfont,wasysym,mathrsfs,latexsym,psfrag,accents,mathtools}
\usepackage{graphicx}
\usepackage{multirow}
\usepackage{dcolumn}
\usepackage{float}
\usepackage{color}
\usepackage{xcolor}
\usepackage{comment}
\usepackage{bbold}
\usepackage{hyperref}
\usepackage{blindtext}

\newcolumntype{L}[1]{>{\raggedright\arraybackslash}p{#1}}
\newcolumntype{C}[1]{>{\centering\arraybackslash}p{#1}}
\newcolumntype{R}[1]{>{\raggedleft\arraybackslash}p{#1}}

			\newcommand{\e}[1]{\begin{align}{#1}\end{align}}
			\newcommand{\es}[1]{\begin{align*}{#1}\end{align*}}
			\newcommand{\m}[1]{\begin{multline}{#1}\end{multline}}
		
		\newcommand{\f}[2]{\frac{#1}{#2}}

		\newcommand{\la}[1]{\label{#1}}

		\newcommand{\q}[1]{Eq.\ (\ref{#1})}
		\newcommand{\qq}[2]{Eqs.\ (\ref{#1})-(\ref{#2})}
		\newcommand{\s}[1]{Sec.\ \ref{#1}}
		\newcommand{\fig}[1]{Fig.\ \ref{#1}}		
		\newcommand{\app}[1]{App.\ \ref{#1}}				
		\newcommand{\tab}[1]{Tab.\ \ref{#1}}
		\newcommand{\ocite}[1]{Ref.\ \onlinecite{#1}}

		\newcommand{\eq}{=&\;}

		\newcommand{\R}{\mathbb{R}}
		\newcommand{\Z}{\mathbb{Z}}
		
		\newcommand{\C}{\mathbb{C}}

\newcommand{\var}{\varepsilon}

\newcommand\as{\;\;\;\;}

\newcommand{\hbr}{\hat{\br}}

\newcommand{\ba}{\boldsymbol{a}}

\newcommand{\be}{\boldsymbol{e}}

\newcommand{\bk}{\boldsymbol{k}}

\newcommand{\bp}{\boldsymbol{p}}

\newcommand{\br}{\boldsymbol{r}}

\newcommand{\bt}{\boldsymbol{t}}

\newcommand{\bA}{\boldsymbol{A}}

\newcommand{\bG}{\boldsymbol{G}}

\newcommand{\bR}{\boldsymbol{R}}
\newcommand{\bS}{\boldsymbol{S}}

\newcommand{\bze}{\boldsymbol{0}}

\newcommand{\bvarpi}{\boldsymbol{\varpi}}

\newcommand{\W}{{\cal W}}

\newcommand{\inv}{\mathfrak{i}}
\newcommand{\mir}{\mathfrak{r}}

\newcommand{\matrixtwo}[4]{\begin{pmatrix} #1 & #2 \\ #3 & #4 \end{pmatrix}}

\newcommand{\ins}[1]{\;\;\;\;\text{#1}\;\;\;\;}

\newcommand{\calc}{{\cal C}}

\newcommand{\calh}{{\cal H}}

\newcommand{\cali}{{\cal I}}

\newcommand{\calp}{{\cal P}}

\newcommand{\calt}{{\cal T}}

\newcommand{\noi}[1]{\noindent (#1)}
\newcommand{\imp}{\;\;\Rightarrow\;\;}
\newcommand{\mo}{\text{-}1}

\newcommand{\braket}[2]{\big\langle #1 \big| #2 \big\rangle}
\newcommand{\ketbra}[2]{\big|  #1  \big\rangle \big\langle #2 \big| }

\newcommand{\bra}[1]{\big\langle#1\big|}
\newcommand{\ket}[1]{\big|#1\big\rangle}

\newcommand{\lin}{\notag \\}

\newcommand{\ab}{\alpha\beta}

\newcommand{\bpm}{\begin{pmatrix}}
\newcommand{\epm}{\end{pmatrix}}

\newcommand{\bal}{\begin{align}}

\newtheorem{definition}{Definition}[section]

\begin{document}

\title{Crystallographic splitting theorem for band representations\\
and fragile topological photonic crystals}
 \author{A. Alexandradinata} \affiliation{Department of Physics and Institute for Condensed Matter Theory, University of Illinois at Urbana-Champaign, Urbana, Illinois 61801, USA}
 \author{J. H\"oller} \affiliation{Department of Physics, Yale University, New Haven, Connecticut 06520, USA}  
 \author{Chong Wang} \affiliation{Institute for Advanced Study, Tsinghua University, Beijing, 100084, China}
 \author{Hengbin Cheng} \affiliation{Institute of Physics, Chinese Academy of Sciences/Beijing National Laboratory for Condensed Matter Physics, Beijing 100190, China}
\affiliation{School of Physical Sciences, University of Chinese Academy of Sciences, Beijing 100049, China}
 \author{Ling Lu} \affiliation{Institute of Physics, Chinese Academy of Sciences/Beijing National Laboratory for Condensed Matter Physics, Beijing 100190, China}
\affiliation{Songshan Lake Materials Laboratory, Dongguan, Guangdong 523808, China}

\begin{abstract}
The fundamental building blocks in band theory are {band representations}  --  bands whose infinitely-numbered Wannier functions are generated (by action of a space group) from a finite number of symmetric Wannier functions centered on a point in space. This work aims to simplify questions on a multi-rank band representation by splitting it into unit-rank bands, via the following \textit{crystallographic splitting theorem}: being a rank-$N$ band representation is {equivalent} to being splittable into a finite sum of bands indexed by $\{1,2,\ldots,N\}$, such that each band is spanned by a single, analytic Bloch function of $\bk$, and any symmetry in the space group acts by permuting $\{1,2,\ldots,N\}$. We prove this theorem for all band representations (of crystallographic space groups) whose Wannier functions transform in the integer-spin representation; in the half-integer-spin case, the only exceptions to the theorem exist for three-spatial-dimensional space groups with cubic point groups. Applying this theorem, we develop computationally efficient methods to determine {whether} a given energy band (of a tight-binding or Schr\"odinger-type Hamiltonian) is a band representation, and, if so, how to numerically construct the corresponding symmetric Wannier functions. Thus we prove that rotation-symmetric topological insulators in Wigner-Dyson class AI are fragile, meaning that the obstruction to symmetric Wannier functions  can be removed by addition of band representations to the filled-band subspace.  An implication of fragility is that its boundary states, while robustly covering the bulk energy gap in finite-rank tight-binding models, can be destabilized if the Hilbert space is expanded to include all symmetry-allowed representations. These fragile insulators have photonic analogs that we identify; in particular, we prove that an existing photonic crystal  built by Yihao Yang et al. [Nature 565, 622 (2019)] is fragile topological with removable boundary states, which disproves a widespread perception of `topologically-protected' boundary states in time-reversal-invariant, gapped photonic/phononic crystals.  As a final application of our theorem, we derive various symmetry obstructions on the Wannier functions of topological insulators; for certain space groups, these obstructions are proven to be equivalent to the nontrivial holonomy of Bloch functions.
\end{abstract}

\date{\today}\maketitle

{\tableofcontents \par}

\newpage
\section{Introduction}

Solid-state physicists have predominantly held that to know a band is to specify its properties in the space of crystal momentum $\bk$.\cite{Fermi_fermigas,Dirac_theoryofQM,Bloch1929,Brillouin_book} The crystallographic space-group symmetry of a band is specified by the different representations of little groups (in $\bk$-space),\cite{tinkhambook,kpbook} their compatibility relations,\cite{bouckaert_theoryofBZ,heinebook,bandcombinatorics_kruthoff,TQC,Po2017} and  associated energy degeneracies.\cite{Hund_crystalsymmetry,Herring_trsband,Herring_accidentaldeg,connectivityMichelZak,Bradlyn_newfermions} \\

As pioneered by Zak,\cite{Zak1979,Zak1981} a real-space formulation of bands specifies how a space group $G$ transforms an infinite set of exponentially-localized Wannier functions distributed over a real-space lattice. Zak proposed that the fundamental building blocks of bands are \emph{band representations}: bands whose infinitely-numbered Wannier functions are generated (by action of $G$) from a finite number of symmetric Wannier functions centered at a point in space (known as a Wyckoff position).  An intuitive example of a band representation is  the Hilbert space of any tight-binding lattice model. Unfortunately, it is generally difficult to identify if an energy band (of a tight-binding or Schr\"odinger-type Hamiltonian) is a band representation, because one would not a priori know the Wyckoff position or the symmetry representation  of the  Wannier functions.\\

Such an identification would confer the following advantages: (i) one may utilize comprehensive databases for the $\bk$-space symmetry representations and compatibility relations of band representations, which have been tabulated in the Bilbao crystallographic server,\cite{elcoro_EBRinBilbao} (ii) some band representations exhibit symmetry-fixed Berry-Zak phases\cite{AA2014,TBO_JHAA} which are measurable  in transport\cite{TBO_JHAA} and cold-atomic experiments,\cite{TBO_JHAA,atala2013,Li2016} and (iii) conversely, \textit{not} being a band representation manifests in various physical implications, which may include nontrivial $\bk$-space holonomy,\cite{zak1989,fukane_trspolarization,yu2011,AA2014,TBO_JHAA} quantum entanglement,\cite{Turner2010,hughes_inversionsymmetricTI,aa_traceindex,ZhoushenHofstadter,Po2017,Bradlyn2019} and robust boundary states.\cite{kane2005A,kane2005B,fidkowski2010,Hsieh_SnTe,AAchen,ChaoxingNonsymm,Hourglass,Cohomological} \\


Following Zak's \emph{real-space} definition of band representations, one may heuristically test if an energy band -- given numerically by a set of Bloch functions on a $\bk$-mesh -- is a band representation. Namely, one would postulate trial Wannier functions with a certain symmetry representation and Wyckoff position, then compute the overlap of these trial Wannier functions with the Bloch functions.\cite{Satpathy1988,marzari1997,soluyanov2011,alexey2011,alexey_smoothgauge,Winkler2015a}\footnote{Essentially this is the same method used by Po et al in \ocite{fragile_po,Po2019}, where trial Wannier functions are validated by a symmetric, gapped interpolation.} Unfortunately, the possible symmetry representations and Wyckoff positions are numerous in  complicated space groups; even if they are correctly chosen for a given band representation, it is still possible that  a trial Wannier function has zero overlap with a given Bloch function on the $\bk$-mesh.
(It is worth interjecting that several groups have claimed to prove band-representability based on $\bk$-space symmetry representations and/or $\bk$-space holonomy;\cite{Bouhon2018,Ahn2018a,Wang2019} we will explain why these alleged proofs are merely suggestive, and offer a theorem that makes some of these proofs rigorous.) \\



With the goal of determining band-representability \textit{without} trial Wannier functions, we propose to reformulate band representations from a \emph{topological} perspective. This perspective emphasizes the notion of continuity that is fundamental to the topological classification of vector bundles.  Applied to band theory, a rank-$N$ vector bundle  over the Brillouin torus is simply a band comprising $N$ {linearly} independent Bloch functions at each $\bk$, and if such Bloch functions can be made continuous and periodic over the torus, the band is said to be \emph{topologically trivial}. In two spatial dimensions, being topologically nontrivial is in one-to-one correspondence with  a nontrivial first Chern class,\cite{Chern1946} which leads to a quantized Hall conductance for band insulators.\cite{Thouless1982} \\

Our topological formulation of band representations can be formalized by the following  \textit{\textbf{crystallographic splitting theorem}}:  being {a rank-$N$ band representation} is equivalent to being splittable into a sum of $N$ unit-rank bands (indexed by $\{1,2,\ldots,N\}$) which are each topologically trivial, such that any symmetry in the space group symmetry  acts by permuting $\{1,2,\ldots,N\}$. Alternatively stated, being a rank-$N$ band representation is equivalent to being splittable into $N$ independent sets of exponentially-localized Wannier functions, such that each set is obtained by Bravais-lattice translations of a single Wannier function, and any space-group symmetry acts by permuting these sets. \\

Our splitting theorem applies to  any band representation (of crystallographic space groups) whose Wannier functions transform in the integer-spin representation. For half-integer-spin band representations, the equivalence applies for any space group in two spatial dimensions;  exceptions to this equivalence exist only for three-spatial-dimensional space groups with cubic point groups. All the above statements generalize to time-reversal-invariant band representations (in Wigner-Dyson\cite{Dyson1962} symmetry classes AI and AII), with the semantic replacement of `space group' by `magnetic space group'.\\

In comparing our topological formulation with Zak's real-space formulation, specifying the space group action on a finite set of topologically trivial, unit-rank bands is  simpler than specifying the group action on an infinite set of Wannier functions. A considerable volume of the manuscript is spent on unpacking the conceptual simplifications and physical implications of the crystallographic splitting theorem, which we summarize in the following section. This summary will also serve as a guide to the structure of the manuscript.

\begin{figure*}[ht]
\centering
\includegraphics[width=14 cm]{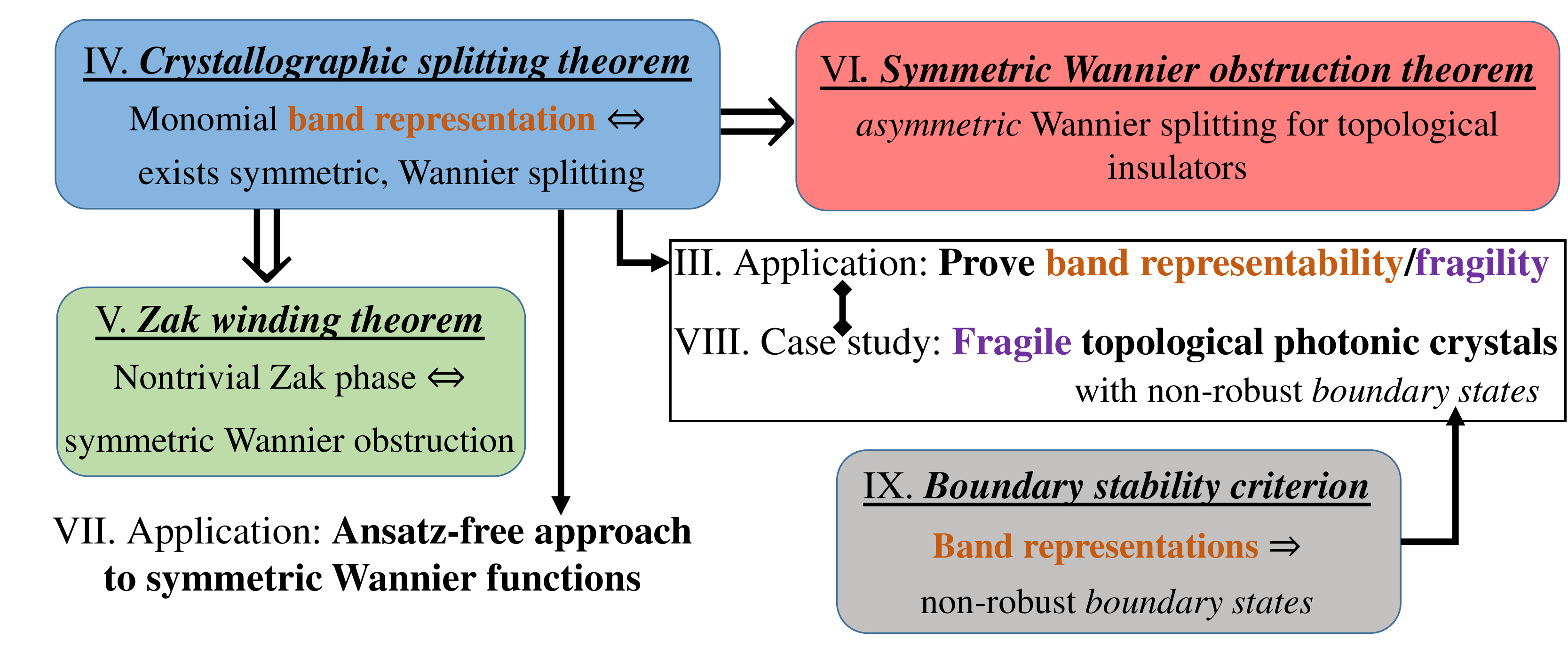}
\caption{Concept map of the sections in this paper, with their section numbers IV.-IX. indicated. \label{fig:conceptmap}}
\end{figure*}

\section{Summary and Outline}\la{sec:outline}

Our main technical accomplishment is a topological formulation of band representations, which is formalized by the crystallographic splitting theorem in \s{sec:equivalencetheorems}. \\

In comparison with the real-space formulation of band representations, the topological formulation is conceptually closer to recent developments in the band-theoretic description of topological insulators.  Indeed, the converse of the topological formulation says that a topologically trivial, rank-$N$ band that is \emph{not} band-representable is spanned by $N$ sets of exponentially-localized Wannier functions which cannot all be permuted by space-group symmetry. (Being \textit{band-representable} is a convenient shorthand for being a band representation.)
For this reason, we call a space-group-symmetric band -- that is \textit{not} band-representable -- an \textit{obstructed representation}. The full implications of this obstruction are explored in \s{sec:wannierobstruction}. In particular, we will derive three types of constraints on the Wannier functions of topological insulators: \\

\noi{i} Wannier functions cannot be localized to a single tight-binding lattice site, \\

\noi{ii}  Wannier functions in Wigner-Dyson symmetry class AII cannot be fully spin-polarized (analogously, Wannier functions in class AI cannot be fully pseudospin-polarized), and \\

\noi{iii} Wannier functions cannot form a representation of certain symmetries in the stabilizer of their Wyckoff position. \\

\noindent (i-iii) are readily observable in numerical constructions of Wannier functions for topological insulators, as will be exemplified by {topological insulators old and new}. \\

Our topological formulation may be applied to determine  if a given energy band is band-representable. Our proposed method involves diagonalizing a projected symmetry operator that is a $\bk$-dependent Hermitian matrix; if the eigenbands of the projected symmetry operator are eigenvalue-nondegenerate  and have trivial first Chern class, then  the given energy band is guaranteed to be band-representable, in accordance with our crystallographic splitting theorem. The advantage of our method is that it can be carried out without having to deal with Wannier functions at all. \\

For demonstration, we prove in \s{sec:demonstration} that the filled band -- of rotation-invariant topological crystalline insulators\cite{fu2011,AAchen} (TCI)  in Wigner-Dyson symmetry class AI -- is a \textit{fragile obstructed representation}. By `fragile obstructed', we mean that the filled  band  has an obstruction to symmetric Wannier functions, but this obstruction is removable by addition of a band representation to the filled-band subspace. Once  removed, the filled-band subspace is symmetrically deformable to a tight-binding (or `atomic') limit, which is incompatible with boundary states that robustly cover the bulk energy gap; this statement is separately proven as the \textbf{\textit{symmetric tight-binding limit theorem}} and the \textbf{\textit{boundary stability criterion}} in \s{sec:norobustsurface}. This means that if a fragile obstructed representation is accompanied by boundary states that robustly cover the bulk gap in finite-rank, tight-binding models (as exemplified by the above rotation-invariant TCI's), then these boundary states can be destabilized if the Hilbert space is expanded to include all symmetry-allowed representations -- we refer to this as a \textit{representation-dependent stability} of boundary states. It is worth remarking that the filled band of this TCI is identical to a band representation with regard to its $\bk$-space symmetry representations, which cautions against generally inferring fragile obstructions or band representability from $\bk$-space representations alone.\footnote{As another case in point, spacetime-inversion-symmetric bands can also have a fragile obstruction that is not identifiable by $\bk$-space representations.\cite{Po2019}}\\

While the above rotation-invariant TCI's have thus far \textit{not} been realized in solid-state materials, we prove in \s{sec:photonic} that their photonic analogs exist in a three-spatial-dimensional tetragonal photonic crystal designed by Tetsuki Ochiai,\cite{Ochiai2017} and in an \emph{existing} hexagonal photonic crystal built by Yihao Yang et al.\cite{Yang2019}  While previous theoretical works\cite{Yang2019,Slobozhanyuk_3DalldielectricphotonicTI} have identified the hexagonal photonic crystal as an analog of the \textit{non-fragile} $\Z_2$ topological insulator in Wigner-Dyson class AII, our  group-theoretic analysis  identifies it properly as an analog of the \textit{fragile}  $\Z$ TCI\cite{AAchen} in class AI. The hexagonal photonic crystal is then a materialization of fragile topology with removable boundary states.
These three-spatial-dimensional tetragonal and hexagonal crystals complement two recently-designed, two-spatial-dimensional photonic crystals\cite{Wang2019,DePaz2019} which have been claimed to be fragile based on different crystallographic symmetries that we specify below. \\




While we have advertised that band-representability can be proven without constructing Wannier functions, sometimes these functions are intrinsically desirable for other practical reasons, e.g., to analyze the formation of chemical bonds,\cite{marzari1997} to investigate the electronic polarization of disordered/distorted insulators,\cite{marzari1997}  or to construct a lower-rank, tight-binding model which  possibly includes many-body interactions. Thus motivated, we present in \s{app:constructWFs} a numerical algorithm  to construct symmetric Wannier functions -- for any band representation that satisfies the crystallographic splitting theorem.
The advantage of our method is that it is ansatz-free, that is, it does not require the user to guess a set of trial Wannier functions, unlike many existing methods.\cite{Satpathy1988,marzari1997,souza2004,soluyanov2011,alexey2011,alexey_smoothgauge,Winkler2015a} \\

As a final application of the topological formulation of band representations, we prove in \s{sec:zakphasewind} an equivalence between the obstruction of symmetric Wannier functions and  nontrivial $\bk$-space holonomy; the latter is a geometric property of Bloch functions that is encoded in the Zak phase. This equivalence holds for point groups which are generated by time reversal and/or spatial inversion.  As cases in point, a fragile obstruction against spatial-inversion-symmetric Wannier functions was explored theoretically in \ocite{AA2014,fragile_po,TBO_JHAA,Cano2018,Bradlyn2019,Else2019}, and may even have a photonic analog;\cite{Wang2019,DePaz2019} a stable obstruction against time-reversal-symmetric Wannief functions characterizes bands with $\Z_2$ Kane-Mele topological order\cite{alexey2011,TQC}; a fragile  obstruction against spacetime-inversion-symmetric Wannier functions\cite{Bouhon2018,Bradlyn2019,Ahn2019} is possibly realized by the nearly-flat bands of twisted bilayer graphene.\cite{Po2019,Song2019}
Despite these examples having been studied extensively from the dual perspectives of $\bk$-space holonomy and symmetric Wannier obstructions, the equivalence of both perspectives is established for the first time here. \\

We have chosen to discuss the fragility of TCIs [cf.\ \s{sec:demonstration}]  before the formal statement of the crystallographic splitting theorem [cf.\ \s{sec:equivalencetheorems}]. This order of consumption is recommended for  physically motivated readers who are versed in the theory of topological band insulators. However, a mathematically-oriented reader who is less interested in our idiosyncratic application  may skip to the splitting theorem in \s{sec:equivalencetheorems}, which is written to be self-contained. Almost every other section should be consumed after having read \s{sec:equivalencetheorems}. One possible exception is our case study  of fragile topological photonic crystals in \s{sec:photonic}, which is the recommended starting point for members in the photonics community.\\

This completes the summary of our results. For the reader's convenience,  we have drawn in \fig{fig:conceptmap} a concept map for the various sections of this paper. The main results are recapitulated in the final Discussion section of \s{sec:discussion}, where we also provide an outlook. Included in \app{sec:preliminaries} is a review of basic notions in band theory, space groups and bundle theory; this review may also be used as a glossary of specialized terms, which the reader may refer to when needed.

\section{Case study: fragile topological crystalline insulators in class AI} \la{sec:demonstration}

In \s{sec:futci}, we will first give a pedagogical introduction to three-dimensional, rotation-invariant topological crystalline insulators (TCI's) in Wigner-Dyson symmetry class AI, focusing on aspects that identify them as obstructed representations. One particular aspect -- having boundary states with a representation-dependent stability -- will be a recurrent theme in the subsequent sections \s{sec:photonic} and \s{sec:norobustsurface}. \\

Underlying the proof of fragility for this TCI is the crystallographic splitting theorem, which we will introduce casually in \s{sec:casualintrosplitting} with a simple example. After these preliminaries, the proof begins properly in \s{sec:outlineprooffragile}.




\subsection{Topological crystalline insulators as obstructed representations of space groups}\la{sec:futci}

As theoretically proposed by Liang Fu in \ocite{fu2011}, the first-known TCI has the space group $G_4=\calt_3 \rtimes C_{4v}\times\Z_2^T$, which is also the symmetry of the tetragonal photonic crystal. In general, $\calt_d$ denotes a translational group of a $d$-dimensional crystal, $C_{nv}$ ($n=2,3,4,6$) denotes a point group generated by an $n$-fold rotation and a mirror plane that contains the rotational axis, and $\Z_2^T$ is the order-two group generated by time reversal $T$; $T^2$  equalling the identity means we are in Wigner-Dyson class AI.  The semidirect product $\rtimes$ structure of $G_4$ reflects that $G_4$ is a symmorphic space group, as briefly reviewed in \app{sec:spacegroupwignerdyson}.\\

We will focus on known qualities of the TCI that identify its filled band  as an \textit{obstructed representation of} $G_4$. By `obstructed representation', we mean that the projector (denoted $P_{OR}$) to the filled band   is invariant under all elements of $G_4$, but the filled band is  \emph{not} a band representation of $G_4$. \\

A tight-binding model for the $G_4$-symmetric TCI was first  proposed by Liang Fu on a tetragonal lattice.\cite{fu2011} The tight-binding  vector space consists of Wannier functions
defined over two sublattices indexed by $\alpha=1,2$. On each sublattice, the Wannier functions transform as a rank-two band representation (BR) of $G_4$. By Zak's standard definition, a BR is an induced representation of a space group, as briefly reviewed in \app{sec:zakdefinesbr}. Here we will describe what induction (in our case study) entails: \\

\noi{a} Begin with a pair of Wannier functions $\{W_{+,\alpha,\bze},W_{-,\alpha,\bze}\}$ centered at the $C_{4v}$-invariant Wyckoff position $\bvarpi_a$, with $W_{\pm,\alpha,\bze}$ having the symmetry of a $p_{\pm}:=(p_x\pm ip_y)$-orbital; these orbitals transform in the irreducible two-dimensional `vector' representation $E$ of $C_{4v}$, the {site stabilizer}  of $\bvarpi_a$; the site stabilizer of a Wyckoff position is the group consisting of all elements of a space group (here, $G_4$) that preserve the Wyckoff position. Here and henceforth, it should be understood  that any `Wannier function' is exponentially-localized, i.e., decaying at least as fast as an exponential function. \\

\noi{b} We then generate an infinite set of Wannier functions $\{W_{+,\alpha,\bR},W_{-,\alpha,\bR}\}_{\bR \in BL}$ by Bravais-lattice translations. Throughout this work, we use $\bR$ to denote a vector in the Bravais lattice.\\

\noindent With regard to its symmetry properties, a BR $(G,\bvarpi,D)$ is fully specified by  a space group $G$, Wyckoff position $\bvarpi$, and a representation $D$ of the corresponding site stabilizer. Our illustrative BRs  are denoted as $(G_4,\bvarpi_a,E;\alpha)$, with $\alpha=1,2$ an additional sublattice index. The rank of a BR is the number independent Wannier functions in one unit cell -- two for each of $(G_4,\bvarpi_a,E;\alpha)$.\\

Suppose we began with a tight-binding Hamiltonian having zero matrix elements between tight-binding Wannier functions centered on distinct lattice sites.  We introduce an on-site potential that distinguishes between $(G_4,\bvarpi_a,E;1)$ and $(G_4,\bvarpi_a,E;2)$, so that they are separated by an energy gap throughout the Brillouin zone. \\

By cleverly tuning the hopping parameters [cf.\ Eq.\ (1) in \ocite{fu2011}], Liang Fu induced a momentary touching between low- and high-energy bands, after which the energy gap (at all $\bk$) is re-established. Let $P_{OR}$ be the projector to the resultant low-energy band. In terms of the symmetry representations of the little group of wavevectors,\cite{tinkhambook}  $P_{OR}$ is identical to both $(G_4,\bvarpi_a,E;\alpha)$.\footnote{This can be inferred from the following observation: intermediate between the two gapped phases is a Weyl-semimetallic phase,\cite{AAchen} where the energy gap closes at generic and mirror-invariant wavevectors. In the latter case, the gap closing is between Bloch states in the same mirror representation\cite{AAchen} } Nevertheless there are indications that $P_{OR}$ is 
\emph{not} band-representable: (i) if the tight-binding Hamiltonian is diagonalized with Dirichlet (`open boundary') conditions that model a  $\calt_2 \rtimes C_{4v}\times\Z_2^T$-symmetric surface, eigen-solutions exist which are localized to the surface and whose eigen-energies robustly cover the bulk gap.\cite{fu2011} (ii) $P_{OR}$ also manifests nontrivial holonomy\cite{berryphaseTCI} which is incompatible\footnote{The incompatibility is proven in \s{sec:prelimzakphase}.} with a BR. \\ 

One aspect of the boundary states distinguishes the TCI phase from the well-known $\Z_2$ Kane-Mele topological insulator. While the TCI boundary states cannot be removed from the bulk gap by continuous  deformations of the given tight-binding Hamiltonian (that maintain both symmetry and the bulk gap), the TCI boundary states can be removed from the bulk gap if the given tight-binding Hilbert space is enlarged --  by inclusion of a boundary-localized band transforming as a unit-rank BR of $\calt_2 \rtimes C_{4v}\times\Z_2^T$ (the symmetry \textit{in the presence of the boundary}).\cite{fu2011} There are four such unit-rank BRs, corresponding to the four one-dimensional, real representations of $C_{4v}\times \Z_2^T$. In contrast, the TCI boundary states would be robust against the addition of BRs corresponding to the two-dimensional vector representation of $C_{4v}\times \Z_2^T$.  \\

We see that an obstructed representation can be accompanied by boundary states which robustly cover the bulk gap of a finite-rank tight-binding model with a restricted set of symmetry representations, however such boundary states can be destabilized by expanding the Hilbert space to include all symmetry-allowed representations. This notion of a \textit{representation-dependent stability} for boundary states is  reminiscent of (but \textit{not} equivalent to) the defining property\cite{fragile_po} of a \textit{fragile obstructed representation} (FOR). Namely, a FOR of $G_4$ is an obstructed representation of $G_4$ with the property that a BR of $G_4$ exists, such that the direct sum of this BR with FOR is a higher-rank band representation. Schematically, FOR$\oplus$BR=BR'. We emphasize that all objects in this equality are  representations of $G_4=\calt_3 \rtimes C_{4v}\times\Z_2^T$, the space group of a  three-dimensional crystal \textit{without boundaries}; moreover, the projector to each of $\{$FOR,BR,BR'$\}$, if restricted to a wavevector $\bk$, is  an analytic function (of $\bk$) throughout the Brillouin zone.\footnote{For a tight-binding Hamiltonian whose real-space matrix elements decay exponentially, the projector (to a spectrally isolated band) is analytic\cite{Panati2007,Read2017}} In contrast, the TCI boundary states have the reduced symmetry $\calt_2 \rtimes C_{4v}\times\Z_2^T$, and if we insist on distinguishing filled and unfilled boundary states, then the filled states cannot continuously be defined in the boundary Brillouin zone.\footnote{This reflects the anomalous nature of TCI boundary states, namely that they cannot be continuously deformed to the energy eigenstates of a Hamiltonian defined over 2D real space.} \\

Proving that $P_{OR}$ is a fragile obstructed representation will occupy \s{sec:outlineprooffragile} to \ref{sec:fu_diagnosefragility}. The proof might have been simple, if hypothetically the unfilled band (of Liang Fu's tight-binding model) transforms as a BR of $G_4$ -- this would imply FOR$\oplus$BR=BR', with BR' corresponding to the tight-binding vector space. In fact, the unfilled band is also an obstructed representation,\cite{AAchen} which motivates a more general methodology to proving fragility.\\

Before we begin the proof, we remark that both a nontrivial $\bk$-space holonomy and a representation-dependent stability of boundary states also characterize the $\calt_3 \rtimes C_{3v}\times\Z_2^T$-symmetric TCI, which was theoretically proposed in \ocite{AAchen,berryphaseTCI}. $\calt_3 \rtimes C_{3v}\times\Z_2^T$ is also the symmetry of the hexagonal photonic crystal.



\subsection{Casual introduction to the crystallographic splitting theorem}\la{sec:casualintrosplitting}

Underlying our proof is a \textit{mathematically equivalent} reformulation of BRs that comprise Wannier functions with integer-valued spin: being a rank-$N$ BR is equivalent to being splittable into $N$ independent sets of exponentially-localized Wannier functions (denoted $\{P_1,\ldots,P_N\}$), such that (a)  each set is derived by Bravais-lattice translations of a single Wannier function, and (b) any symmetry in the space group acts to permute $\{P_1,\ldots,P_N\}$. We shall refer to a splitting satisfying (a) as a \textit{Wannier splitting}, satisfying (b) as a \textit{symmetric splitting}, and satisfying both (a-b) as a \textit{symmetric Wannier splitting}.\\

This equivalence is formalized as the crystallographic splitting theorem in \s{sec:equivalencetheorems}, and proven in \app{app:permthm}; also discussed therein is the partial generalization to Wannier functions with half-integer-valued spin. While not essential to our proof, we now offer a simple example of a symmetric Wannier splitting to develop intuition. \\

\begin{center}
\textit{Example: symmetric Wannier splitting of  BR$(G_4,\bvarpi_a,E)$}
\end{center}

\noindent Let $P_{a,E}$ be the rank-two projector of BR ($G_4,\bvarpi_a,E$); presently we omit the sublattice index. As shown in \s{sec:futci}, $P_{a,E}=\sum_{j=\pm}\sum_{\bR}\ketbra{W_{j\bR}}{W_{j\bR}}$ is spanned by a set of Wannier functions transforming (on each site) in the $p_x\pm ip_y$ representation of $C_{4v}$. \\

\noindent Consider the Wannier splitting  $P_{a,E}= P_++P_-$, with $P_{\pm}=\sum_{\bR}\ketbra{W_{\pm, \bR}}{W_{\pm, \bR}}$ corresponding to $p_x\pm ip_y$ orbitals on each site.\footnote{An equally natural Wannier splitting is given by $P_{a,E}=P_x+P_y$ corresponding to $p_x$ and $p_y$ orbitals on each site. One may also verify that each $g\in G$ permutes $\{P_x,P_y\}$, though a trivial permutation in the $P_{x,y}$ splitting may be nontrivial in the $P_{\pm}$ splitting. } By construction, each  unit-rank projector consists of Wannier functions related by Bravais-lattice translations, hence any translation $\in \calt_3$ trivially permutes $\{P_+,P_-\}$. What remains is to determine the permutation actions for the generators of the point group $C_{4v}\times \Z_2^T$. In the $p_x\pm ip_y$ basis, the two-dimensional matrix representation of four-fold rotation ($C_4$) is diagonal, while that of time reversal ($T$) and reflection ($\mir_x:(x,y,z)\rightarrow (-x,y,z)$) are off-diagonal. It follows that all point-group symmetries act as permutations: 
\e{[\hat{C}_4,P_{\pm}]=0, \;\; \hat{T}P_{+}\hat{T}^{-1}=P_{-},\;\;\hat{\mir}_xP_{+}\hat{\mir}_x^{-1}=P_{-}, \la{symmetricdecompositionC4T} } 
meaning that $P_{a,E}=P_++P_-$ is a symmetric Wannier splitting.\\

\noindent The permutation relations in \q{symmetricdecompositionC4T} are deducible from a general observation:
for any Wannier splitting of a rank-$N$ representation of space group $G$, if there exists  $N$ representative Wannier functions  which are permuted by $g\in G$ (up to a $U(1)$ phase), then  $g$ would similarly permute the $N$ unit-rank projectors corresponding to that Wannier splitting.\footnote{The proof is elementary: if there exists  Wannier functions $\{W_{1\bze},\ldots,W_{N\bze}\}$ that represent $\{P_j=\sum_{\bR}\ketbra{W_{j\bR}}{W_{j\bR}}\}$ in the Wannier splitting $P=\sum_{j=1}^NP_j$, and if $\hat{g}\ket{W_{j\bze}}=\lambda_{g;j}\ket{W_{\sigma_g(j),\bze}}$, with $\lambda$ a phase factor and $\sigma$ a permutation on $\{1,\ldots,N\}$, then $\hat{g}P_j\hat{g}^{-1}=P_{\sigma_g(j)}$.}

\subsection{An outline for the proof of fragility} \la{sec:outlineprooffragile}


Taking the crystallographic splitting theorem as a given [cf.\ \s{sec:equivalencetheorems}], we now begin the proof of fragility, which is split into three subsections:\\


\noi{i} In \s{sec:introprojsymmetry}, we will introduce a systematic method to obtain a symmetric Wannier splitting. This method involves the diagonalization of a projected symmetry operator and will be used in the remainder of the proof. \\

\noi{ii} $P_{OR}$ being an obstructed representation of $G_4$ [cf.\ \s{sec:futci}] means there must exist an obstruction to a symmetric Wannier splitting, which we illustrate in \s{sec:obstructP}. \\

\noi{iii} Finally in \s{sec:fu_diagnosefragility}, we prove  that a symmetric Wannier splitting exists for the sum of $P_{OR}$ with a unit-rank BR -- this would prove that $P_{OR}$ is a fragile obstructed representation of $G_4$.

\subsection{Symmetric Wannier splitting via projected symmetry operators}\la{sec:introprojsymmetry}


In proving the fragility of $P_{OR}$, we hypothesize the existence of a BR such that $P_{OR}\oplus P_{BR}$  is a band representation BR'. A priori, we would neither know what is the Wyckoff position of BR', nor know the representation of the site stabilizer -- without these information, one would not know how   $P_{OR}\oplus P_{BR}$ decomposes into a symmetric Wannier splitting. What is desirable is a systematic method to deduce the symmetric Wannier splitting for BRs in any space group. On this front, we have made partial progress that is reported in \app{app:methods_symmetricdecomp}; one of the techniques discussed therein will be applied to the present case study.\\     

To summarize the technique, we propose to diagonalize a symmetry operator that is projected to a hypothesized BR. (Our approach may be viewed as a space-group generalization of the projected spin operator proposed by Prodan in \ocite{prodan_spinchern}.) The symmetry in our case study is the four-fold rotation $C_4$.  If a Hermitian matrix representation $\tilde{C}_4$ of this symmetry is chosen, then the \textit{projected symmetry operator} is  a $\bk$-dependent Hermitian operator  distinct from the original tight-binding Hamiltonian. The projected symmetry operator can be chosen so that its eigenbands (assumed nondegenerate in eigenvalue) are permuted by all elements of the space group. Thus if each eigenband is determined to have trivial first Chern class, there must exist a basis of exponentially-localized Wannier functions for each eigenband, and the corresponding Wannier splitting is symmetric by construction.

\subsection{Obstruction to symmetric Wannier splitting for the filled band of the TCI}\la{sec:obstructP}

While not strictly necessary for the proof of fragility, it is instructive to diagonalize the projected $\tilde{C}_4$ operator for the obstructed representation $P_{OR}$, for which a symmetric Wannier splitting does \textit{not} exist. How this obstruction manifests (as a nodal-line degeneracy in the projected symmetry spectrum) will help us identify which BR should be summed with $P_{OR}$, such that their sum becomes band-representable.  \\

In more detail, the Hermitian representation of $C_4$ is given by $\tilde{C}_4=(-i\pi/2)\,\text{log}\,\hat{C}_4$, with $\hat{C}_4$ the unitary matrix representation of $C_4$ in the tight-binding basis of Wannier functions. $\tilde{C}_4$ has two eigenvalues $\pm 1$ which distinguish the  $p_x\pm ip_y$ basis vectors; each eigenvalue is doubly degenerate due to the presence of two sublattices. The projected symmetry operator is $\tilde{C}_{4\bk}:=p(\bk)\tilde{C}_4p(\bk)$, with $p(\bk)=\sum_{i=1}^2\ketbra{u_{i\bk}}{u_{i\bk}}$ the rank-two projector to the low-energy band of Fu's tight-binding Hamiltonian $h(\bk)=\sum_{n=1}^4\var_{n\bk}\ketbra{u_{n\bk}}{u_{n\bk}}$ [cf.\ Eq.\ (2) of \ocite{fu2011}]. Like $h(\bk)$, $\tilde{C}_{4\bk}$ is periodic under reciprocal-lattice translations, and has a four-fold symmetry
\e{\hat{C}_4\tilde{C}_{4\bk}\hat{C}^{-1}_4=\tilde{C}_{4,C_4\circ\bk}; \as C_4\circ \bk =(-k_y,k_x,k_z). \la{convC4}}
However, time reversal and spatial reflection act unconventionally as  antisymmetries:
\e{\hat{\mir}_x\tilde{C}_{4\bk}\hat{\mir}_x^{-1}=-\tilde{C}_{4,{\mir}_x\circ \bk},\as
\hat{T}\tilde{C}_{4\bk}\hat{T}^{-1}=-\tilde{C}_{4,-\bk}. \la{unconvT}}
The action of $\hat{T}$ is analogous to that of particle-hole conjugation in a Bogoliubov-de Gennes Hamiltonian.  \\

It is vanishingly improbable for the spectrum of $\tilde{C}_{4\bk}$ to be degenerate -- except on a set of $\bk$ with measure zero. If the spectrum were  nondegenerate throughout the Brillouin zone, then the two eigenbands would be permuted trivially by $C_4$ [cf.\ \q{convC4}], and  permuted nontrivially by $T$ and $\mir_x$ [cf.\ \q{unconvT}]. Furthermore, if each nondegenerate eigenband were to have trivial first Chern class (that is, the first Chern number
vanishes in any 2D cut of the 3D Brillouin zone), then each eigenband has a basis of exponentially-localized Wannier functions\cite{Brouder2007,Panati2007} -- the eigenbands would then give a symmetric Wannier splitting, in contradiction with  $P_{OR}$ projecting to an obstructed representation. This means one of our assumptions must break down: either (i) the wave function is non-analytic at a zero-measure set of $\bk$ where the spectrum (of $\tilde{C}_{4\bk}$) is degenerate, or (ii) if the spectrum were everywhere nondegenerate, the first Chern class must be nontrivial.  Alternatively stated, for an obstructed representation, the projected symmetry operator  \textit{must be the Hamiltonian of either a topological `semimetal' or a Chern `insulator'.} \\

\begin{figure}
\centering
\includegraphics[width=1.0\columnwidth]{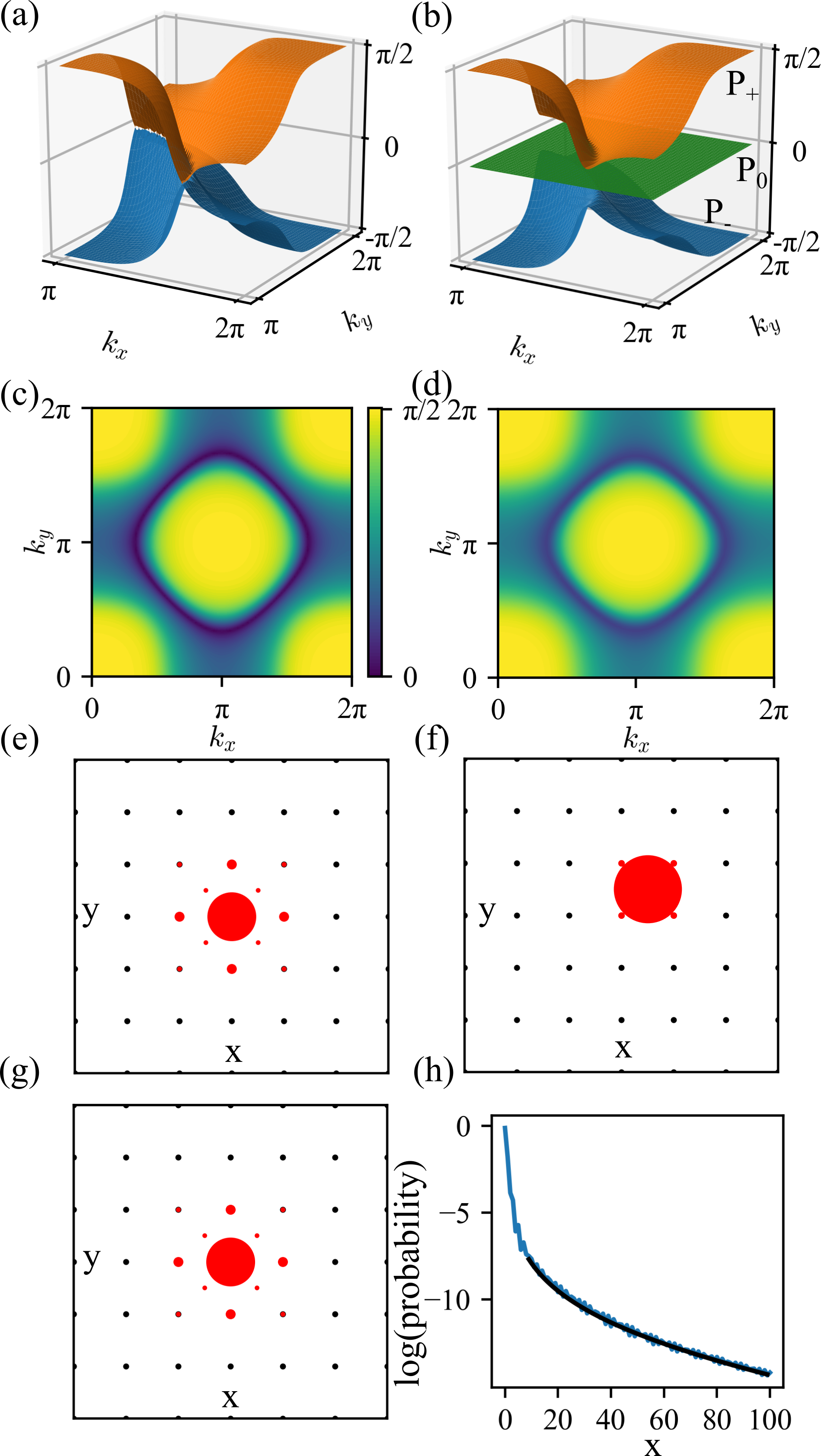}
\caption{Removing the symmetric Wannier obstruction for the obstructed representation $P_{OR}$ of $\calt_3\rtimes C_{4v}\times \Z_2^T$. Focusing on the $k_z=\pi$ slice of the Brillouin zone, we illustrate the spectra of the projected rotation operator of $P_{OR}$ [(a),(c)] and $P_{OR}+P_{BR}$ [(b),(d)], with $P_{BR}$ a unit-rank BR. 
(a) and (b) plot the band dispersions over a rotation-reduced quarter of the $k_z=\pi$ slice. (c) shows that the spectral gap (scaled by a factor of half) vanishes over a circular nodal line, while (d) shows the nonvanishing spectral gap between the lowest ($P_-$) and middle ($P_0$) band.  The Wannier functions constructed for $P_+,P_0$ and $P_-$ are shown in (e), (f) and (g), respectively. The size of the red dots represents the  probability  of a representative Wannier function on each tight-binding lattice site (indicated by black dots). For a representative Wannier function in $P_-$,  (h) is the plot of its probability distribution {(blue curve)} along a real-space line; this line is parametrized by $x$ and begins from the Wannier center ($x=0$). The tail of this curve is fitted to the exponential function $-2.45562\exp(-0.0175546x)/x^{2.1928}$, which is plotted as a black curve. \label{fig:Fu_phase}}
\end{figure}

For this TCI, the  obstruction (to a symmetric Wannier splitting) manifests  as a nodal-line spectral degeneracy confined to the $k_z=\pi$ slice of the Brillouin torus, as illustrated in  \fig{fig:Fu_phase}(a). To explain the robustness of this nodal line, the group of any wavevector in this slice contains $C_2T$ symmetry -- the composition of two-fold rotation and time reversal.  Acting as an antitunitary antisymmetry, $C_2T$ constrains $\tilde{C}_{4\bk}$ to be skew-symmetric under transpose; there being only one Pauli matrix that is skew-symmetric, the codimension of a two-fold eigenvalue-degeneracy (for $\tilde{C}_{4\bk}$) is unity. This means that the nodal-line degeneracy  is at least stable (within the $k_z=\pi$ slice) against symmetric perturbations of $\tilde{C}_{4\bk}$. In fact, the nodal-line degeneracy is even stable against any continuous deformation of $P_{OR}$ that preserves symmetry and analyticity (in $\bk$). This is because the nodal line is not contractible -- it encircles a $C_4$-invariant $\bk$-line, where the spectrum is necessarily gapped due to Bloch states having distinct $C_4$ eigenvalues. 

\subsection{`Breaking' the obstruction by adding a band representation}\la{sec:fu_diagnosefragility}

The codimension argument for the stability of the nodal-line degeneracy relied not just on $C_2T$ symmetry, but also on $P_{OR}$ having rank two. The codimension is generally greater for a three-fold eigenvalue degeneracy than it is for a two-fold degeneracy. \\

This suggests the following scenario that is  illustrated in  \fig{fig:Fu_phase}: we introduce  an additional zero-eigenvalue band \textit{without} interband hybridization, so as to enhance the degeneracy of the nodal line [cf.\ \fig{fig:Fu_phase}(a)]; this triple degeneracy is then unstable upon hybridization of bands [cf.\ \fig{fig:Fu_phase}(b)]. A zero-eigenvalue band of the projected rotation operator is simply a unit-rank BR induced from a trivial representation of $C_{4v}\times \Z_2^T$, e.g., an $s$ orbital on a four-fold invariant Wyckoff position.\\

To outline the procedure: (i) we expand the tight-binding vector space to include this unit-rank, $s$-like BR. Initially the $s$ band is introduced below the bulk energy gap (of the tight-binding Hamiltonian), with zero $s$-$p$ hybridization. (ii) This hybridization is then introduced by way of additional tight-binding matrix elements (detailed in App. \ref{app:numerics}), taking care that $G_4$ symmetry is preserved and the bulk energy gap never closes.  (iii) We then re-diagonalize  the projected rotation operator $\tilde{C}_{4\bk}$, with $\tilde{C}_4=(-i\pi/2)\,\text{log}\,\hat{C}_4$ now having an additional zero eigenvalue, and  $p(\bk)$ now a rank-three projector. $\tilde{C}_{4\bk}$ still satisfies the symmetry constraints of \qq{convC4}{unconvT}, with an appropriate generalization of $\hat{T}$ and $\hat{\mir}_x$.\\ 

The resultant spectrum shows three  bands which we numerically verify to be nondegenerate (throughout the Brillouin zone) and to have trivial first Chern class.\footnote{It is sufficient to calculate  the first Chern number on three independent slices of the Brillouin zone: $k_x=0$, $k_y=0$, and $k_z=0$. There are standard numerical techniques to calculate the Chern number, e.g., by calculating the winding of the Zak phase.}  The projectors to the top ($P_+$) and bottom bands ($P_-$) are still permuted as in \q{symmetricdecompositionC4T}, while the projector $P_0$ to the zero-eigenvalue band is invariant under all symmetries. In combination, all symmetries in $G_4$ act as a permutation on $\{P_+,P_0,P_-\}.$ Thus $P_{OR}\oplus P_{BR}=P_+\oplus P_0 \oplus P_-$ is a symmetric Wannier splitting, and must be a BR of $G_4$ according to the crystallographic splitting theorem  in \s{sec:state_equivthm}.\\

To recapitulate, we have proven that the filled band of the Fu TCI, while transforming as a rank-two obstructed representation of $\calt_3 \rtimes C_{4v}\times\Z_2^T$, becomes a rank-three BR upon addition of a unit-rank BR -- this means that the Fu TCI phase is fragile obstructed. In essentially identical fashion, we have proven that the TCI with $\calt_3 \rtimes C_{3v}\times\Z_2^T$ symmetry is also fragile obstructed; the details are given in \app{app:provefragileC3v}. Our \textit{rigorous} proofs of fragility are consistent with the topological classification by Zhida et al,\cite{Song2018} which has predicted that all obstructed representations in class AI are fragile, based on an \textit{argument} of adiabatic continuity to a `topological crystal'. \\

We remark that the projected symmetry operator provides an alternative method to numerically construct symmetric  Wannier function without need for trial Wannier functions. Given a symmetric Wannier splitting for a BR (e.g., $P_+\oplus P_0 \oplus P_-$), half the work is already done. What remains is to numerically construct a basis of Wannier functions for each of $\{P_+,P_0,P_-\},$ such that each Wannier function transforms in a definite representation of $C_{4v}\times \Z_2^T$ on each lattice site. This is accomplished by  a numerical algorithm that is described in {\s{app:constructWFs}}, and we present the final result for our case study in  \fig{fig:Fu_phase}(e-h).


\section{Topological formulation of  band representations} \la{sec:equivalencetheorems}

Our topological formulation applies to a class of band representations (BRs) that are monomial. The notion of \emph{monomial band representations} -- which will be introduced in \s{sec:monomialbr}  -- naturally generalizes the standard notion of monomial representations in finite-order groups to representations of infinite-order space groups. As we will prove in \s{sec:whichBRmonomial}, all BRs of space groups in two spatial dimensions ($d=2)$ are monomial; the only exceptions in $d=3$ exist for {double} cubic  point groups.\\

\subsection{From monomial representations of point groups to monomial band representations of space groups}\la{sec:monomialbr}
Let us briefly review some basic notions in the representation theory of finite groups. We shall primarily be concerned with  point groups whose elements are discrete  isometries of two- or three-dimensional space; also of interest are magnetic point groups, whose elements are either spatial isometries, or combinations of spatial isometries with time reversal.\\

A complex, linear representation of a finite group $H$ maps every $h\in H$ {to a finite-dimensional, invertible matrix $U(h)$, which may be taken to be unitary without any loss of generality.\cite{tinkhambook}}
A \textit{monomial representation} of a finite group $H$ is defined to be a representation of $H$ induced from a one-dimensional representation of a subgroup of $H$. (We review the notion of induction in \app{app:finitegroup}; a subgroup $H'$ of $H$ is denoted as $H' < H$.) A direct sum of monomial representations will also be referred to as a monomial representation. Equivalently, a representation of $H$ is monomial if and only if there exists  a basis (for the representation space) where every element  of $H$ is mapped to a complex permutation matrix  (a permutation matrix whose nonzero matrix elements are generalized to unimodular complex numbers). The proof of this equivalence is provided in \app{app:finitegroup}. \\

If all irreducible representations (irreps) of $H$ are monomial, then $H$ is referred to as a \textit{monomial group}. As we will see in \s{sec:whichBRmonomial}, the great majority of point groups are monomial.\\

\noindent \textit{Example of monomial representation of the point group $C_{4v}\times \Z_2^T$.} Let $E$ be a two-dimensional representation spanned by $p_x\pm ip_y$ orbitals. The generators of  $C_{4v}\times \Z_2^T$ are {represented as:} 
\e{C_4\rightarrow\matrixtwo{+i}{0}{0}{-i};\,\, \mir_y, T \rightarrow \matrixtwo{0}{1}{1}{0} .}
$C_4$ is mapped to a complex generalization of the trivial permutation matrix, while 
$\mir_y$ and $T$ are mapped to the same {transposition matrix.}\\

\noindent  A \textit{monomial band representation} of a space group $G$ is a BR($G,\bvarpi,D$) for which $D$ is a {monomial representation} of the site stabilizer $G_{\varpi}$. \\

\noindent \textit{Example of monomial band representation.} Consider BR($G{=}\calt_3\rtimes C_{4v}\times \Z_2^T,\bvarpi_a,E$), which make up the tight-binding basis in a model considered in \s{sec:futci}. As described in the previous example, $E$ is a two-dimensional monomial representation of the site stabilizer $G_{\varpi a}=C_{4v}\times \Z_2^T$, and therefore the corresponding BR is monomial.

\subsection{The crystallographic splitting theorem}\la{sec:state_equivthm}

We propose an equivalent formulation of a monomial BR that emphasizes the topological perspective:\\

\noindent \textbf{Crystallographic splitting theorem} Let $P$ be a rank-$N$ representation of $G$. $P$ is a monomial BR of $G$ if and only if there exists a splitting $P=\oplus_{j=1}^N P_j$ satisfying:

\noi{A} each $P_j$ is analytic (throughout the Brillouin torus) {and has} trivial first Chern class, and

\noi{B} $G$ acts as a permutation on $\{ P_j \}_{j=1}^N$, i.e., for all $g\in G$, $g:P_j\rightarrow P_{\sigma_g(j)}$ with $\sigma_g$ a permutation on $\{1,\ldots,N\}$.\\

Having trivial first Chern class means being topologically trivial as a complex vector bundle, as reviewed in \app{sec:bandbundle}.  Being \textit{analytic throughout the Brillouin torus} means that the restriction  of $P_j$ to $\bk$ is an analytic function of $\bk$ (for all $\bk$ in the Brillouin zone), and is periodic in reciprocal-lattice translations. All $g$ in the translational subgroup of $G$ always acts as the trivial permutation on {the indices} $\{1,\ldots,N\}$.  This theorem is proven in \app{app:permthm}. 

\subsection{Which band representations are monomial?}\la{sec:whichBRmonomial}

\begin{figure}
	\centering
	\includegraphics[width=0.9\columnwidth]{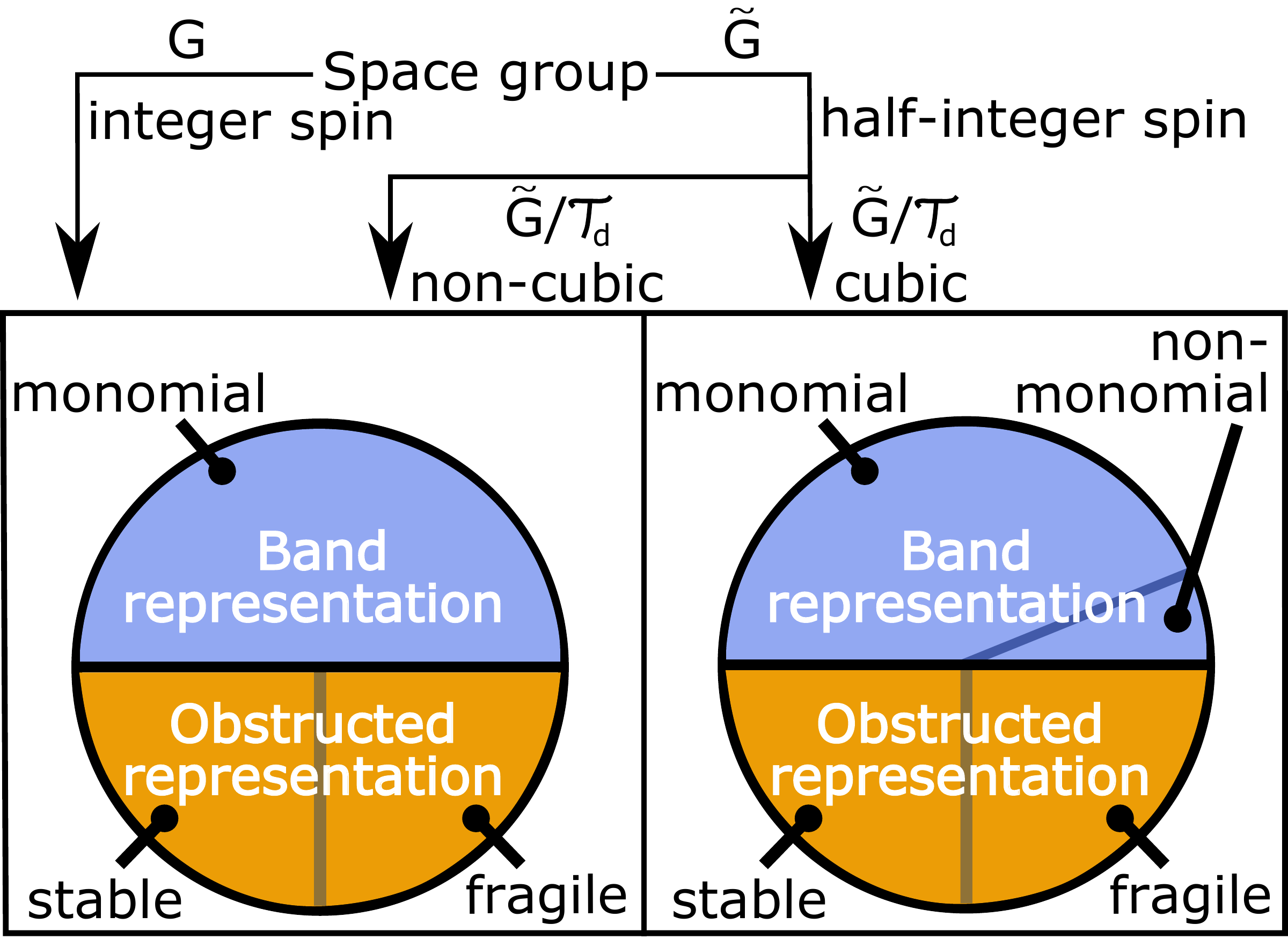}
	\caption{  Flow chart for the categorization of rank-$N$ bands with analytic projector  and space-group symmetry, in spatial dimension  $d=1,2$ and $3$. Included in this chart are the crystallographic space groups, the grey magnetic space groups, as well as their double covers (which apply to particles with half-integer spin).
	The translational subgroup of double space group $\tilde{G}$ is denoted as $\calt_d$. For $d=3$, the point group ($\tilde{G}/\calt_d$) of $\tilde{G}$ is subdivided as cubic vs non-cubic; for $d\leq 2$, all point groups are non-cubic. Rank-$N$ bands with nontrivial first Chern class 
	fall under the category of stable obstructed representations; for rank $N=1$, having trivial first Chern class is equivalent to being a BR.\cite{nogo_AAJH}   \la{fig:Pie}
	}
\end{figure}

The applicability of the crystallographic splitting theorem depends on the generality of monomial BRs, which we summarize in \fig{fig:Pie} and 
explain in the following three remarks:\\

\noi{i} All unit-rank BRs are monomial BRs. The reason is that a one-dimensional representation (of a site stabilizer) is automatically a monomial representation. Thus if $P$ has unit rank, then the splitting theorem simplifies to: $P$ is a BR of $G$ if and only if $P$ is analytic with trivial first Chern class. Condition (B) is trivially satisfied. This unit-rank statement was previously proved by {two of} us in \ocite{nogo_AAJH}.\\

\noi{ii} All BRs of crystallographic space groups (and grey magnetic space groups) are monomial. By \textit{crystallographic space group} (denoted $G$), we mean a group of spatial isometries for $(d\leq 3)$-dimensional crystals. A \textit{grey magnetic space group}, denoted $G_T$, is a direct product of any crystallographic space group $G$ with $\Z_2^T$, the order-two group generated by time reversal $T$ in Wigner-Dyson class AI.   By a `BR of $G$ or $G_T$', we restrict ourselves to linear (i.e., integer-spin) representations of the corresponding site stabilizer. That all BRs of $G$ (and  $G_T$) are monomial follows from a result that we prove in \app{app:pointgroupsmonomial}: the 32 crystallographic point groups ($\calp$), as well as the 32 grey magnetic point groups ($\calp \times \Z_2^T$), are monomial groups. Indeed for any $G$ or $G_T$, any site stabilizer  must be one of the 32 (magnetic) point groups, which are all monomial groups; thus for any BR($G$ or $G_T,\bvarpi,D)$,  $D$ must be a monomial representation. \\

\noi{iii} In spatial dimension $d=2$, all BRs of double space groups $\tilde{G}$, as well as type-1 magnetic double space groups $\tilde{G}_T$ (class AII), are monomial BRs. (The double groups $\tilde{G}$ and $\tilde{G}_T$ are the double covers of $G$ and $G_T$ respectively, as reviewed in \s{sec:spacegroupwignerdyson}. We shall only concern ourselves with the half-integer-spin representations of the double groups.) In $d=3$, there exists  BRs (of $\tilde{G}$ or $\tilde{G}_T$) which are not monomial BRs, owing to the existence of non-monomial irreducible representations of the \textit{cubic} double point groups (comprising the three tetrahedral groups  and the two octahedral groups); we prove     in  \app{app:pointgroupsmonomial} that all other double point groups (numbering $32-5=27$) are monomial groups. Note the non-cubic double point groups of three-dimensional crystals include all double point groups of two-dimensional crystals. Further discussion of the non-monomial BRs is postponed to \s{sec:discussion}.\\

\noindent \emph{Example of non-monomial band representation of the double space group $G=P23$.} The point group of this space group is the double cover $\tilde T$ of the tetrahedral group, which is isomorphic to the alternating group of four elements -- a standard example of a non-monomial group. A BR of $G=P23$ that is induced from the 
two-dimensional representation $\bar E$ of the site stabilizer $\tilde G_{1a} \approx \tilde T$ is non-monomial, as we show in \app{app:nonmononial}.\footnote{We thank Barry Bradlyn for pointing us to this example.}

\subsection{Applications of the crystallographic splitting theorem}\la{sec:applications_equivthm}

\noi{a} The splitting theorem may be applied to prove that a given band $P$ is a (monomial) BR. One approach would be to first decompose $P=\oplus_{j=1}^N P_j$ into unit-rank projectors satisfying the symmetry condition (B), namely,  that for all $g\in G$, $g:P_j\rightarrow P_{\sigma_g(j)}$ with $\sigma_g$ a permutation on $\{1,\ldots,N\}$. We define this as a \emph{symmetric splitting of $P$ with respect to $G$}. 
Having a symmetric splitting, we would then verify (A), e.g., by numerical computation of the winding number of the Zak phase.  We have illustrated this approach for fragile obstructed insulators in \s{sec:demonstration}; a systematic methodology for symmetric splitting   will be described in \app{sec:diagnoseBR_projsymm}.\\

\noi{a'} In complementarity with (a), an alternative approach (to proving $P$ is a monomial BR)  is to first decompose $P=\oplus_{j=1}^N P_j$ into unit-rank projectors satisfying condition  (A), namely that each $P_j$ is analytic (throughout the Brillouin torus) and has trivial first Chern class. Such a splitting will be referred to as a \emph{Wannier splitting of $P$ with respect to $G$}, because condition (A) guarantees\footnote{This equivalence is valid for $d=2,3$,\cite{Brouder2007,Panati2007} which is assumed throughout this work.}
that each $P_j$ has a basis of  exponentially-localized Wannier functions. Given this Wannier splitting, we would then verify (B). While this alternative approach is possible in principle, we do not know if it is practical. 
Given the above definitions,  $P=\oplus_{j=1}^N P_j$ (which satisfies both (A) and (B)) shall also be called a \emph{symmetric Wannier splitting of $P$ with respect to $G$}.\\

\noi{b} The crystallographic splitting theorem implies that any representation of a space group that is not a monomial BR cannot simultaneously satisfy conditions (A-B). In particular,  (A-B) cannot simultaneously hold for \textit{obstructed representations} -- defined as representations of a space group which are not band-representable.\\

\noi{b-i} Suppose (A) holds, giving a set of Wannier functions that span $P$, then [not (B)] manifests as an obstruction to symmetry conditions of the Wannier functions, as  we elaborate in \s{sec:wannierobstruction}.\\

\noi{b-ii} If instead (B) holds, with $P=\oplus_{j=1}^NP_j$ a symmetric splitting, then [not (A)] manifests as an obstruction to an exponentially-localized Wannier basis for $P_j$. This obstruction may manifest as a non-analyticity of $P_j$, as exemplified by the `nodal-line semimetal' in the case study of  \s{sec:fu_diagnosefragility}. Alternatively, $P_j$ may be analytic but has nontrivial first Chern class  -- this has nontrivial implications for the Zak phase of $P_j$ that is elabroated in    \s{sec:zakphasewind}.

\section{Zak phase of monomial band representations and obstructed representations}\la{sec:zakphasewind}

The crystalline generalization of Berry's phase\cite{berry1984} is known as the Zak phase\cite{zak1989} -- it encodes the holonomy of Bloch functions around loops in the Brillouin torus.  A rank-$N$ band, which consists of $N$ independent Bloch functions at each $\bk$, may then be characterized by $N$ Zak phases for each cycle. The Zak phase has increasingly been used as a diagnostic of obstructed representations -- bands which are not band representable.  \\

A priori, there is no direct relation between $\bk$-space holonomy (a geometric property of Bloch functions) and band representability (a symmetry condition on exponentially-localized Wannier functions).  For a band whose projector is analytic throughout the Brillouin zone, it is known that the non-existence of exponentially-localized Wannier functions is a necessary and sufficient condition for the nontriviality of the first Chern class;\cite{Brouder2007,Panati2007} this nontriviality {also} manifests as a nontrivial Zak phase.\cite{AA2014}   The goal of this section is to prove an analogous relation for obstructed representations with a trivial first Chern class; our proof will rely on the crystallographic splitting theorem of \s{sec:state_equivthm}.\\ 

Generally, if the Zak phase is nontrivial (in a manner that will be made precise), it is guaranteed that the band is not band-representable; this point of view has been advocated by Topological Quantum Chemistry.\cite{TQC,Cano2018,Bradlyn2019}
However, the converse statement, namely that \textit{an obstructed representation must have a nontrivial Zak phase}, has not been proven. This will be proven in \s{sec:resultzakphase} for certain space groups to be specified. Before  this result is presented, we  review basic properties of the Zak phase in \s{sec:prelimzakphase}, and also clarify the distinction between trivial vs. nontrivial Zak phases.

\subsection{Preliminaries on the Zak phase}\la{sec:prelimzakphase}

Let $\calc$ denote a loop (in the Brillouin torus) with base point $\bk$, end point $\bk+\bG$, and $\bG$ a reciprocal vector. Two $\bk$-loops which are continuously deformable into each other are said to be equivalent under homotopy. A homotopy class $[\calc]$ of $\bk$-loops is specified by the reciprocal lattice vector $\bG$ that connects the base and end points -- for any representative of $[\calc]$.\\

Given a rank-$N$ $P$ that is analytic throughout the Brillouin torus, it is always possible\footnote{Here we applied that there is no topological obstruction to analytic Bloch functions over the base space $S^1$.} to choose a basis for the Bloch functions $\{\psi_{n\bk}\}_{n=1\ldots N}$ that is (i)
analytic for all $\bk$ in the Brillouin zone, and (ii) periodic under translation by the reciprocal vector $\bG$ specifying $[\calc]$.\\

Defining ${u_{n,\bk}}(\br)=e^{-i\bk\cdot\br}\psi_{n\bk}(\br)$ as the cell-periodic component of the Bloch function, the non-abelian Berry connection is given by
\e{ \big[ \bA(\bk) \big]_{j'j} = \braket{ u_{j'\bk}}{ i \nabla_{\bk} u_{j\bk} }_{\mathrm{cell}}, \la{Bconn} }
where in $\langle \cdot | \cdot\rangle_{\mathrm{cell}}$, we integrate (or sum) over the coordinates in one unit cell. The Wilson loop of the Berry gauge field is given by path-ordered integration of $\bA$ over $\calc$:
\e{ \W(\calc) = \mathcal{P} \mathrm{exp} \big[ i \oint_{\calc} \bA(\bk) \cdot \mathrm d\bk \big]. \la{definewilsonloop}}
The spectrum of the Wilson loop is given by
\e{ \text{spec}\W(\calc) = \{ e^{i2\pi x_j(\calc)}\}_{j=1}^N,\la{definezakphase}}
with $2\pi x_j$ defined as the \textit{Zak phase}. {In general, $x_j$ depends on $\calc$ and not just on $[\calc]$.} \\

Given $P$ of rank $N$, and a homotopy class  of loops (specified by $\bG$), we say that the \textit{Zak phase of $(P,\bG)$ is trivial} if $x_j(\calc)$ is independent of the  representative choice for $[\calc]$, for all $j=1\ldots N.$  \\

\noindent \textit{Example of trivial Zak phase.} For a category of BRs that has been termed strong BRs,\cite{TBO_JHAA} their projected position operators $\{PxP,PyP,PzP\}$ mutually commute in the symmetric tight-binding limit, which would imply that the Zak phase of  $(P,\bG)$  is trivial in this limit, for $\bG=2\pi \be_x$, $2\pi \be_y$ and $2\pi \be_z$; $\be_a$ here denotes the unit vector in the $a$ direction. {Strong BRs include all BRs having only a single Wannier function on each Wannier center.} \\

If $P$ can be continuously deformed (while preserving analyticity and symmetry) such that $x_j$ is representative-independent, we  say that the Zak phase of $(P,\bG)$ is \textit{trivializable}{; in particular, a trivial Zak phase is trivializable}. A \textit{nontrivial Zak phase} is {not trivializable.} \\


For simplicity of presentation, we henceforth assume a  rectangular real-space lattice and set all lattice periods to unity. (All  results in \s{sec:zakphasewind} hold also for non-orthogonal lattices, if  one  replaces $(k_x,k_y)$ with $(\bk\cdot \bR_1,\bk\cdot \bR_2)$,  $\bR_j$ being a primitive Bravais-lattice vector.)  
To diagnose a nontrivial Zak phase for $P$ of rank $N$, we introduce the notion of winding numbers for the Zak phase. Let $[\calc]$ be specified by $\bG=2\pi \be_x$; a set of representatives for $[\calc]$ is given by $\{\calc(k_y)\}_{k_y}$; for the {straight} $\bk$-loop $\calc(k_y)$,  $k_x$ is varied while fixing  $k_y$. From \q{definezakphase}, we obtain $N$ Zak phases parametrized by $k_y$: $\{ 2\pi x_j(k_y) \}_{j=1}^N$. Since $P(\bk):=\sum_{n=1}^N\ketbra{\psi_{n\bk}}{\psi_{n\bk}}$ is analytic and periodic over the Brillouin torus, each $e^{i2\pi x_j(k_y)}$ is a smooth function in $k_y$, and when $k_y$ is advanced by $2\pi$ there is generally a permutation $\Sigma_{\bG}$ in the Zak-phase index: $e^{i2\pi x_j(k_y+2\pi)}=e^{i2\pi x_{\Sigma_{\bG}(j)}(k_y)}$. Let us define the smallest positive integer $Z_{\bG}$ such that $\Sigma_{\bG}^{Z_{\bG}}=$ identity as the \textit{Zak permutation order}; examples of which are illustrated in \fig{fig:zak}. Generally, the phase  $2\pi x_j$ may wind as $k_y$ is advanced by $Z_{\bG}$ periods; focusing on $Z_{\bG}=1$, we define the \textit{Zak winding number} $W_{j,\bG}$ through
\e{ x_j(k_y+2\pi)-x_j(k_y)= W_{j,2\pi \be_x} \in \Z. \la{windingnumber}} 

If $N=2$, we say that the Zak phase has a \textit{relative winding} if $W_{1,\bG}=-W_{2,\bG}\neq 0$. If in addition, $W_{1,\bG}=-W_{2,\bG}$ is not reducible to zero by an analytic, $G$-symmetric deformation of $P$, then we say that the relative winding is \textit{robust}.
If $W_{1,\bG}$ is odd, the Zak phase has an \textit{odd relative winding}.  \\

\begin{figure}
	\centering
	\includegraphics[width=\columnwidth]{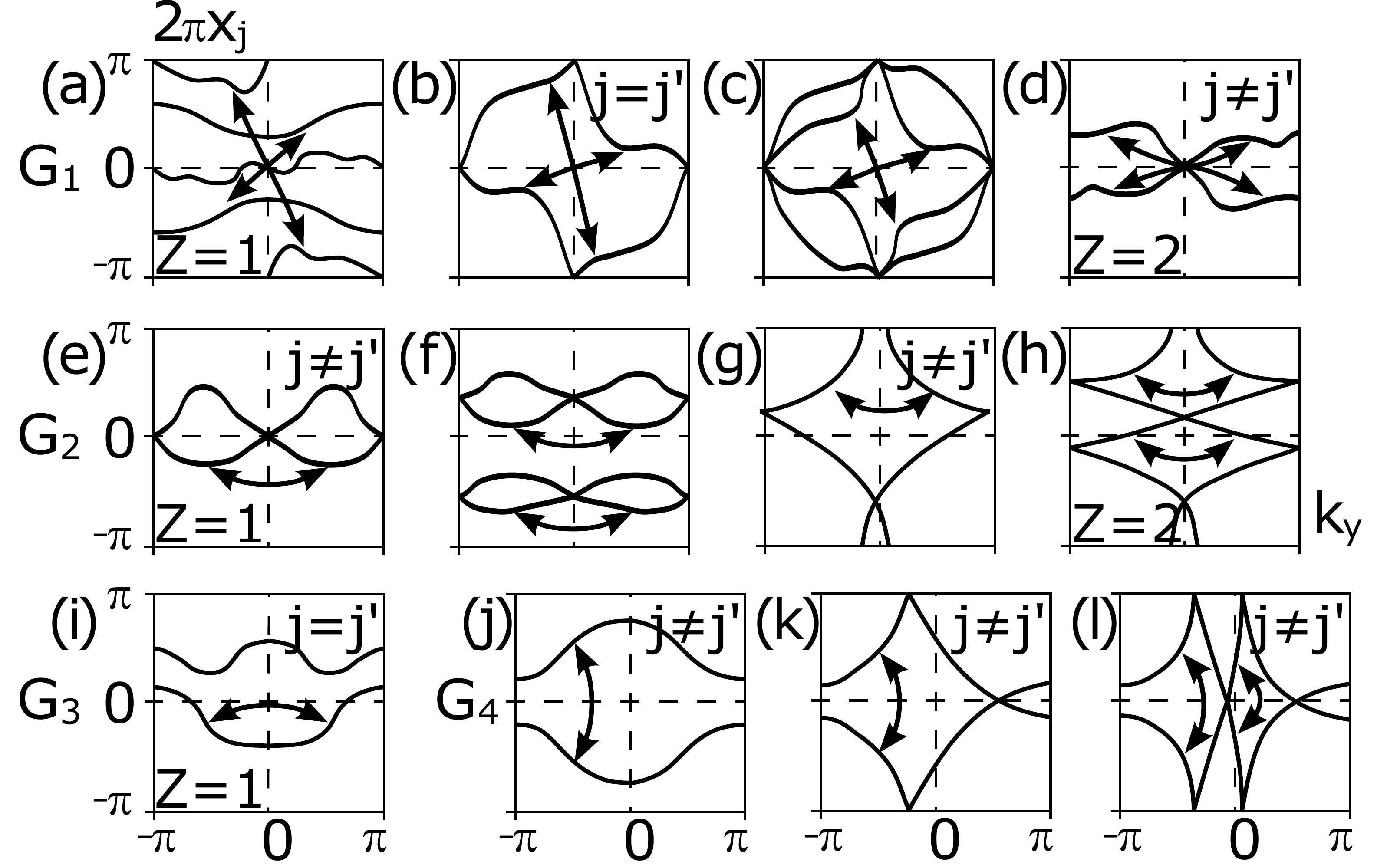}
	\caption{ Representative examples of the Zak phase for the family of $\bk$-loops $\{\calc(k_y)\}_{k_y}$, and with the space-group symmetry $G_{1,2,3,4}$, as defined in the main text. The first row has $G_1$ symmetry,  the second row has $G_2$, while in the last row the left most panel is $G_3$-symmetric, while the right three panels are $G_4$-symmetric. In each panel, the vertical axis is $2\pi x_j$ (Zak phase) and horizontal axis is $k_y$.   The nontrivial point-group symmetry in $G_j$ constrains $g_{\parallel}\circ x_{j}(k_y)=x_{j'}(s_g\check{g}_{yy}k_y)$, which is indicated by double-headed arrows. 
	For panels describing rank-two {bands}, $j,j'\in \{1,2\}$, and we distinguish $j=j'$ vs $j\neq j'$.
	We also indicate the Zak permutation order  $Z\in \{1,2\}$; each panel without an indicated $Z$ has the same value (for $Z$) as the panel to its left.
	\la{fig:zak}}
\end{figure}

Since every BR of $G$ has a symmetric tight-binding (or `atomic') limit [as proven in \s{sec:instabilitycriteria}], then $P$ being a BR implies that $W_j$ must either be zero, or reducible to zero by an analytic, $G$-symmetric deformation of $P$. This follows from the following lemma that we prove in \app{app:zakwannier}:\\

\noindent \textit{Lemma for Zak phases of tightly-bound band representations.} In the tight-binding limit of any BR, $x_j(k_y)$ becomes independent of $k_y$, for all $j$.\\

\noindent Conversely, if $W_{j,\bG}$ is neither zero nor reducible to zero, then the Zak phase is not trivializable, and $P$ cannot be a BR of $G$. This fact is used throughout this work for proving that certain $P$ are obstructed representations.  \\

Finally, we review the spectral equivalence between the Wilson loop and the projected position operator: 
\e{ \big(PxP-x_j(k_y)-R\big)\ket{h_{j,k_y,R}}=0, \as j=1\ldots N,\; R\in \Z. \la{projectedpositioneigen}}
The eigenfunctions of $PxP$ are hybrid functions that are extended in $y$ as a Bloch wave (with crystal wavenumber $k_y$), and exponentially localized in $x$ as a Wannier function (with unit cell coordinate $R$).\cite{Maryam2014} Modulo lattice translations in $x$ (with unit lattice period), the eigenvalues of $PxP$ are in one-to-one correspondence\cite{AA2014} with the Zak phases; cf.\ \q{definezakphase}. If  $x_j(k_y)$ is nondegenerate, one can uniquely define a unit-rank projector 
\e{ P^{x}_j:= \sum_{R\in \Z}\int \f{dk_y}{2\pi}\ketbra{h_{j,k_y,R}}{h_{j,k_y,R}}, \la{pjpxp}}
which gives a splitting of $P=\oplus_{j=1}^N P^x_j$. Even if  a degeneracy $x_j(k_y)=x_{j'}(k_y)$ exists at isolated $k_y$, the assumed condition $Z_{2\pi \be_x}=1$ means that we can still uniquely define $P^x_j$  by imposing that $\ketbra{h_{j,k_y,R}}{h_{j,k_y,R}}$ is smooth in $k_y$. We will refer to $P^x_j$ as the projector to a \textit{band of the projected position operator} $PxP$, and $x_j(k_y)$ as the corresponding \textit{dispersion} (assumed smooth in $k_y$).\\

Analogous to the above discussion, we may also consider a family of $\bk$-loops represented by $\{\calc'(k_x)\}_{k_x}$, and the corresponding Zak phases $\{ 2\pi y_j(k_x) \}_{j=1}^N$.  If the Zak permutation order $Z_{2\pi \be_y}=1$, then the winding numbers $W_{j,2\pi \be_y}$ are well-defined by \q{windingnumber} with $x\leftrightarrow y$, and a splitting $P=\oplus_{j=1}^N P^y_j$ is given by \qq{projectedpositioneigen}{pjpxp} also with $x\leftrightarrow y$.

\subsection{Relating the winding of the Zak phase to the crystallographic splitting theorem}\la{sec:resultzakphase}

Here we will show how the crystallographic splitting theorem constrains the winding numbers of the Zak phases, as defined in \qq{Bconn}{windingnumber}; in turn, the Zak-phase winding is related to a winding in the dispersion of the projected position operator [cf.\ \qq{projectedpositioneigen}{pjpxp}].\\

To recapitulate, the splitting theorem states a necessary and sufficient condition for a monomial BR, namely that there must exist a symmetric Wannier splitting. (We remind the reader that all BRs of two-spatial-dimensional space groups are monomial; cf.\ \s{sec:whichBRmonomial}.) We will find that a symmetric (but not necessarily Wannier) splitting is given by the bands of the projected position operator, for certain space groups that are identified by the following lemma. \\

\noindent \textbf{Symmetric splitting lemma} Let $G$ be a space group such that all $g\in G$ satisfy two conditions:

\noi{i} the action of $g$ on $\br$ decomposes as $g\circ (x,y)=(g_{\parallel}\circ x, g_{\perp}\circ y)$, such that both $g_{\parallel}$ and $g_{\perp}$ are one-dimensional isometries.

\noi{ii} $g$ does not enforce a degeneracy $x_j(k_y)=x_{j'}(k_y)$ {for $j \neq j'$}, except possibly at isolated $k_y$. 

\noindent Then $P=\oplus_{j=1}^N P^x_j$, with $P^x_j$ defined through \qq{windingnumber}{pjpxp}, is a symmetric splitting with respect to $G$. Moreover, each $P^x_j$ is analytic  in $\bk$ over the Brillouin torus. \\


The above lemma also holds with $x \leftrightarrow y$.\\

To clarify condition (i), any symmetry of a two-spatial-dimensional space group acts on spacetime as $g\circ \br = \check{g}\br+\bt_g$ and $t\rightarrow s_g t$, with $\check{g}$ a two-by-two orthogonal matrix acting on a two-component vector $(x,y)$, and $s_g=-1$ if $g$ reverses time; for a general review of space groups, we refer the reader to \app{sec:spacegroupwignerdyson}. If $\check{g}$ is a diagonal matrix with on-diagonal elements $(\check{g}_{xx},\check{g}_{yy})$ being either of $\pm 1$, then $g_{\parallel}\circ x = \check{g}_{xx} x+t_{g,x}$ and $g_{\perp}\circ y = \check{g}_{yy} y+t_{g,y}$ indeed act as one-dimensional isometries. The corresponding action on $\bk$ would also decompose into one-dimensional isometries: $k_x\rightarrow s_g\check{g}_{xx} k_x$ and $k_y\rightarrow s_g\check{g}_{yy}k_y$. We list a few representative examples of space groups satisfying  conditions (i-ii):\\

\noindent \textit{Example 1.} $G_1=\calt_2 \rtimes \Z^{\inv}_2$, with $\Z^{\inv}_2$ an order two-group generated by the spatial inversion $\inv$ [which maps $(x,k_y)\rightarrow (-x,-k_y)]$, and $\calt_2$ the translational subgroup of a two-dimensional crystal. \\

\noindent \textit{Example 2.} $G_2=\calt_2\rtimes \Z^T_4$, with $\Z_4^T$ an order-four group generated by $T$ symmetry [$(x,k_y)\rightarrow (x,-k_y)]$. $T$ squares to a $2\pi$ rotation  which is distinct from the identity element; this corresponds to Wigner-Dyson class AII.\\

\noindent \textit{Example 3.} $G_3=\calt_2\rtimes \Z^T_2$, with $\Z_2^T$ generated by $T$ symmetry; this corresponds to Wigner-Dyson class AI.\\

\noindent \textit{Example 4.} $G_4=\calt_2\rtimes \Z_2^{C2T}$, with $\Z_2^{C2T}$ generated by the composition of two-fold rotation $C_2$ with time reversal [$(x,k_y)\rightarrow (-x,k_y)]$.\\

\noindent \textit{Proof of symmetric splitting lemma.} Condition (iii) allows for $P^x_j$ to be uniquely defined, as shown in \s{sec:prelimzakphase}. Conditions (i-ii) imply that $P=\oplus_{j=1}^N P^x_j$ is a symmetric splitting; this follows from an elementary argument, which is simple to write for rank $N=2$:  for any $g\in G$, we have assumed that $x \mapsto g_{\parallel}\circ x$ and $k_y \mapsto s_g\check g_{yy} k_y$ are isometries. This implies that $P (g_{\parallel}\circ x) P\big|_{k_y}$ is unitarily equivalent to $P x P\big|_{s_g\check g_{yy} k_y}$, thus its eigenvalues satisfy $g_{\parallel}\circ x_j(k_y) \equiv x_{j'}(s_g\check g_{yy} k_y)$ (modulo integer)  with $j,j'\in \{1,2\}$.  If $j=j'$, then $g$ trivially permutes $\{P_1^x,P_2^x\}$ (the bands of $PxP$); if $j\neq j'$, then the permutation is nontrivial. Both cases are illustrated in \fig{fig:zak} for the space groups $G_{1,2,3,4}$.  It follows that any $g\in G$ acts as a permutation, hence    $P=\oplus_{j=1}^{N=2} P^x_j$ is a symmetric splitting. (The generalization of the above argument for rank $N>2$ is straightforward, and illustrated for a few examples in \fig{fig:zak}.) The analyticity of $P_j^x$,   is proven in \app{sec:projposition}. \\





\noindent \textbf{Zak winding theorem}
Assume $P$ is a rank-$N$ representation of the space group $G$, with $G$ satisfying conditions (i-ii) in the \textit{symmetric splitting lemma}, and $Z_{\bG}=1$ for either $\bG=2\pi \be_x$ or $2\pi \be_y$. Then  $P$ is a BR of $G$ if and only if all Zak winding numbers  $W_{j,\bG}=0$, or are reducible to zero by an analytic, $G$-symmetric deformation of $P$. \\ 

\noindent \textit{Proof.} If each $W_{j,\bG}=0$ then each $P_x^j$ has a trivial Chern class. The \textit{symmetric splitting lemma} implies that $P=\oplus_{j=1}^N P^x_j$ is a symmetric Wannier splitting; consequently, all conditions in the splitting theorem are met for $P$ to be a monomial BR.
To prove the converse statement, we apply the  symmetric tight-binding limit theorem [cf.\ \s{sec:instabilitycriteria}]  and the \textit{Lemma for Zak phases of tightly-bound band representations} [cf.\ \s{sec:prelimzakphase}]; together they imply that all Zak winding numbers (for BRs) are zero or reducible to zero. \\

A useful corollary of the Zak winding theorem states: \\

\noindent \textbf{Relative winding corollary}  Let  $P$ with trivial first Chern class be a  rank-two representation of space group $G$, with $G$ satisfying conditions (i-ii) in the \textit{symmetric splitting lemma}, and $Z_{\bG}=1$ for either $\bG=2\pi \be_x$ or $2\pi \be_y$. Then $P$ is an obstructed representation of $G$ if and only if there is a robust relative winding for the Zak phase of $(P,\bG)$.\\

\noindent Indeed, if $P$ is obstructed with $Z_{\bG}=1$, then $\{W_{1,\bG}, W_{2,\bG}\}$ cannot both vanish according to the Zak winding theorem. Since $P$ has trivial first Chern class, $W_{1,\bG}=-W_{2,\bG}\neq 0$, implying a relative winding of the Zak phase. 
\\

The Zak winding theorem does not say that an obstructed representation of $G$ [satisfying (i-ii) and with $Z_{\bG}=1$] always exists. If it does exist, the theorem does not say what winding numbers $W_{j,\bG}$ are allowable or robust -- these numbers can only be determined by further symmetry analysis of the Wilson loop matrix,\cite{AA2014,Cohomological,AALG_100,Wang2019a} as will be exemplified by several applications in the subsequent \s{sec:applyrelativewinding}.

\subsection{Applications of the Zak winding theorem}\la{sec:applyrelativewinding}

We briefly outline the remainder of this \s{sec:zakphasewind}:\\

\noi{i} In \s{sec:nogoAI} we will apply the Zak winding theorem to prove that no obstructed representations (fragile or stable) exist for $G_3=  \calt_2\times \Z^T_2$ (Wigner-Dyson class AI).  \\

\noi{ii} For class AII, we will prove in \s{sec:equivAII} that having $Z_2$ Kane-Mele topological order is equivalent to being an obstructed representation of $G_2=\calt_2 \times \Z^{T}_4$. \\


\noi{iii} The obstructed representations of $G_1=\calt_2\rtimes \Z^{\inv}_2$ and $G_4=\calt_2\rtimes \Z^{C2T}_2$ are discussed subsequently in \s{sec:zakinv} and \s{sec:zakc2t}, with emphasis on the possible Zak winding numbers. In \s{sec:zakinv}, we will also exemplify how the Zak winding theorem may be used as an alternative method to prove band representability, or to prove fragility for  an obstructed representation.\\

\noi{iv} We end this section by discussing the limitations of the Zak winding theorem in \s{sec:limitzakthm}, with an outlook toward possible generalizations. 

\subsubsection{Wigner-Dyson class AI}\la{sec:nogoAI}

The Zak winding theorem can be used to prove that there exists no obstructed representations of certain space groups. Indeed, for a \textit{subset} of space groups satisfying conditions (i-ii) in the \textit{symmetric splitting lemma}, it is guaranteed that $Z_{\bG}$  is reducible to unity by an analytic, symmetric deformation, and $W_{j,\bG}$ is also reducible to zero. \\

In general, $Z_{\bG}>1$ being robust requires at least one symmetry-protected degeneracy for the Zak phase, as illustrated in \fig{fig:zak}(d). If the first Chern class is trivial, $W_{j,\bG}\neq 0$ being robust also requires symmetry-protected degeneracies, because the net winding number must vanish. Whether such degeneracies exist can be determined by a symmetry analysis of the Wilson loop matrix.\cite{AA2014,Cohomological,TBO_JHAA,AALG_100,Wang2019a}  \\

Applying this analysis to  $G_3=\Z^T_2 \times \calt_2$, we find that there is no symmetry-enforced degeneracy of the Zak phase, hence the Zak permutation order is always reducible to unity. Moreover, $G_3$ ensures that the first Chern class is trivial, hence if any $W_{j,\bG}$ is nonzero, there must be other nontrivial windings such that the net sum vanishes. If two winding numbers have opposite sign, their corresponding Zak-phase functions must  necessarily be degenerate at isolated wavevectors. But we have just claimed that such degeneracies are never protected by $G_3$ alone, hence all $W_{j,\bG}$ are eventually reducible to zero. We are led to the following no-go theorem:\\

\noindent \textbf{No-go theorem for Wigner-Dyson class AI} In spatial dimension $d=2$, there exists no  obstructed representation of  $G_3=\Z^T_2 \times \calt_2$. \\

\noindent While it is known that there is no stable obstructed representation of $G_3$ from K-theoretic approaches,\cite{kitaev_periodictable} our no-go theorem goes further to say there is no fragile obstructed representation of $G_3$. Our no-go theorem is consistent with the absence of `non-stable' topological insulators in class AI, that has been derived from the equivariant homotopy properties of Real vector bundles.\cite{DeNittis_classifyAI} 

\subsubsection{Wigner-Dyson class AII}\la{sec:equivAII}

One special feature in Wigner-Dyson class AII is that the Zak permutation order $Z_{\bG}$ ($\bG =2\pi \be_x,2\pi \be_y)$ is always reducible to unity for any BR of $G_2=\calt_2 \times \Z^{T}_4$, but not necessarily for any obstructed representation of $G_2$.\\


This follows from a symmetry analysis of the Wilson-loop matrix,\cite{AA2014,yu2011} which shows that all Zak phases are pair-wise degenerate at time-reversal-invariant wavevectors ($k_y=0,\pi$ for $\bG=2\pi \be_x$, and $k_x=0,\pi$ for $\bG=2\pi \be_y$); there are no $G_2$-protected degeneracies at generic $k_x$ and $k_y$. This implies that $x_j(k_y)$
can always be reduced to two classes of graphs illustrated in \fig{fig:zak}(e-h). One class of graphs corresponds to a splitting into a direct sum of rank-two projectors with unit Zak permutation order; the trivial Zak winding then implies that $P$ is a BR of $G_2$, according to the Zak winding theorem.  The second class of graphs has a robust zigzag connectivity that has been described as a `switching of Kramers partners'\cite{fu2011} -- such a nontrivial Zak phase implies that $P$ is an obstructed representation of $G_2$, according to the symmetric tight-binding limit theorem [cf.\ \s{sec:instabilitycriteria}]  and the \textit{Lemma for Zak phases of tightly-bound band representations} [cf.\ \s{sec:prelimzakphase}]. Combining these results leads to the following theorem:\\


\noindent \textbf{Zak winding theorem for Wigner-Dyson class AII} $P$ is a band representation of $G_2=\calt_2\times \Z_4^T$ if and only if all Zak winding numbers are reducible to zero by an analytic, $G_2$-symmetric deformation of $P$. \\ 

It has been established that the two classes of Wilson-loop graphs are in one-to-one correspondence with the $\Z_2$ Kane-Mele topological invariant.\cite{yu2011,GSHI} Combining this correspondence with the  above Zak winding theorem, we derive that having $Z_2$ Kane-Mele topological order is equivalent to being an obstructed representation of $G_2$. The latter equivalence is already widely believed, but -- to our knowledge -- {our present} work presents the first proof.

\subsubsection{With spatial inversion symmetry}\la{sec:zakinv}

\begin{figure}
	\centering
	\includegraphics[width=0.8\columnwidth]{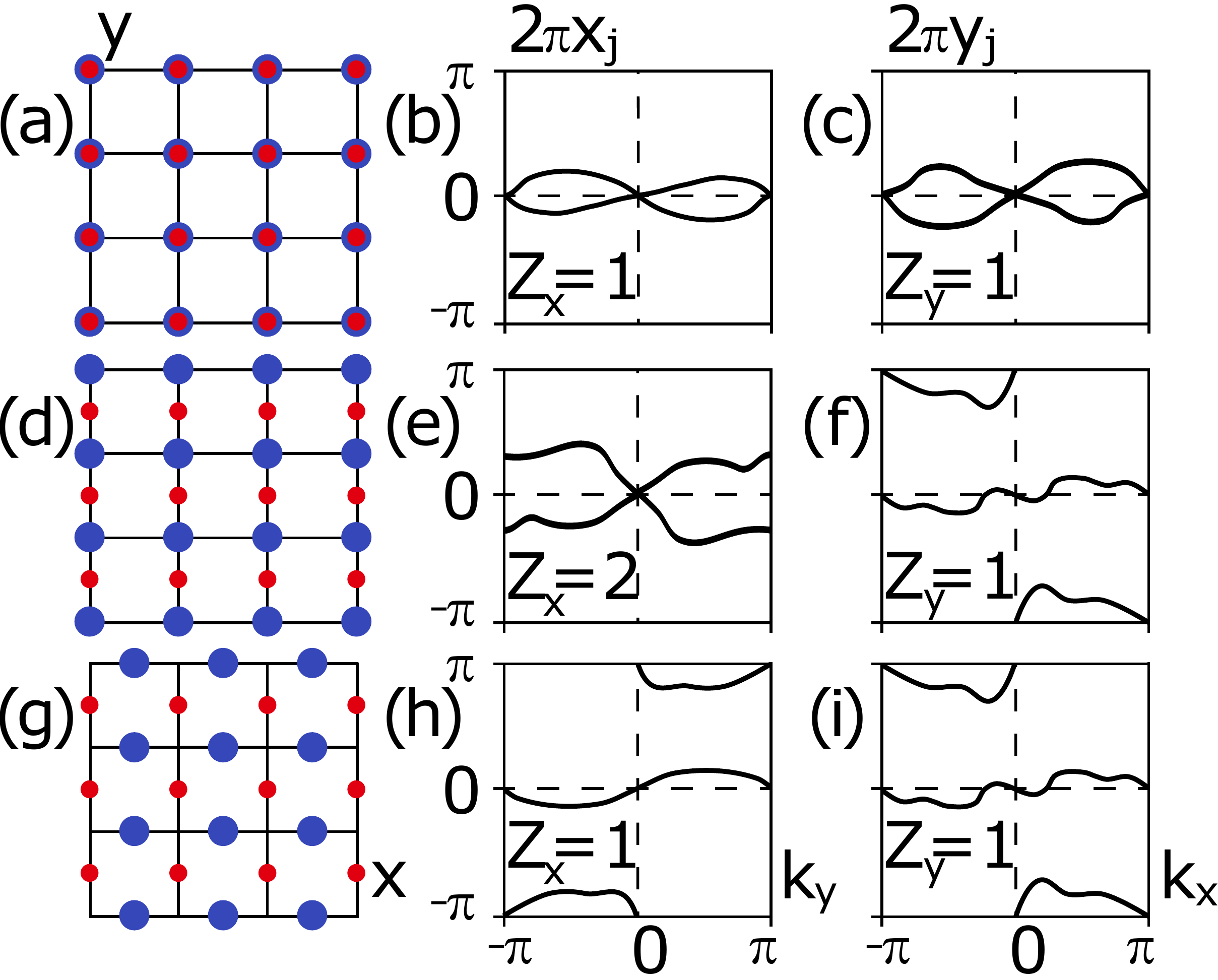}
	\caption{ For three BRs of space group $G_1$, we illustrate their real-space distribution of Wannier  centers (left column), and their corresponding Zak phases $2\pi x_j(k_y)$ and $2\pi y_j(k_x)$ (middle and right columns). Red and blue dots in the first column indicate Wannier centers for linearly independent Wannier functions with even parity. All Wannier centers lie on inversion-invariant Wyckoff positions on a rectangular lattice. 
	\la{fig:WFreal}}
\end{figure}

For the space group $G_1=\calt_2\rtimes \Z^{\inv}_2$, it is possible for the Zak phase to be symmetry-fixed to an integer multiple of $\pi$ at inversion-invariant wavevectors ($k_y=0,\pi$ for $\bG=2\pi \be_x$, and $k_x=0,\pi$ for $\bG=2\pi \be_y$); the multiplicity  of the symmetry-fixed eigenvalue depends on the symmetry representation of $P$ at $\inv$-invariant wavevectors in the Brillouin torus.\cite{AA2014,TBO_JHAA}  There are no $G_1$-protected degeneracies of the Zak phase at generic $k_x$ and $k_y$. \fig{fig:WFreal}(a-f) illustrates the Zak phases of two BRs of $G_2$, one with Zak permutation order $Z_{2\pi \be_{x}}=Z_{2\pi \be_{y}}=1$ and the other with $Z_{2\pi \be_x}=2, Z_{2\pi \be_{y}}=1.$\\

Obstructed representations of $G_1$ exist for any even rank,\cite{AA2014,Cano2018,Wieder2018,Else2019,DePaz2019} and are characterized by the Zak windings illustrated in \fig{fig:zak}(a-d) for rank $N=2$ and $4$ respectively.\\

Focusing on the case $N=2$ with an odd relative winding [cf.\ relative winding corollary], we now show the corresponding rank-two $P$ is fragile obstructed. Fragility is proven by adding a rank-two BR to $P$ and recomputing the Zak phase for the resultant rank-four subspace. The required rank-two BR is given by the Wannier representation and Zak phases in \fig{fig:WFreal}(g-i). Applying a theorem for symmetry-protected Zak phases\cite{AA2014,TBO_JHAA}, {which takes as input the $\inv$-symmetry eigenvalues in \tab{tab1}}, we derive that the four Zak-phase functions are reducible to a graph with unit permutation order and trivial winding, as illustrated in \fig{fig:zak}(a). This means that the rank-four band is a BR of $G_1$, according to the Zak winding theorem.   

\begin{table}[h!]
$\begin{array}{|c|c|c|}
\hline
  \inv(\Gamma) & \inv(X)=\inv(Y) & \inv(M) \\
  \hline \hline
  -1, -1 & 1, 1 & 1, 1 \\ 
  \hline
  1, 1 & -1, 1 & -1, -1 \\  
  \hline
\end{array}$
 \caption{For the obstructed representation of $G_1$, we give the  $\inv$ eigenvalues (of Bloch functions at $\inv$-invariant $\bk$-points $\Gamma,X,Y,M$) in the upper row. The symmetry obstruction can be removed by adding a BR with $\inv$ eigenvalues that are given in the lower row. \la{tab1} }
\end{table}

\subsubsection{With spacetime inversion symmetry}\la{sec:zakc2t}

For a rank-two obstructed representation of the space group $G_4=\calt_2\rtimes\Z_2^{C2T}$ (and also {its double cover} $\tilde{G}_4$), it is possible that the Zak winding number robustly takes on any integer value; the case of $W_{1,2\pi \be_x}=-W_{2,2\pi \be_x}=2$ is illustrated in \fig{fig:zak}(l). This robust winding follows from irremovable degeneracies (of the Zak phase) that are movable along the $k_{x}$ (or $k_y$) axis.\cite{Bouhon2018,Bradlyn2019,Wang2019a} The integer winding number has also been related to the Euler class of rank-two bundles with $C_2T$ symmetry.\cite{Ahn2019}




\subsection{Generalizations and limitations of the Zak winding theorem}\la{sec:limitzakthm}

As stated, the Zak winding theorem applies directly to space groups which satisfy conditions (i-ii) in the \textit{symmetric splitting lemma}. What of space groups (denoted $G'$) not satisfying conditions (i-ii), but containing a space subgroup $G < G'$ that does? Our theorem may then be used, in combination with a Zak-phase calculation,  to determine whether a representation $P'$ of $G'$ subduces to a  band  representation of $G$. However, it would not be possible to deduce if $P'$ is a band representation of $G'$ from a Zak-phase calculation, contrary to the illogical procedures in \ocite{Bouhon2018,Wang2019}.
This is because the splitting given by the projected position operator is symmetric under $G$ but not under $G'$.\\


In all cases of robust Zak winding\cite{yu2011,AA2014,TBO_JHAA,Bouhon2018,Cano2018,Ahn2019,Ahn2018a} that we know (some of which have been discussed in the previous \s{sec:applyrelativewinding}), the space group $G$ of the obstructed representation either satisfies (i-ii), or contains a space subgroup  that satisfies (i-ii) and is also bigger than the translational subgroup  $\calt_2 < G$. This suggests that robust Zak windings can always be rationalized by  the existence of a symmetric splitting  by the projected position operator. \\

Our Zak winding theorem is agnostic of obstructed representations ($P''$) of $G''$, if the only space subgroup of $G''$ that satisfies (i-ii) is the translational subgroup $\calt_2 < G''$. We are not aware of any robust winding of the Zak phase of $(P'',\bG)$, for any $G''$. In spite of this, it is possible that Zak windings for other families of $\bk$-loops may diagnose the obstruction in $P''$. 
As a case in point, a family of contractible, hexagonal $\bk$-loops can be used to diagnose  an obstructed representation of $G''= \calt_2 \times \Z_3^{C_3}=P3$,\footnote{The  Zak phase of the hexagonal $\bk$-loops defined in \ocite{Bradlyn2019} has a nonzero relative winding that is illustrated in the {top-left corner of} Fig.\ 8 in \ocite{Bradlyn2019}. The illustrated  Zak-phase degeneracies have unit codimension\cite{cwaa_landauquantization} owing to a Wilsonian $C_3$-rotation symmetry\cite{Cohomological} {(visualized in bottom-left corner of Fig.\ 8 in \ocite{Bradlyn2019})}. Another way to deduce the obstruction is by verifying that the band's symmetry representation in $\bk$-space is incompatible with any BR of $\tilde P3$\cite{zhida_fragileaffinemonoid}} which was previously studied  in \ocite{Bradlyn2019} with an additional reflection symmetry.

\section{Wannier functions of obstructed representations}\la{sec:wannierobstruction}

The topological triviality of an analytic band projector $P$ is equivalent to the existence of a \textit{Wannier basis}, i.e., an infinite set of exponentially-localized Wannier functions which span $P$. (In spatial dimension $d=2$ or $3$, having trivial first Chern class is a necessary and sufficient condition for topological triviality in the category of complex vector bundles. This condition is assumed henceforth in this section.)  $P$ being a representation of a space group $G$ means that the \textit{complete} set of Wannier functions is invariant under any element of $G$. (In this section we will not use the previously-developed notation which distinguishes  the different categories of space groups: crystallographic vs magnetic, integer- vs half-integer spin. Unless otherwise specified, a `space group $G$' includes all said categories.)\\

By definition, an obstructed representation of $G$ is \textit{not} a band representation of $G$, that is to say, it is {not} induced from a finite set of Wannier functions centered on a Wyckoff position $\bvarpi$ and transforming in a representation of the site stabilizer $G_{\varpi}$.  Our goal is to unpack the physical implications of this definition,   by utilizing the new perspective afforded by the crystallographic splitting theorem.  Though there is no obstruction to the {existence of} Wannier functions that are $G$-invariant as a complete set  spanning $P$, there is a subtler obstruction to $G$ permuting translation-invariant subsets of Wannier functions, as   encapsulated by the following theorem.\\

\noindent \textbf{Symmetric Wannier obstruction theorem}  Let  $P$ be a rank-$N$,  obstructed representation  of a space group $G$. Suppose $P=\oplus_{j=1}^NP_j$ is a Wannier splitting. Then the following cannot hold true, namely for all $g\in G$, $g:P_j\rightarrow P_{\sigma_g(j)}$ with $\sigma_g$ a permutation on $\{1,\ldots,N\}$. \\

\noindent The symmetric Wannier obstruction theorem follows directly from the splitting theorem of \s{sec:state_equivthm}.\\

\noindent \textit{Application to Wigner-Dyson class AII:} Suppose $P=P_1\oplus P_2$ were a rank-two BR of $\calt_2\times \Z_4^T$. Owing to our splitting theorem, time reversal $T$ must
permute $\{P_1,P_2\}$. This permutation must be nontrivial owing to the Kramers degeneracy at time-reversal-invariant wavevectors. 
If instead  $P=\sum_{j=1}^2\sum_{\bR}\ketbra{W_{j\bR}}{W_{j\bR}}$ were an obstructed representation of $\calt_2\times \Z_4^T$, then   one must relax  the nontrivial permutation condition and allow for $TP_1T^{-1}P_1\neq 0$. This relation, in combination with the Kramers orthogonality of $\hat{T}\ket{W_{1\bze}}$ and $\ket{W_{1\bze}}$,\footnote{This follows from $\hat{T}^2=-I$.} implies that time reversal has a nonlocal action on the unit-cell coordinate $\bR$ of Wannier functions: $\braket{W_{1\bR\neq \bze}}{\hat{T}W_{1\bze}}\neq 0$; in contrast, time reversal has a local action on the continuous spatial coordinate.  \\

For a rank-$N$ obstructed representation of $G$, our symmetric Wannier obstruction theorem establishes that the entirety of $G$ cannot permute $\{P_j\}_{j=1}^N$. However, it would be possible that a proper subgroup $H< G$ permutes $\{P_j\}_{j=1}^N$, if $P$ subduces to a BR of $H$. (Alternatively said, if $P$ becomes band representable when the group $G$ is relaxed to $H$, then $H$ may permute $\{P_j\}_{j=1}^N$.) Such $H$ would determine the symmetry properties of Wannier functions for an obstructed representation. Depending on $G$, the choice of $H$ may not be unique and becomes a matter of preference. \\

\begin{figure}
	\centering
	\includegraphics[width=1.0\columnwidth]{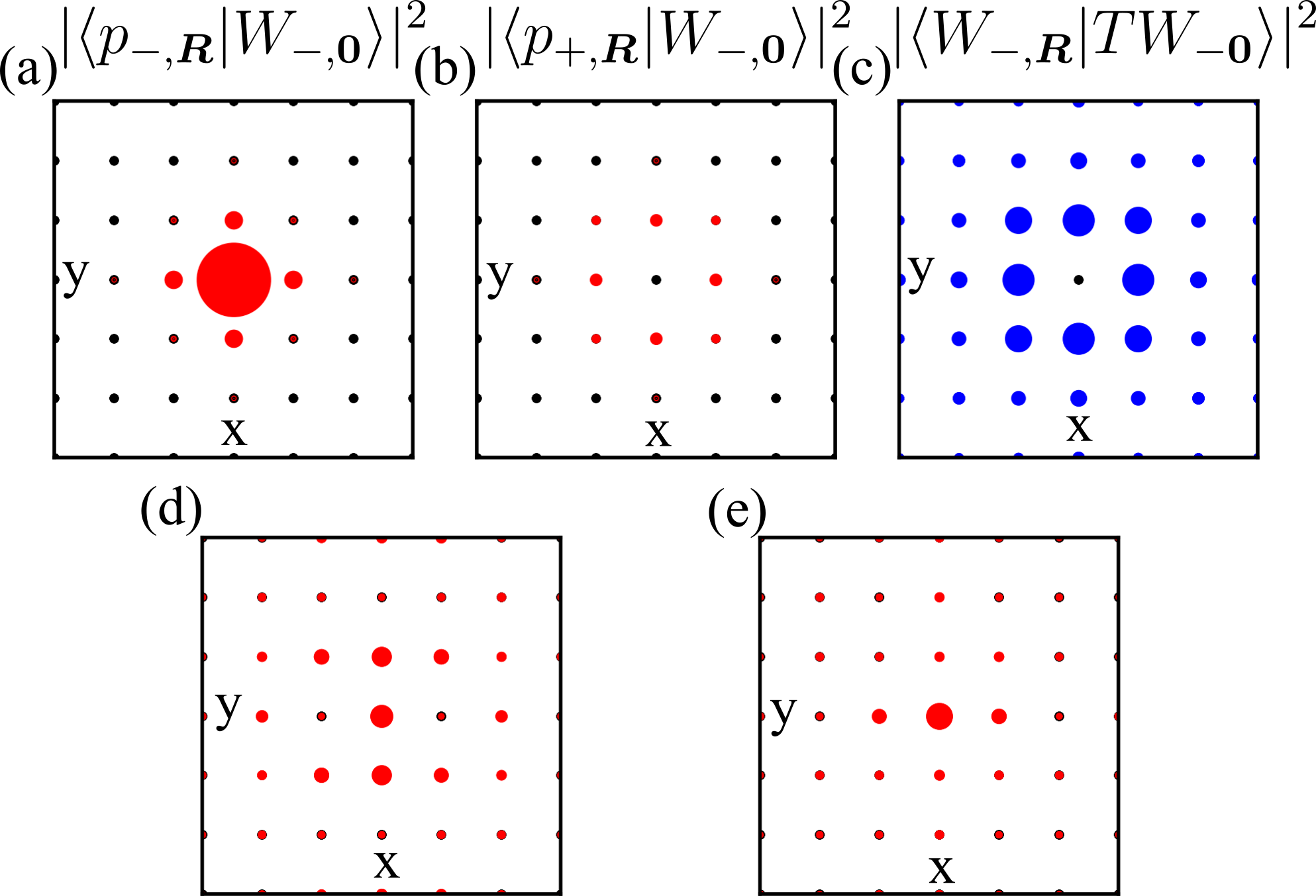}
	\caption{Panels (a-b) illustrate the four-fold symmetric Wannier function $W_{-,\bR=\boldsymbol{0}}$ constructed for the $P_-$ band of the projected symmetry operator. To illustrate the lack of pseudospin polarization [cf.\ \s{sec:spin_frustration}], the overlap of $W_{-,\boldsymbol{0}}$ with the $p_-=p_x-ip_y$ orbital [resp.\ $p_+=p_x+ip_y$] on each lattice site is indicated by the radii of red dots in (a) [resp.\ (b)]. To illustrate the nonlocal action of time reversal ($T$) symmetry on the unit-cell coordinate, the overlap between $TW_{-,\boldsymbol{0}}$ and $W_{-,\boldsymbol{R}}$ ($\bR$ being the unit-cell coordinate) is indicated by the radii by blue dots in panel (c). Alternatively, real-valued Wannier functions ($\{W_{x,\bR}\}_{\bR}$ and $\{W_{y,\bR}\}_{\bR}$)  can be constructed for the same obstructed representation; the  probability distributions of $W_{y,\bze}$ and $W_{x,\bze}$ are illustrated with red dots in (d) and (e), respectively. By inspection, the two distributions are neither individually four-fold invariant, nor mutually related by  a four-fold rotation.\label{fig:Fu_wannier}
}
\end{figure}

\noindent \textit{Example of symmetry-distinct Wannier bases for the same obstructed representation.} The non-uniqueness of $H$ applies to our case study of rotation-invariant TCIs in class AI [cf.\  \s{sec:demonstration}]. The obstructed representation ($P_{OR}$) of $\calt_3\rtimes C_{4v}\times \Z_2^T$ subduces either to a BR of $\calt_3\rtimes C_{4}$ or to a BR of $\calt_3\times \Z_2^T$. The two possible subductions correspond to two \textit{symmetry-distinct} Wannier bases for the \textit{same} obstructed representation of $\calt_3\rtimes C_{4v}\times \Z_2^T$, as we illustrate in  \fig{fig:Fu_wannier}.  \fig{fig:Fu_wannier}(a-b) shows our numerical simulation for the former type of Wannier splitting $P_{OR}=P_+\oplus P_-$, where  $P_{\pm}=\sum_{\bR}\ketbra{W_{\pm,\bR}}{W_{\pm, \bR}}$ projects to Wannier functions transforming in the vector representation of $C_4$:  $\hat{C}_4W_{\pm,\bze}=\pm i W_{\pm,\bze}$. While the four-fold rotation  acts as the trivial permutation: $[\hat{C}_4,P_{\pm}]=0$, time reversal does \textit{not} act as a nontrivial permutation. The latter implies that $T$ has a nonlocal action on the unit-cell coordinate [cf.\ \fig{fig:Fu_wannier}(c)], which rules out Wannier functions that are localized to a single lattice site; the theme of localization is explored more generally in \s{sec:localization_obstruction}. In comparison, \fig{fig:Fu_wannier}{(d-e)} illustrates the {real-valued}  Wannier functions ($W_{x,\bze}$ and $W_{y,\bze}$) of a   symmetry-distinct Wannier splitting for $P_{OR}=P_x\oplus P_y$, where time reversal acts as a trivial permutation ($[\hat{T},P_{x,y}]=0$)  but four-fold symmetry fails to act as any permutation. \\

Additionally, we describe how the symmetric Wannier obstruction theorem is applied to constrain three properties of Wannier functions -- namely their real-space localization [cf.\ \s{sec:localization_obstruction}],  their spin (or pseudospin) polarization [cf.\ \s{sec:spin_frustration}], and their  symmetry representations of the site stabilizers [cf.\ \s{sec:symmetry_frustration}]. We hope these constraints serve to guide the numerical construction of Wannier functions for topological insulators in any space group, as pioneered for the Kane-Mele topological insulator by Soluyanov and Vanderbilt.\cite{alexey2011,alexey_smoothgauge,winkler_smoothgauge}

\subsection{Localization obstruction}\la{sec:localization_obstruction}

The tension of  localizing  Wannier functions in topologically nontrivial bundles  is a recurrent theme in topological band theory.\cite{bandanalytic_cloizeaux,Thouless1984,Nenciu1991,Chen2014,Budich2014,Read2017} It is well-known that the exponential localization of Wannier functions is in one-to-one correspondence with topological triviality as a complex vector bundle.\cite{Brouder2007,Panati2007,Panati2013} \\

For \textit{tight-binding Wannier functions} which are defined in a tight-binding lattice model, one may consider a stricter form of localization, namely, that the functions vanish everywhere except on a finite number of points.  
Such Wannier functions will be said to have  \textit{compact support}. 
In recent works on the tenfold classification of topological insulators and superconductors,\cite{Dubail2015,Read2017} it was found that the only nontrivial bands that can be spanned by compactly-supported Wannier functions are those with a nontrivial winding that occurs in the same symmetry class in one spatial dimension. Their result was derived assuming discrete translational symmetry, but not assuming any other crystallographic spatial symmetry. \\    

Our symmetric Wannier obstruction theorem allows us to formulate an analogous obstruction -- to localization --  that applies to bands with crystallographic symmetry. We consider an even stricter form of localization for tight-binding Wannier functions, namely \textit{one-site localized} Wannier functions that have support only on a single tight-binding lattice site [cf.\ \s{sec:norobustsurface}]. \\

\noindent \textbf{Localization obstruction lemma} Suppose an obstructed representation  of a space group has a basis of exponentially-localized Wannier functions.  Then it is not possible for all Wannier functions to be one-site localized.\\

\noindent Postponing a general proof of the lemma to \app{app:locobslemma}, we offer here an  elementary version of the proof -- for a specific space group -- to develop intuition.\\

\noindent \textit{Example: $\Z_2$ Kane-Mele topological insulator.} Let $P$ be an obstructed representation of $\calt_2\times \Z_4^T$. Suppose on the contrary that $P=\sum_{j=1}^2\sum_{\bR}\ketbra{W_{j\bR}}{W_{j\bR}}$ has a Wannier basis in which all Wannier functions are one-site localized. Since the representation $\hat{T}$ of time reversal squares to minus identity,  $\hat{T}{W_{1\bR}}$ must be orthogonal to ${W_{1\bR}}$. Since all Wannier functions are one-site localized, and time reversal is a spatially-local operation, $\hat{T}{W_{1\bR}}$ must have zero overlap with any Wannier function $W_{1\bR'\neq \bR}$ -- hence $\hat{T}{W_{1\bR}}$ must equal $W_{2\bR}$ up to a phase. This being true for all $\bR$ implies that $T$ nontrivial permutes $P_1$ and $P_2$, hence $P=P_1\oplus P_2$ is a symmetric Wannier splitting -- in contradiction with $P$ being an obstructed representation. \\

 
A few remarks are in order. \\

\noi{i} The impossibility of one-size localization (for all Wannier functions of obstructed representations) allows for the possibility of spectrally robust boundary/domain-wall states [cf.\ \s{sec:norobustsurface}], as exemplified by the Kane-Mele topological insulator. \\

\noi{ii} One may relax the one-site localization condition  to a less stringent condition that all Wannier functions have compact support, with no two Wannier functions (centered on different positions) having intersecting support. In fact the localization obstruction lemma also holds with this generalized localization condition, as can be proven by following essentially the same steps in the proof of \app{app:locobslemma}.

\subsection{Spin and pseudospin frustration }\la{sec:spin_frustration}

\subsubsection{Wigner-Dyson class AII}

Let $\Z_4^T$ be the order-four group generated by time reversal (Wigner-Dyson class AII), and $\calt_d$ the translational subgroup of a $d$-dimensional crystal ($d=2,3$).\\

\noindent \textbf{Spin frustration corollary} Let $P$ be a rank-two, obstructed representation  of $\calt_{d}\times \Z_4^T $. Then for any Wannier basis of $P$, it is not possible that a Wannier function is fully spin polarized (along any spin quantization axis).  \\

\noindent \textit{Proof of spin frustration corollary.} Let $P=\sum_{j=1,2}P_j=\sum_{j=1,2}\sum_{\bR}\ketbra{W_{j\bR}}{W_{j\bR}}$ satisfy all premises stated in the corollary. Suppose $W_{1\bze}$ were fully spin polarized, then by translational symmetry any Wannier function $W_{1\bR}$ in $P_1$ is likewise fully spin polarized. Since time reversal $T$ inverts spin (whichever the quantization axis), $TP_1T^{-1}$ must be orthogonal to $P_1$. Since $P$ is a representation of $\calt_{d}\times \Z_4^T $ (which includes $T$ symmetry), $TP_1T^{-1}$ must belong in $P$. Given that $P$ is rank-two, we may identify $P_2=TP_1T^{-1}$, hence $T$ symmetry acts as a nontrivial permutation on $\{P_1,P_2\}$. Our splitting theorem then states that $P$ must be a BR of $\calt_{d}\times \Z_4^T $, which contradicts the premise in the corollary.\\

One implication of the spin frustration corollary may be deduced from an elementary argument, if one 
assumes that $\calt_{d=2}\times \Z_4^T $-symmetric $P$ has additionally a $U(1)$ symmetry for the conservation of the spin component  $S_z$.
We present this argument to develop intuition, as well as to establish a relation with the  `spin Chern number', as formulated for an infinite sample without boundaries.\cite{prodan_spinchern}
As proven in \s{sec:applyrelativewinding}, an obstructed representation  of $\calt_{2}\times \Z_4^T $  must have $\Z_2$ Kane-Mele topological order. With the addition of $S_z$ symmetry, the Kane-Mele phase can be split into two unit-rank bands with opposite $S_z$ and opposite Chern numbers (which are necessarily odd); the latter are known as \textit{spin Chern numbers}.\cite{prodan_spinchern} Due to the topological nontriviality of each unit-rank band in the $S_z$ basis, a Wannier basis can only be constructed from linearly combining Bloch functions with different $S_z$. We emphasize that our spin frustration corollary makes a stronger statement in three regards: (i) if only one spin component (e.g., $S_z$) is conserved, the Wannier function cannot be polarized along any spin quantization axis, and not just $S_z$. This spin frustration (ii)  holds even if not one spin component is conserved, and (iii) applies also to the three-spatial-dimensional $\Z_2$ topological insulators. \\

We offer a physical interpretation for spin frustration. It is often said that the $\Z_2$ Kane-Mele obstructed representation requires spin-orbit coupling. (Indeed, if such coupling were absent, spin $SU(2)$ and time-reversal symmetries enforce that the spin Chern number vanishes, which implies the trivial phase in the $\Z_2$ classification.) In solids, spin-orbit coupling is predominantly described in the $\bk$-space perspective,\cite{winklerbook} with reference to how the spin of a Bloch state is locked to its momentum.\cite{Hsieh_tunable} In complementarity, we may view spin frustration  as a manifestation of the topology-enforced spin-orbit coupling -- in the real-space, Wannier perspective.  \\

There is a second interpretation of the spin frustration corollary that emphasizes a relation with the mirror Chern insulator.\cite{teo2008} We consider the mirror operation $\mir$ that maps the spatial coordinate $(x,y,z)\rightarrow (x,y,-z)$, and rotates spin by a $\pi$ angle about $z$. If restricted to the $z=0$ plane, $\mir=e^{-i\pi S_z/\hbar}=-iS_z$ becomes a spatially-local operation {in $x$ and $y$}. This means  that the spin frustration corollary can, in spatial dimension $d=2$, be viewed as the impossibility for a Wannier function to transform in a definite representation of $\mir$. Such an obstruction is already known in case $\mir$ is a symmetry of $P$, i.e., if $P$ is the filled band of a mirror Chern insulator.\cite{nogo_AAJH} The novel implication of our corollary is that this obstruction persists even where $\mir$ is not a symmetry.

\subsubsection{Wigner-Dyson class AI}

We present an analog of the spin frustration corollary that applies to integer-spin representations of time reversal (Wigner-Dyson class AI), as well as to grey magnetic space groups with a nontrivial crystallographic point group. We remind the reader that a grey magnetic space group is expressible as $G\times \Z_2^T$, with $G$ a crystallographic space group (without time-reveral symmetry) and $\Z_2^T$ an order-two group generated by time reversal. \\


To formulate an analog of spin polarization in class AI, we  utilize Wigner's seminal classification\cite{Wigner_ontheoperationoftimereversal,wignerbook,tinkhambook} of crystallographic point-group representations as real, complex and quaternionic; this classification  is briefly  reviewed in \app{typeI}. A one-dimensional representation is real or complex; if real, it is $T$-invariant; if complex, it is not $T$-invariant, and must be paired up with its complex-conjugate representation in the presence of $T$ symmetry, i.e., the pair forms a two-dimensional (`pseudospin') representation. \\

\noindent\textit{Example of pseudospin.} As we have  encountered in \s{sec:futci}, the two-dimensional irreducible representation of the point group $C_{4}\times \Z_2^T$ is the direct sum of two complex representations, which transform like $p_x\pm ip_y$ orbitals.\\

Let us formulate a notion of pseudospin polarization for  Wannier functions in a tight-binding model. The tight-binding vector space  is generally spanned by  one-site localized Wannier functions transforming as a BR of $G\times \Z_2^T$; for simplicity we consider all basis Wannier functions (in one unit cell) to be one-site localized on a single position $\bvarpi$, with associated site stabilizer $G_{\varpi}$; by applying the translational subgroup $\calt_d < G$ on $\bvarpi$ (the Wyckoff position), we generate  the \textit{tight-binding lattice}. Let $D$ be  a one-dimensional complex representation of the site stabilizer $G_{\varpi}$ of a crystallographic space group $G$. We say that a tight-binding Wannier function $W$ is \textit{polarized with respect to} $(G,\bvarpi,D)$, if for all  sites $\{g\circ \bvarpi| g\in \calt_d\}$ related to $\bvarpi$ by Bravais-lattice translations, the restriction of $W$ to $g\circ \bvarpi$ transforms in a representation of $G_{g\circ \bvarpi}$ that is isomorphic to $D$; note  $G_{g\circ \bvarpi} \cong G_{\bvarpi}$ are isomorphic as groups. \\

\noindent \textbf{Pseudospin frustration corollary} Let  $P_{\calh}$  project to  a tight-binding vector space, which transforms as a band representation of $G\times \Z_2^T$ with the Wyckoff position $\bvarpi$. Let $P\subset P_{\calh}$ be a rank-two obstructed  representation  of $G\times \Z_2^T$, with the Wannier splitting $P=P_1\oplus P_2$. Then it is not possible that $P_1=\sum_{\bR}\ketbra{W_{1\bR}}{W_{1\bR}}$  represents $G$ with $W_{1\bze}$ that is polarized with respect to $(G,\bvarpi,D)$, for any $D$ that is a one-dimensional complex representation of $G_{\bvarpi}$. \\  

\noindent \textit{Proof of corollary.} If $W_{1\bze}$ were {polarized with respect to} $(G,\bvarpi,D)$, then any Wannier function $W_{1\bR}$ (in $P_1$) is likewise polarized, owing to the translational symmetry of $P_1$. Since time reversal maps each $D$ representation to its complex conjugate $\bar{D}$, each Wannier function in  $TP_1T^{-1}$ must be polarized with respect to $(G,\bvarpi,\bar{D})$. Therefore $TP_1T^{-1}$ must be orthogonal to $P_1$, further implying that $T$ acts as a nontrivial permutation on $\{P_1,P_2\}$. Given that both $P_1$ and $P$ represent $G$, so must $P_2$, hence any $g\in G$ acts as the trivial permutation on $\{P_1,P_2\}$. In combination, all $g\in G\times \Z_2^T$ acts as a permutation on $\{P_1,P_2\}$, which implies $P$ is a BR of $G\times \Z_2^T$ -- in contradiction with our premise.\\

\noindent \textit{Application to fragile obstructed crystalline insulator.} Let $P_{OR}$ be an obstructed representation of $\calt_3\rtimes C_4 \times \Z_2^T$. A tight-binding model with a $C_4$-invariant Wykcoff position $\bvarpi$ was first proposed by Liang Fu\cite{fu2011}, and is reviewed in \s{sec:futci}. Applying the pseudospin frustration corollary, we find there does not exist a Wannier splitting $P_{OR}=P_+\oplus P_-$ with $P_{\pm}=\sum_{\bR}\ketbra{W_{\pm \bR}}{W_{\pm \bR}}$ representing $\calt_3\rtimes C_4$, and $W_{j\bze}$ being polarized with respect to $(\calt_3\rtimes C_4,\bvarpi,D)$, where $D$ is the complex representation (e.g., $p_x+ip_y$) of the site stabilizer $C_4$. For illustration, we decomposed the Wannier function of $P_-$  into $p_x-ip_y$ and $p_x+ip_y$ orbitals, in \fig{fig:Fu_wannier}(a) and (b) respectively.

\subsection{Symmetry frustration }\la{sec:symmetry_frustration}

Certain symmetry representations of site stabilizers are impossible for the Wannier functions of obstructed representations -- we refer to this as a \textit{symmetry frustration} for Wannier functions.\\

\noindent \textit{Example 1: inversion-symmetric fragile obstructed insulator.} As a case in point, consider the space group $G_1=\calt_2 \rtimes \Z^{\inv}_2$, with  $\Z^{\inv}_2$ being the order-two group generated by spatial inversion $\inv$  symmetry, and $\calt_2$ the translational subgroup of a 2D lattice. A rank-two obstructed representation ($P'_{OR}$) of $G_1$  was proven in \s{sec:applyrelativewinding} to have odd relative winding of the Zak phase. The symmetry frustration manifests in the following way: for any Wannier basis of $P'_{OR}$, it is not possible for any single Wannier function to represent a site stabilizer that is isomorphic to $\Z_2^{\inv}$. This result is an application of the following corollary.\\

\noindent \textbf{Symmetry frustration corollary.} Let $P$ be a rank-$N$, obstructed representation  of a space group $G$. Assume $P$ has a tight-binding Wannier basis where the $N$ linearly-independent Wannier functions in one unit cell are centered at $\{ \br_j\}_{j=1}^N$, with each site stabilizer $G_{\br j}$ being isomorphic to the point group of $G$. Then the following cannot hold for any order-$(N-1)$ subset of  $\{1\ldots N\}$, namely that the Wannier function centered at $\br_j$ transforms in a one-dimensional representation of $G_{\br j}$. \\

\noindent \textit{Proof of corollary.} Given $P=\sum_{j=1}^N\sum_{\bR}\ketbra{W_{j\bR}}{W_{j\bR}}$ and $J$ that is an order-$(N-1)$ subset of  $\{1\ldots N\}$,  suppose on the contrary that for $j\in J$,  $W_{j\bze}$ that is centered at  $\br_j$ transforms in a one-dimensional representation of $G_{\br j}$. Since $G_{\br j}$ is isomorphic to the point group of $G$, the extension of $G_{\br j}$ by the translational subgroup $\calt_d < G$ simply gives $G=\calt_d\rtimes G_{\br j}$.\footnote{In more detail, $G_{\br j}$ being isomorphic to $G/\calt_d$ means that each element of $G_{\br j}$ is a  representative of an equivalence class in the coset $G/\calt_d$; distinct elements of  $G_{\br j}$ correspond to different equivalence classes in the coset. Therefore, any element in $G_{\br j}$ is the composition of an element in  $G_{\br 1}$ with a Bravais-lattice translation; moreover, the extension of $G_{\br j}$ (by $\calt_d$) is identical to the extension of $G_{\br 1}$; this holds independent of $j$.}  
It follows that {for $j\in J$}, $P_j=\sum_{\bR}\ketbra{W_{j\bR}}{W_{j\bR}}$ is invariant under all elements of $G$.\footnote{To prove this, it is convenient to set the spatial origin at $\br_j=\bze$, such that any element in $G_{\br j}$  is a point-preserving transformation $(\bze|\check{g})$ without any translational component. Any element of $G=\calt_d\rtimes G_{\br j}$ can then be written as $(\bR|\check{g})$ for some $(\bze|\check{g})\in G_{\br j}$ and some $\bR$ in the Bravais lattice. It being assumed that $(\bze|\check{g})\ket{W_{j\bze}}=\lambda_{\check{g}}\ket{W_{j\bze}}$ with $\lambda_{\check{g}}\in U(1)$, and applying further the translational property $\ket{W_{j\bR}}=(e|\bR)\ket{W_{j\bze}}$, it follows that $(\bR|{\check{g}})\ket{W_{j\bR'}}=\lambda_{\check{g}}\ket{W_{j,\bR+\bR'}}$. Finally we obtain the desired result: 
$(\bR|{\check{g}})P_j(\bR|{\check{g}})^{-1}=\sum_{\bR'}\ketbra{W_{j,\bR+\bR'}}{W_{j,\bR+\bR'}}= P_j$}
Since by assumption this invariance holds also for $P$, it must be that $G$ acts as the trivial permutation on $\{P_1,\ldots,P_N\}$, implying $P=\oplus_{j=1}^N P_j$ is  BR of $G$, and contradicting our premise.\\


\noindent \textit{Example 2: rotation-symmetric fragile obstructed insulators.} In $d=2$, two-fold rotation $C_2$ and spatial inversion $\inv$ act identically on integer-spin representations, hence the conclusions in \textit{Example 1} carry forward with $\inv$ replaced by $C_2$.  (However, the conclusions of \textit{Example 1} are more generally applicable to half-integer-spin representations.) 
A rank-two, obstructed representation of $\tilde P3=\calt_2\rtimes \tilde C_3$ exists, with the symmetry-frustration property that its Wannier functions  cannot represent a site stabilizer isomorphic to $\tilde C_3$. This obstructed representation has been realized by tight-binding models with symmetry that is higher than $\tilde P3$, namely $\tilde P6mm$\cite{fragile_po}, $\tilde P31'$\cite{Bouhon2018} and  $\tilde P3m1$\cite{Bradlyn2019}. However, the additional symmetries are superfluous to the $C_3$-symmetry obstruction for Wannier functions, as proven through a holonomy argument in \s{sec:limitzakthm}. It can further be shown that the obstructed representation of $\tilde P_3$ is fragile, by the {numerical procedure used} in \ocite{fragile_po}.

\section{Ansatz-free approach to symmetric Wannier functions} \la{app:constructWFs}

Given $P$ that is a {monomial} band representation of a space group $G$, we would like to construct a locally-symmetric Wannier basis for $P$, without having to postulate trial Wannier functions. (What it means for a Wannier basis to be locally-symmetric is reviewed in \app{app:locallysymmetricWannierbasis}.)\\

We first obtain a symmetric Wannier splitting $P=\oplus_{i=1}^NP_j$, which is guaranteed to exist by the crystallographic splitting theorem. Depending on $G$, such a splitting may be obtained from bands of the projected symmetry or position operator, as described in \s{sec:introprojsymmetry}, \s{sec:resultzakphase} and \app{app:methods_symmetricdecomp}. The symmetries of each $P_j$ form a group that we denote as $G_j:=\{g\in G|[\hat{g},P_j]=0\}$. \\

The next step is to find a Bloch function $\psi_{j\bk}$ that spans $P_j$ at each $\bk$, with the property that $\psi_{j\bk}$ is periodic over and analytic throughout the Brillouin torus. Such a Bloch function is guaranteed to exist because each $P_j$ (of a Wannier splitting) is analytic and has trivial first Chern class. Such a Bloch function can be obtained by the parallel-transport procedure described in \ocite{alexey_smoothgauge}, where it is described as a `smooth gauge'.\\

The last step is to perform a $U(1)$ phase transformation $\psi_{j\bk}\rightarrow \psi_{j\bk}e^{i\varphi_j(\bk)}:=\tilde{\psi}_{j\bk}$,
with $e^{i\varphi_j(\bk)}$ that is periodic and analytic in $\bk$,
such that $\tilde{\psi}_{j\bk}$ becomes \textit{canonically symmetric}. By this, we mean that every element $g=(\bt_g|\check{g})$ in the site stabilizer $G_{j,\bvarpi_j} :=\{g\in G_j|g\circ\bvarpi_j=\bvarpi_j\}$ acts on the
Bloch function as\footnote{Any element in $G_j$ can be expresses as  a product of an element of $G_{j,\bvarpi_j}$ and a lattice translation. Therefore, Eq.\ \ref{definemanifestlysymmetricbloch} implies $\hat{g}\tilde{\psi}_{j\bk} = \mathrm{e}^{-i s_g \check g \bk \cdot (g \circ \bvarpi_j - \bvarpi_j)} \rho_{g,j} \, \tilde{\psi}_{j s_g \check g \bk}$ for $g\in G_j$, which has been shown in \ocite{nogo_AAJH} to be a sufficient condition for band representability.}
\begin{equation}
\hat{g}\tilde{\psi}_{j\bk} = \rho_{g,j} \, \tilde{\psi}_{j s_g \check g \bk},\label{definemanifestlysymmetricbloch}
\end{equation}
where $s_g=-1$ if $g$ inverts time, and otherwise $s_g=+1$.  $\bvarpi$ can be determined, modulo Bravais-lattice translations, by computing the Brillouin-zone average of the Berry connection, in accordance with the geometric theory of polarization.\cite{kingsmith1993} $\rho_{g,j}$ is a $U(1)$ phase factor determined by the action of $g$ on the Wannier function obtained by Fourier transform of $\tilde{\psi}_{j\bk}$: 
\begin{equation}
\tilde{W}_{j\bR}:= \int_{BZ} d\bk \f{e^{-i\bk \cdot \bR}}{\sqrt{|BZ|}}  \tilde{\psi}_{j\bk}, \;\;\;\; \hat{g}\tilde{W}_{j\boldsymbol{0}} = \rho_{g,j}\tilde{W}_{j\boldsymbol{0}},\label{definemanifestlysymmetricwannier}
\end{equation}
with $|BZ|$ the volume of the Brillouin zone. 
The advantage of canonically symmetric Bloch functions is that the Wannier functions $\{\tilde{W}_{j\bR}\}_{\bR \in BL}$ form a locally-symmetric Wannier basis for  a band representation of $G_j$,\cite{Zak1979,nogo_AAJH} thus $\{\tilde{W}_{j\bR}\}_{j\in \{1\ldots N\},\bR \in BL}$ gives the desired locally-symmetric Wannier basis for $P$, a monomial band representation of $G$.\\

The existence of a canonically symmetric  Bloch function [cf.\  \q{definemanifestlysymmetricbloch}] has been rigorously proven in \ocite{nogo_AAJH}, for any unit-rank band with analytic projector, trivial first Chern class, and the symmetry of a symmorphic space group. (A symmorphic space group is a semidirect product of its translational subgroup and its point group, as reviewed in \app{sec:spacegroupwignerdyson}.) 
We are not aware that any nonsymmorphic space group allows for unit-rank bands,\cite{Michel1999,Parameswaran2013,Po2016,Watanabe2017a} so  we assume henceforth that $G_j$ is symmorphic; it is not necessary, however, to assume $G$ is symmorphic.  While \q{definemanifestlysymmetricbloch} exists in principle, we now present an algorithm that 
inputs an analytic, periodic Bloch function $\psi_{j\bk}$, and outputs a Bloch function $\tilde{\psi}_{j\bk}$ that is analytic, periodic and canonically symmetric.\\

\noindent \textbf{Symmetrization algorithm} For any $g\in G_{j,\varpi_j}$ we define
\e{\tilde{\psi}_{j\bk}= \f{1}{|G_{j,\bvarpi_j}|}\sum_{g \in G_{j,\bvarpi_j} } \rho_{g,j}^{-1} \, \hat{g} \psi_{j s_g \check g^{-1} \bk}, \la{symmetrization}}
with $|H|$ denoting the order of a finite group $H$. One may verify that \q{symmetrization} indeed satisfies \q{definemanifestlysymmetricbloch} for all $g \in G_{j,\bvarpi_j}$. \\

Applying the symmetrization algorithm to the three bands of the projected rotation operator [cf.\ \s{sec:fu_diagnosefragility}], we obtain a locally-symmetric Wannier basis for the rank-three band representation of $\calt_3\rtimes C_{4v}\times \Z_2^T$, which is illustrated in Fig. \ref{fig:Fu_phase}.

\section{Fragile topological photonic crystals}\la{sec:photonic}

Time-reversal-invariant topological photonic and phononic crystals (with a full energy gap) have recently emerged that emulates the spin-orbit-coupled Kane-Mele $\Z_2$ topological insulator.\cite{Hafezi2011,Khanikaev2012,tzuhsuan_guidingEMwaves,Slobozhanyuk_3DalldielectricphotonicTI,Wu2015,Wang2019} By exploiting an analogy between the electronic spin and a photonic pseudospins, much progress has been made in the design and construction of fully-gapped topological photonic crystals.
The practical success of this analogy has obscured the correct topological classification of these photonic crystals, which relies on a precise group-theoretic treatment of photonic band structure.\\

Photons transform in the integer-spin representation of  crystallographic spacetime symmetries. Therefore, time-reversal-invariant photonic crystals lie in Wigner-Dyson symmetry class AI  and not AII. This distinction is crucial: in class AII, there exists electronic topological insulators whose filled bands transform as obstructed representations (of a space group $G$), regardless  of the addition of any BR (of $G$) to the filled-band subspace. More generally stated, these are obstructed representations which are \textit{not} fragile obstructed; they will be referred to as \textit{stable obstructed}. (A stable obstructed representation is nontrivial in the stably-equivalent classification of $G$-equivariant K-theory.\cite{kitaev_periodictable,FreedMoore_twistedequivariantmatter,shiozaki_review,bandcombinatorics_kruthoff})  A  paradigmatic example is the Kane-Mele $\Z_2$ topological insulator. In contrast,  all \textit{known} topological insulators in class AI are fragile obstructed. Moreover, it has been argued that every topological insulator is adiabatically deformable to a `topological crystal',\cite{Song2018} which would imply that \textit{all} topological insulators in class AI are fragile obstructed.\\

The distinction between fragile vs stable is not just academic. If a fragile obstructed representation (FOR) is accompanied by in-gap boundary states, the possibility of FOR $\oplus$ BR=BR' makes the in-gap boundary states less robust than might naively be expected -- from tight-binding or $\bk\cdot \bp$ methods. (We shall be concerned with the possibility of \textit{spectrally robust} boundary states, that are irremovable from the energy gap by any continuous deformation that preserves both gap and symmetry.\footnote{What is meant by preserving the symmetry is clarified in \s{sec:instabilitycriteria}.}) In practice, this means that a great majority of topological insulators and gapped photonic crystals (in class AI) do not have spectrally robust boundary states -- a perspective that we explore generally in  \s{sec:norobustsurface} and more specifically in   \s{sec:boundaryftifphc}.\\

While \s{sec:norobustsurface} contains general arguments for the non-robustness of boundary states, more specific arguments  have been given for fragile obstructed representations of space group $\calt_3\rtimes C_{nv} \times \Z_2^T$ ($n=3,4$), whose accompanying boundary states manifest a representation-dependent stability [cf.\ \s{sec:futci}].
In the present section we prove that (a) a tetragonal photonic crystal designed by Ochiai realizes the fragile obstructed representation of $\calt_3 \rtimes C_{4v}\times\Z_2^T$ [cf.\ \s{sec:tetragonal}], and (b) a hexagonal photonic crystal built by Yihao et. al. realizes the  fragile obstructed representation of $\calt_3 \rtimes C_{3v}\times\Z_2^T$ [cf.\ \s{sec:hexagonal}]. Finally in \s{sec:hexagonaldomainwall}, we prove the  spectral non-robustness of the observed domain-wall states\cite{Yang2019} of the hexagonal photonic crystal.

\subsection{Topological classification of tetragonal photonic crystal} \la{sec:tetragonal}

\begin{figure}
	\centering
	\includegraphics[scale=1.0]{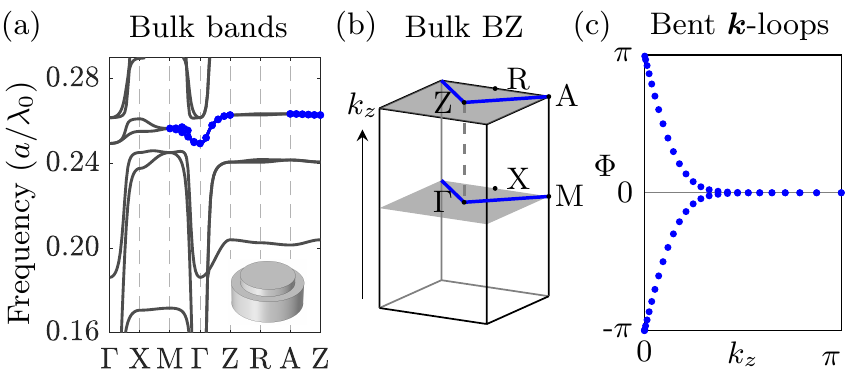}
	\caption{(a) Bulk band structure of the tetragonal photonic crystal, in which $a$ is the lattice constant, $\lambda_0$ is the vacuum wavelength and $a/\lambda_0$ is the dimensionless normalized frequency. The middle rank-two band, indicated by blue dots, is denoted $Q_{phc}$. The inset in panel (a) illustrates the high-refractive-index pillar in one real-space unit cell of the photonic crystal. (b) The bulk Brillouin zone of the tetragonal photonic crystal. The bent $\bk$-loops $\calc(k_z)$, for $k_z=0$ and $\pi$, are illustrated as blue lines. (c) Zak phases ($\Phi$) of $Q_{phc}$ for the family of bent $\bk$-loops. \label{fig:tetragonal_phc}}
\end{figure}

The 3D tetragonal photonic crystal  designed by Tetsuyuki Ochiai is composed of an array of circular pillars with high refractive index. A geometrical anisotropy of the pillar  breaks spatial inversion symmetry ($\inv$) and reduces the space group to $\calt_3 \rtimes C_{4v}\times\Z_2^T$.\cite{Ochiai2017} A secondary effect of the anisotropy is to introduce an energy gap between the lowest rank-three band, and an energetically-isolated  rank-two band ($Q_{phc}$)  illustrated in the middle of \fig{fig:tetragonal_phc}(a). \\

If the photonic crystal is terminated by a $\calt_2 \rtimes C_{4v}\times\Z_2^T$-symmetric surface with the boundary condition of a perfect electric conductor (zero surface-parallel electric field), Ochiai found evanescent eigen-solutions to Maxwell's equations which are localized to the surface. For the specific termination chosen by Ochiai, the eigen-energies of these surface states  cover the bulk energy gap below $Q_{phc}$, and their energy-momentum dispersion is qualitatively equivalent to Liang Fu's prediction for the rotation-invariant TCI, as was reviewed in \s{sec:futci}.  \\


However, the stability of these surface states are representation-dependent, which raises some doubt as to the analogy with the TCI. For a conclusive proof it is desirable to have a bulk diagnostic that is insensitive to the choice of surface termination.  One approach is  to calculate the $\Z_2$ bulk topological invariant originally formulated by Liang Fu,\cite{fu2011} and equivalently reformulated (by one of the present authors) in terms of Zak phases;\cite{berryphaseTCI} the latter formulation is simpler for numerical computation. For a general review of Zak phases, we refer the reader to \s{sec:prelimzakphase}.\\

\noindent \textit{Summary of Zak-phase diagnostic of $\Z_2$ invariant}.  Suppose $P$ is  a rank-two energy band that is energetically isolated, and carries the same symmetry representations (in $\bk$-space) as a BR of $\calt_3 \rtimes C_{4v}\times\Z_2^T$, induced from  the two-dimensional irreducible  representation of $C_{4v}\times \Z_2^T$. To diagnose if $P$ is nontrivial in the $\Z_2$ classification, we would numerically diagonalize the Wilson loop of the non-abelian Berry gauge field [cf.\ \q{definewilsonloop}] for a family of bent $\bk$-loops $[\calc(k_z)]$ illustrated in \fig{fig:tetragonal_phc}(b). For each loop $\calc(k_z)$, $k_z$ is fixed and $(k_x,k_y)$ varied along a path with an orthogonal kink at each $C_4$-invariant wavevector. Since $P$ has rank two,  the Wilson loop matrix has two eigenvalues $\{e^{i\Phi_1(k_z)},e^{i\Phi_2(k_z)}\}$, with $\Phi_j$ the Zak phase. Due to the four-fold symmetry, $\Phi_1(k_z)\equiv-\Phi_2(k_z)$  (mod $2\pi$) and it suffices to consider just $\Phi_1$. At $k_z=0$ (and also $k_z=\pi$), the symmetries of time reversal and four-fold rotation result in the Zak phase being fixed either to $\Phi_1=0$ or $\pi$. Then $\Phi_1(0)\equiv \Phi_1(\pi)$ vs $\Phi_1(0)\not\equiv \Phi_1(k_z)$ correspond respectively to the trivial vs nontrivial   $\Z_2$ class.\\

The above diagnostic cannot be applied to the  rank-three subspace below the gap,\cite{berryphaseTCI} but can be applied to the rank-two subspace $Q_{phc}$ just above the gap. We plot how the Zak phase of $Q_{phc}$ disperses with respect to $k_z$  in \fig{fig:tetragonal_phc}(c), thus confirming its nontriviality in the $\Z_2$ classification.\\

We remark that the same obstructed representation of $\calt_3 \rtimes C_{4v}\times\Z_2^T$ can in principle be realized by a 3D tetragonal lattice of dielectric cavities embedded in an artificial metallic plasma.\cite{yannopapas_photonicTCI} Based on a tight-binding Hamiltonian description of weakly-coupled plasmons (associated to the surfaces of dielectric cavities), Yannopapas proposed to realize Liang Fu's tight-binding model of the TCI; however, this remains a hypothesis in the absence of a concrete design. If ever such a design is conceived, it would be interesting to explore the implications of fragility in a setting that differs from Ochiai's.


\subsection{Topological classification of hexagonal photonic crystal} \la{sec:hexagonal}

The photonic crystal by Yihao Yang et al. consists of metallic split-ring resonators arranged in a 3D hexagonal array with symmetry of $\calt_3 \rtimes C_{3v}\times\Z_2^T \equiv P31m$, and has been claimed to be the first experimental realization of a topological band gap in three spatial dimensions.\cite{Yang2019} \\

The design principle for this hexagonal photonic crystal (and related crystals\cite{Khanikaev2012,tzuhsuan_guidingEMwaves}) has been to emulate the spin-orbit-coupled $\Z_2$ Kane-Mele topological insulator. That is, by fine-tuning the crystalline structure, Yinghao et al. have designed a photonic band touching at the $K$ point, which is described by the following  $\bk\cdot \bp$ Hamiltonian
\e{ H = v_{\parallel}\tau_0{\otimes} (k_x\sigma_x{+}k_y\sigma_y) +m\tau_x{\otimes}\sigma_z +v_zk_z\tau_y{\otimes}\sigma_z.\la{effham}} 
This Hamiltonian is identical (as a $\bk$-dependent matrix) with that of the critical point of the spin-orbit-coupled Kane-Mele model. (Above, $\sigma_{i=0,x,y,z}$ and $\tau_{i=0,x,y,z}$ are distinct  sets of Pauli matrices. For concreteness, we have shown the form of the Hamiltonian, but postpone its technical description.) \\

Despite being identical as matrices, the bases of the two $\bk\cdot \bp$ Hamiltonians  differ -- the photonic basis forms an integer-spin representation of crystallographic point-group symmetries, while the electronic basis forms a half-integer-spin representation. The  difference in bases will not matter to the \emph{existence} of Jackiw-Rebbi soliton eigen-solutions\cite{Jackiw1976} of \q{effham}, which  are localized to a mass domain wall -- this is how Yinghao et. al. (and related works) justify their experimentally-observed domain-wall states that disperse as a Dirac cone.  However, the difference in bases will matter to the \emph{robustness} of these domain-wall states -- unlike time-reversal-invariant topological insulators, the domain-wall states of the time-reversal-invariant hexagonal photonic crystal is not spectrally robust, as we prove  in \s{sec:hexagonaldomainwall}. \\

\begin{figure}[ht]
	\centering
    \includegraphics[scale=1.0]{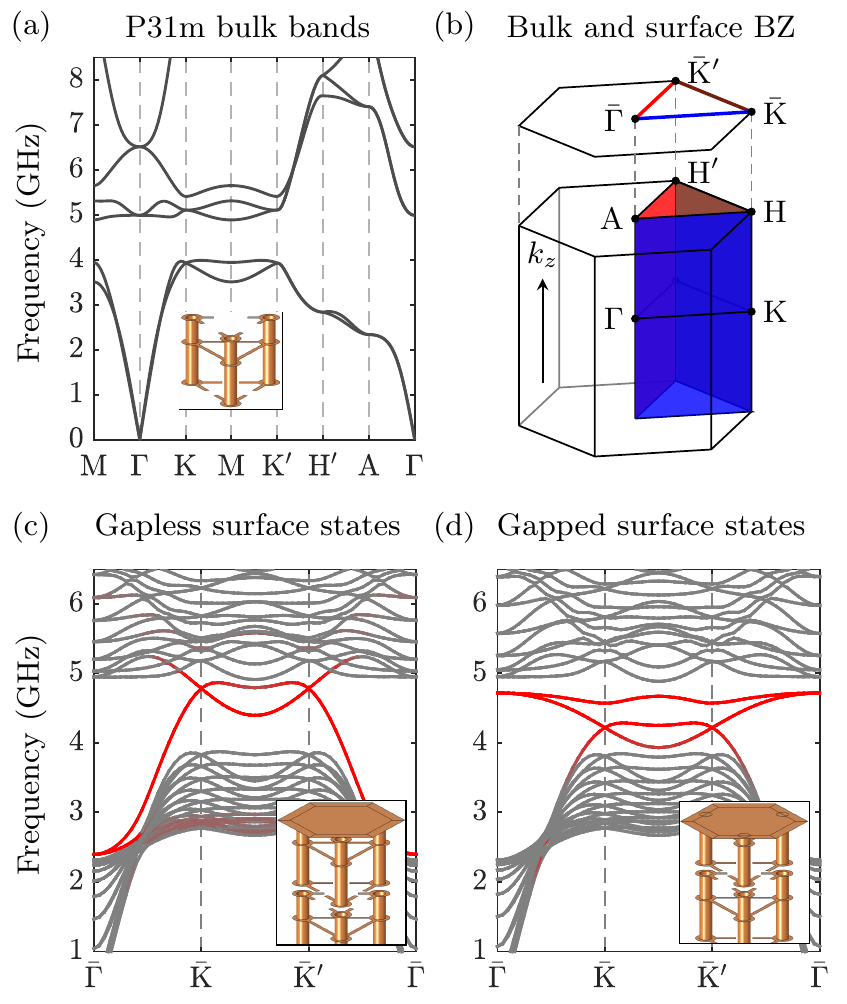}
    \caption{ (a) Bulk band structure of the hexagonal photonic crystal, with the split-ring resonator illustrated in the inset. Our simulated crystal is deformed from the experimental system of Yihao Yang et al., so that the crystal is closer to the critical point of the topological phase transition. This deformation preserves both symmetry and the relevant bulk gap. Bottom of (b) panel: bulk Brillouin zone of hexagonal photonic crystal; the halved mirror plane is colored blue. Top: 001 surface Brillouin zone.  (c) and (d) illustrate the spectrum of Maxwell's equation with perfect-electric-conductor boundary condition imposed on the 001 surface, for two different surface terminations; surface-localized states are colored red.
    \label{fig:hexagonal_phc}}
\end{figure}

There is yet another motivation for a proper group-theoretic analysis of the Hamiltonian in \q{effham}. Ultimately, photonic bands cannot realize the Kane-Mele $\Z_2$ topological invariant; the appropriate topological invariant for three-fold-invariant photonic crystals in Wigner-Dyson class AI  has been identified (by one of the present authors) as the \emph{halved-mirror chirality} $\chi$,\cite{berryphaseTCI} so named because it is an \emph{integer} topological invariant  defined over a halved mirror-invariant plane [illustrated by the blue rectangle in \fig{fig:hexagonal_phc}]. We will prove below that the  Hamiltonian  in \q{effham}, when interpreted with the correct photonic basis, describes a topological phase transition where $\chi$ changes by unity. $\chi=1$ indicates a fragile obstructed representation, as we have proven in \app{app:provefragileC3v}. \\

Accompanying this change from $\chi=0$ to $\chi=1$ is the development of in-gap, surface-localized states 
that are illustrated in   \fig{fig:hexagonal_phc}(c); for this figure, we have terminated the crystal with a $\calt_2\rtimes C_{3v}\times \Z_2^T$-symmetric 001 surface, on which is imposed the perfect-electric-conductor boundary condition. The nontrivial connectivity of surface states (colored red) over the high-symmetry line $\bar{\Gamma}\bar{K}\bar{K}'\bar{\Gamma}$ was initially predicted by one of us in \ocite{berryphaseTCI}. However, in principle these surface states are removable from the bulk gap while preserving both gap and symmetry, owing to hybridization with conventional surface states (transforming as a unit-rank BR of $\calt_2\rtimes C_{3v}\times \Z_2^T$).  \fig{fig:hexagonal_phc}(d) illustrates how such conventional gapped surface states may emerge from the continuum of high-energy bands above the bulk gap, owing to a slightly different surface termination that maintains $\calt_2\rtimes C_{3v}\times \Z_2^T$ symmetry.  This is another manifestation of the representation-dependent stability of surface states [cf.\ \s{sec:futci}].\\  

\noindent \textit{Proof that \q{effham} (with $m=0$) is a critical point  for $\chi$}\\

The  Hamiltonian in \q{effham} is a small-$\bk$ expansion around the $K$ point of the hexagonal Brillouin zone; the little group of $K$ is the point group $C_{3v}$, which is generated by the three-fold rotation $C_3$ (about $z$) and a reflection $\mir_y$ that inverts $y\rightarrow -y.$ These symmetries are represented as
\e{ \hat{C}_3\eq\tau_0\otimes e^{i2\pi \sigma_z/3} , \;\;\hat{\mir}_y =\tau_{z}\otimes \sigma_x.       \la{repc3v}}
Two of four basis vectors transform under $C_{3v}$ as  $x\pm iy$, corresponding to circularly-polarized transverse electric modes $E_x\pm iE_y$; the other two basis vectors transform as $\mp i z(x\pm iy)$,
corresponding to circularly-polarized transverse magnetic modes $H_x\pm iH_y$. $\sigma_z=\pm 1$ distinguishes the two circular polarizations, which are inverted under reflection [cf.\ \q{repc3v}].\\

While not crucial to our proof, it is worth clarifying the physical origin of these transverse modes. The design principle of the hexagonal photonic crystal relies on first constructing a $D_{3h}$-symmetric crystal, and then reducing the symmetry to $C_{3v}$ with a structural bi-anisotropy\cite{jinaukong_bianisotropicmedia} that breaks $\mir_z: z \rightarrow -z$ reflection symmetry;\cite{Yang2019} $\mir_z$ is mapped to $\tau_z\otimes \sigma_0$ in the representation space of \q{effham}, 
and the bi-anisotropy is  reflected by the mass term in \q{effham}. The transverse electric and magnetic modes transform respectively in the $E'$ and $E''$ representations of  $D_{3h}$,\footnote{For the $E'$ and $E''$ notations, refer to the character table in the appendix of \ocite{tinkhambook}.} 
but in the same $E$ representation of $C_{3v}$; therefore, the bi-anisotropy allows to  couple electric and magnetic modes.\\

Focusing on the ${r}_y$-invariant plane ($k_y=0$), we perform a unitary transformation $U$ [specified below] such that  $U^\dagger \hat{\mir}_y U$ is diagonal with on-diagonal elements: $1,1,-1,-1$. The first two (resp.\ last two) basis vectors will be said to belong in the $(\eta=+1)$-eigenspace (resp.\ $(\eta=-1)$-eigenspace) of reflection. The Hamiltonian then becomes block diagonal with respect to $\eta$:
\begin{widetext}
\begin{equation}
U^\dagger H U=\left(
\begin{array}{cccc}
m & v_\parallel k_x+i v_z k_z & 0 & 0 \\
v_\parallel k_x-i v_z k_z & -m & 0 & 0 \\
0 & 0 & m & v_\parallel k_x-i v_z k_z \\
0 & 0 & v_\parallel k_x+i v_z k_z & -m \\
\end{array}
\right),
\;\;
U= \f{1}{2} \left(
\begin{array}{cccc}
1 & 1 & 1 & -1 \\
1 & 1 & -1 & 1 \\
1 & -1 & 1 & 1 \\
-1 & 1 & 1 & 1 \\
\end{array}
\right)\label{tranformedH}
\end{equation}
\end{widetext}
On inspection, Eq.\ (\ref{tranformedH}) is a massive Dirac Hamiltonian with opposite chiralities in the different mirror  eigenspaces. When the Dirac mass $m$ changes sign, the integrated Berry curvature $(\int_{k_y=0} F_{\eta})$ in the $\eta$ subspace changes by $\eta\in \{+1,-1\}$.\footnote{This is demonstrated in Chapter 8 of Ref.\ \onlinecite{bernevig_topological_2013}} It follows that $\chi=\int_{HMP}(F_{\eta=+1}-F_{\eta=-1})$, being an integral (over the \textit{halved} mirror plane) of the \textit{differential} Berry curvature,\cite{berryphaseTCI} changes by unity. This completes the proof.



\subsection{Instability of domain-wall states of tetragonal and hexagonal photonic crystals} \la{sec:hexagonaldomainwall}

\begin{figure}[ht]
	\centering
    \includegraphics[scale=1.0]{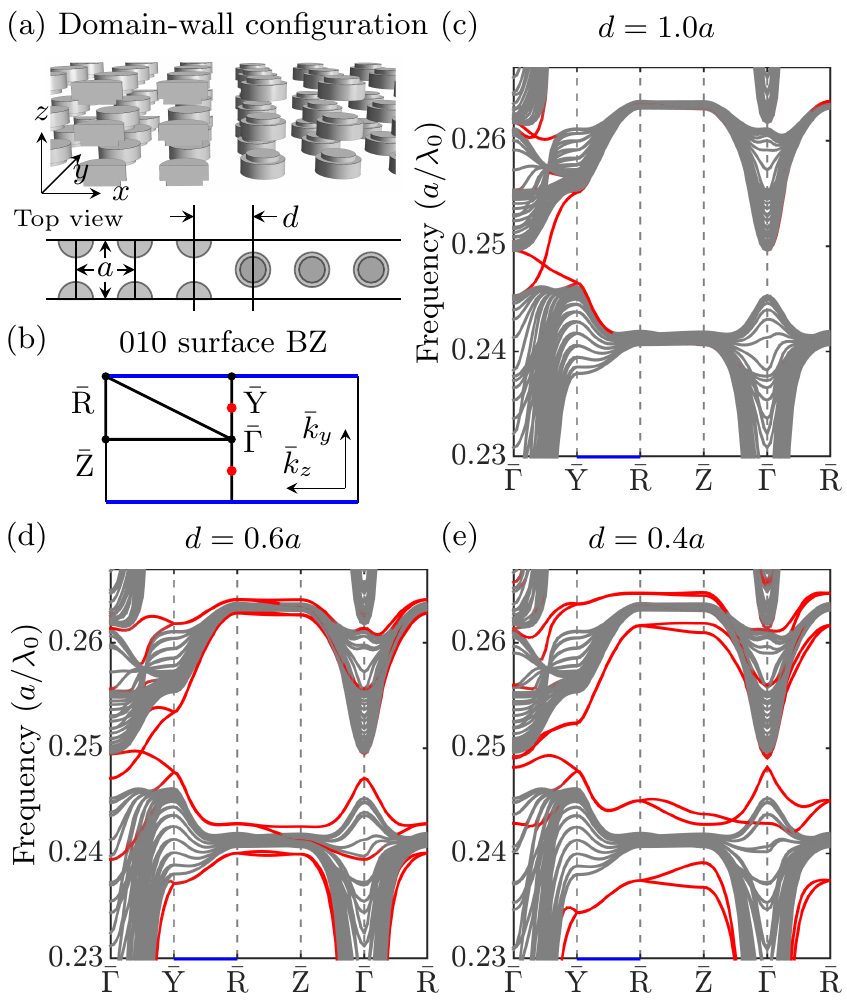}
    \caption{ (a) Screw-symmetric domain-wall configuration for the tetragonal photonic crystal. The bulk lattice period  $a$ and domain-wall thickness $d$ are indicated in the bottom panel. (b) Two-dimensional BZ of the domain-wall configuration. (c-e) illustrate band structure calculations for $d/a=1.0$, $d/a=0.6$, and $d/a=0.4$ respectively. States localized to the domain wall are colored red. The $\bk$-positions of the Dirac points (for $d/a=1.0$) are indicated by red dots in (b). Note that the two-fold degeneracy of domain-wall states along the high-symmetry $\bk$-line $\bar{Y}\bar{R}$ is due to a combination of screw and time-reversal symmetry. 
    \label{fig:domain_wall}}
\end{figure}

Here we investigate the robustness of domain-wall states of fragile topological photonic crystals. A simple example of a two-dimensional domain wall separates two three-dimensional crystals, which  differ only in that one crystal is geometrically reflected relative to the other. A domain-wall  configuration of the tetragonal [resp.\ hexagonal] photonic crystal is illustrated in \fig{fig:domain_wall}(a) here [resp.\ Fig.\ 2(a) of \ocite{Yang2019}]. A domain-wall configuration generally has the symmetry of a crystallographic layer group, and we will find that certain layer groups allow for the existence of Dirac-type domain-wall states. However, we will prove that such domain-wall states are removable from the bulk energy gap by a continuous deformation that preserves both gap and the layer-group symmetry. Though \textit{not} spectrally robust, we will explain that these domain-wall states have a weaker type of robustness that is analogous to topological Dirac-Weyl (semi)metals.\cite{AAchen,Nielsen_ABJanomaly_Weyl,wan2010,halasz2012,chen_multiweyl,Murakami2007B,LMAA} \\

In fact, Dirac-type domain-wall states have been experimentally observed  for the hexagonal case,\cite{Yang2019} and numerically simulated for the tetragonal case in \fig{fig:domain_wall}(c) for a specific domain-wall thickness. To explain the existence of Dirac points, the crucial observation is that both domain-wall configurations (of the tetragonal and hexagonal photonic crystals) have in common a screw symmetry, which is the composition of a two-fold rotation axis (lying parallel to the domain wall), and a half Bravais-lattice translation along this axis. (We see that the layer group can be nonsymmorphic, even though the three-spatial-dimensional space group of either crystal is symmorphic.)  Each Dirac point is then a crossing between two distinct screw representations, of the type first theoretically predicted by one of us for gapless photonic crystals.\cite{LMAA} In particular, these Dirac points originate from a band inversion between two rank-two bands, as illustrated by the middle band in Fig.\ 1(c) of \ocite{LMAA}.  \\

If the screw symmetry of the domain-wall configuration is broken, then the two screw representations are allowed to hybridize, and the Dirac-point degeneracy would generically lift.\\

One may however ask if the Dirac points are robust in the presence of screw symmetry. Here, a nuanced notion of robustness is advantageous. On one hand, the Dirac-type domain-wall states are \textit{not spectrally robust}. Indeed, since the Dirac points result from a band inversion of domain-wall states rather than bulk states, it is possible to reverse the inversion  while preserving both the bulk gap and screw symmetry. Through this reversal, a pair of Dirac points with opposite chirality\cite{AA_glideresolved} would eventually meet (at a time-reversal-invariant $\bk$-point) and mutually annihilate. The result is that the domain-wall states no longer cover the bulk gap  -- this proves the spectral non-robustness of domain-wall states for both tetragonal and hexagonal photonic crystals. This reversal of the band inversion is numerically simulated for  the tetragonal crystal -- by screw-symmetrically decreasing the domain-wall thickness, as illustrated in   \fig{fig:domain_wall}(d-e). \\

On the flip side, one may say that the screw-protected Dirac crossings persist so long as two crossings of opposite chirality do not meet and annihilate; such persistence may be rationalized by a nontrivial Berry-Zak phase of $\pi$, for any screw-symmetric $\bk$-loop that encircles an odd number of Dirac points.\cite{LMAA} This weaker notion of robustness is  closely analogous to a class of nonsymmorphic topological semimetals without spin-orbit-coupling, as was proposed in \ocite{LMAA}. \\
    
We will describe three more examples of domain-wall states in the literature of photonic crystals. In all cases mentioned here, the role of crystallographic symmetry, as well as the spectral non-robustness, has not been appreciated.\\

\noi{i} A different realization of screw-symmetric  domain-wall states can be found for the all-dielectric metamaterial crystal of  \ocite{Slobozhanyuk_3DalldielectricphotonicTI}; see, in particular, their Fig.\ 3. \\

\noi{ii} Not just screw symmetry can protect Dirac-type domain-wall states. For example, the domain-wall configuration in Fig.\ 7 of \ocite{Ochiai_hybridmethod} has a two-fold rotational axis \textit{parallel} to the domain wall, and their simulated Dirac point is a crossing between distinct representations of rotation. \\

\noi{iii} Our last example is the domain-wall configuration in  Fig.\ S6 of \ocite{Slobozhanyuk_3DalldielectricphotonicTI}, which has a two-fold rotational axis \textit{perpendicular} to the domain wall; their simulated Dirac points exist because of the composition of rotation and time reversal, which reduces the codimension of an eigenvalue degeneracy to two, according to the well-known Wigner-von Neumann non-crossing rule.\cite{neumann1929} This case is closely analogous to the Dirac points of graphene.

\section{Band representations are incompatible with robust boundary and domain-wall states}\la{sec:norobustsurface}

Throughout this work, we have employed the crystallographic splitting theorem in various guises to determine if a given band is a band  representation (BR) or an obstructed representation. Here we argue for one utility of such a determination, namely that BRs are incompatible with {spectrally robust} in-gap states -- localized either to a boundary interface between crystal and vacuum, or to a domain-wall interface between two crystals which are relatively inverted. After elaborating on the distinction between a boundary and domain wall in \s{sec:defineboundary}, we will formalize  the above-mentioned incompatibility by proving a  necessary condition for spectrally robust boundary and domain-wall states in \s{sec:instabilitycriteria} (the precise meaning of `spectrally robust' will also be given there). Finally we will apply these results  to the in-gap states of  fragile topological insulators and photonic crystals in \s{sec:boundaryftifphc}. \\


In proving the absence of spectrally robust in-gap states for BRs, we will apply that every BR has a symmetric deformation to an tight-binding (or `atomic') limit. In fact the converse is also true: if an tight-binding limit exists for a band then it must be band-representable. The equivalence between the existence of a symmetric tight-binding limit and band-representability is formalized in a theorem in  \s{sec:instabilitycriteria}.

\subsection{Distinguishing boundary from domain wall}\la{sec:defineboundary}

Here we precisely define a boundary vs a domain wall, and give a casual introduction to the tight-binding method; this preliminary discussion is to prepare us for a subsequent proof of the boundary/domain-wall stability  criterion  in \s{sec:instabilitycriteria}.\\

Suppose we are given a $d$-spatial-dimensional crystalline insulator (or a photonic crystal) with space group $G$, and an energy gap separating the low-energy subspace  and the complementary, high-energy subspace. Any continuum description of crystals formally involves an infinite number of bands; however one is typically interested in physics within a finite energy window, and it is common practice to  truncate the continuum Hilbert space and model the truncated subspace  by a finite-rank tight-binding lattice model. In this manner, we obtain a tight-binding Hamiltonian with an energy gap separating a low- and high-energy subspace, with corresponding finite-rank projectors that are  assumed to be  analytic throughout the Brillouin zone.\\

Let us enclose the crystal by a $(d{-}1)$-spatial-dimensional hypersurface, and ask if there are \textit{evanescent eigenstates} that are exponentially-localized to the hypersurface, with energies lying within the energy gap (defined by the translation-invariant crystal). The answer depends on what lies on both sides of the hypersurface; two scenarios are commonly encountered:\\

\noindent \textit{Definition of boundary.} For any crystal with an energy gap, it is convenient to define a    \emph{crystalline vacuum} that satisfies two properties: its symmetry group  contains the space group (of the crystal) as a subgroup, and energy eigenstates below  a threshold energy $E_{v}$ are forbidden; the minimal bound for $E_v$ is given by the maximal energy of the  gap (of the translation-invariant crystal). We define a \textit{boundary} as a hypersurface that separates the `bulk' crystal from its vacuum. Moving away from the hypersurface and into the crystalline bulk, we assume that the Hamiltonian (or classical mode equation) asymptotically approaches  a form that is locally identical to a translation-invariant crystal; concretely, we insist that the deviation (from the translation-invariant form) decays at least as fast as an exponential function. Moving away from the hypersurface into the crystalline vacuum, we also assume that the Hamiltonian exponentially approaches a form locally identical to a translation-invariant vacuum. \\

In practice, $E_v$ is determined by the specific physical realization of the crystalline vacuum. For example, a metal acts as a vacuum for photons owing to the screening ability of metallic electrons, and $E_v$ is then given by the plasma frequency of the metal.  
In cases where the energy window of interest lies far below $E_v$, one may reasonably take  $E_v$  to infinity and impose idealized boundary conditions, such as the Dirichlet (`open') boundary condition sometimes used to model finite-size, solid-state crystals, or the perfect-electric-conductor boundary condition\cite{Ochiai2017,Yanqinghui_discovernodalchains} often used for photonic crystals. We will not need to assume that the boundary is smooth  -- evanescent eigenstates which are localized to $(d{-}2)$-spatial-dimensional boundary kinks (or `hinges') have been explored in higher-order topological insulators,\cite{zhouyuanjun_surfacepolarizationedgecharge,benalcazar_quantizedmultipole,zhida_dminus2,Schindler_higherorderTI} and form a subclass of what we call {boundary states}.\\

\noindent \textit{Definition of domain wall.} A domain wall is a hypersurface separating two crystals with the same space group and the same bulk energy gap; the two crystals differ only in that  bulk states (in the relevant energy window) are inverted with respect to the center of the bulk energy gap. Such a spatially-inhomogeneous band inversion can be engineered in photonic crystals.\cite{Ochiai2017,Yang2019} Our notion of a domain wall is conceptually similar but not identical to a massive domain wall of the Dirac equation,\cite{Jackiw1976} whose chiral zero modes are anomalous.\cite{callanharvey_anomalies}

\subsection{Necessary criteria for spectrally robust boundary and domain-wall states}\la{sec:instabilitycriteria}

The sense in which robust boundary states are incompatible with band representations is stated in the following criterion.\\

\noindent  \textbf{Boundary stability criterion} Given a crystal with space group $G$ and a bulk energy gap,  a necessary condition for spectrally robust, in-gap boundary states is that the low-energy subspace is an obstructed representation of $G$. \\

\noindent By `spectrally robust', we  mean that the eigen-energies of the boundary states form a band that covers the bulk energy gap, and this covering is insensitive to continuous and ($G_{int},G$)-symmetric deformations that preserve the energy gap.  Here, $G_{int}< G$ is the symmetry of the interface; if the interface has discrete translational symmetry in  two independent directions, $G_{int}$ is generally a layer group. By a ($G_{int},G$)-symmetric deformation, we mean that the deformation is everywhere $G_{int}$-symmetric; moving away from the interface,  the $G$-asymmetric component of the deformation is assumed to decay exponentially. While specifying $G_{int}$ is necessary to uniquely define spectral robustness, actually the boundary stability criterion holds regardless of the choice for $G_{int}$. The following criterion for domain walls also holds regardless of the symmetry of the domain-wall interface.\\

\noindent  \textbf{Domain-wall stability criterion} Given a crystal with space group $G$ and a bulk energy gap,  a necessary condition for spectrally robust, in-gap domain-wall states is that either the low-energy subspace or the high-energy subspace is an obstructed representation of $G$. \\

The proof of the above criteria is based on  two physically intuitive claims: (i) every band representation has a symmetric tight-binding limit, and (ii) the boundary (or domain wall) of a crystal does not have to intersect any tight-binding lattice site. \\

Statement (i) is a widely-believed folklore, and our contribution is a restatement of (i) that is amenable to a rigorous, bundle-theoretic derivation. We define an \textit{tightly-bound BR} of $G$ as a  BR of $G$ with the property that each Wannier function only has support on a single lattice site. (In the tight-binding formalism, each Wannier function is defined over a real-space lattice with a finite-dimensional complex vector space on each lattice site.) Each Wannier function in a tightly-bound BR is said to be \textit{one-site localized}.  \\

\noindent \textbf{Symmetric tight-binding limit theorem} For a $G$-symmetric band, being a band representation of $G$ is equivalent to the existence of a $G$-symmetric homotopy to a $G$-symmetric band spanned by one-site localized Wannier functions.\\


\noindent To qualify this statement, for $P$ a BR of $G$, the just-stated $G$-symmetric homotopy always exists for a tight-binding model that contains $P$ as the complete, low-energy subspace, and also contains a high-energy subspace with sufficiently large rank -- this will shortly be clarified in \s{sec:proofboundarycriterion}.\\


The forward arrow of the above theorem is proven by showing that a BR$(G,\bvarpi,D)$ and a tightly-bound BR$(G,\bvarpi,D)$ are isomorphic as $G$-vector bundles [cf.\ \app{app:atomicBR}], and then applying the  universal $G$-bundle theorem\cite{Atiyah1989} to prove existence of the $G$-symmetric homotopy.\footnote{The universal bundle theorem in \ocite{Atiyah1989} applies to $\calp$-vector bundles with finite group $\calp$. For  $G$-vector bundles with $G$ a space group, we observe that the translation subgroup of $G$ acts trivially on the bundle, hence we  may directly apply the universal bundle theorem with $\calp$ identified as the (finite) point group of $G$.} Such a homotopy will be referred to as {a \textit{deformation to the symmetric tight-binding limit}}, and is exemplified numerically by the adiabatic deformation in \ocite{fragile_po}. The backward arrow is proven in \s{sec:localization_obstruction}, where we also discuss its implications for the Wannier functions of topological insulators.\\

The remaining argument will separately treat boundaries and domain walls. 

\subsubsection{Proof of boundary stability criterion}\la{sec:proofboundarycriterion}

We will argue for the contrapositive restatement of the boundary stability criterion. Our strategy is  to exclude boundary states for a  model  tight-binding Hamiltonian $H_{b}$ (to be specified below) whose low-energy subspace is band-representable. This would imply the absence of spectrally robust boundary states for any Hamiltonian that is continuously deformable to $H_{b}$, while preserving the bulk energy gap and $(G_{int},G)$ symmetry, {for any $G_{int}< G$}. \\

Our model Hamiltonian is
\e{H_{b} = PB_{b}P +Q+2Q',}
where $P$ (resp.\ $Q$) is the analytic, finite-rank projector to the low-energy (resp.\ high-energy) band of the crystal.\footnote{Except for the term $2Q'$, $H_{b}$ is identical to a model of an insulator-vacuum interface proposed in \ocite{fidkowski2011}.} While $P$ and $Q$ lie within the energy window of interest, a formal proof will additionally require $Q'$ which projects to energy bands lying above $Q$ on the energy axis; such bands always exist because we are approximating a continuum description of crystals. $Q'$ has the following properties: (a) $Q'$ has finite rank and is orthogonal to both $P$ and $Q$; $I=P+Q+Q'$, (b) $Q'$ is analytic over the Brillouin torus, and (c) $Q'$ transforms as a BR of $G$.  While $P,Q$ and $Q'$ have the symmetry of the space group $G$, $G$ is not a symmetry of  $H_{b}$ owing to the leftmost term involving $B_{b}$ -- a spatial-bipartition operator that equals  $-1$ within the crystalline bulk  and $+1$ without.  Since matrix elements of $P$ have exponential decay in real space, $H_{b}$ exponentially approaches the form $2Q'+Q-P$ within the crystalline bulk; this form models a crystalline Hamiltonian with spectrally-flattened bulk bands, and with the relevant bulk energy gap in the interval $(-1,+1)$; outside of the bulk region, $H_{b}$ exponentially approaches $Q'+I$, which models a crystalline vacuum with threshold energy $E_v=+1$. \\

If $P$ is a BR, then by the symmetric tight-binding limit theorem, $P$ is continuously deformable to a tightly-bound band representation by a $G$-symmetric homotopy. The corresponding one-site-localized Wannier functions are all eigenstates of $H_{b}$, assuming that the $G_{int}$-symmetric boundary hypersurface does not intersect any tight-binding lattice site. (No matter $G_{int}$, it is always possible to symmetrically deform the hypersurface to satisfy this zero-intersection condition.) Since any state in the orthogonal subspace $Q$ (resp.\ $Q'$) is an eigenstate of $H_{b}$ with eigenvalue $+1$ (resp.\ eigenvalue $+2$), the spectrum of $H_{b}$ is just  $\{-1,+1,+2\}$, with no eigenenergies in the bulk energy gap. \\

We now address a subtlety in the above argument that formally justifies the presence of $Q'$ in $H_{b}$. Supposing $Q'=0$, it is possible that the $G$-symmetric homotopy (between $P$ and a tightly-bound BR) does not exist:\\

\noindent Given a $G$-symmetric tight-binding lattice model with a band subspace $P$  that forms a  band representation of $G$, we say that $P$ has a   \textit{symmetric tight-binding obstruction} if it cannot be deformed to a symmetric tight-binding limit. \\

\noindent A symmetric tight-binding obstruction is an artifact of the tight-binding formalism, and  reflects that a subset of the
Wannier centers of $P$ are rigidly displaced from any tight-binding lattice site (a lattice site is the positional center of a localized, tight-binding basis vector). Such an obstruction  has some conceptual similarities with  the `obstructed atomic limit' formulated by \ocite{TQC,DePaz2019}; one important distinction is that the tight-binding obstruction does \textit{not} apply to continuum crystals. In the proof of the symmetric tight-binding limit theorem, the universal $G$-bundle theorem guarantees the  existence of the $G$-symmetric homotopy if $P$ is a subspace of a tight-binding vector space with \textit{sufficiently large rank}, as explained in \app{ALproof}. A continuum description of crystals might be viewed heuristically as a tight-binding lattice model with an infinitely fine {real-space} mesh -- in this case the symmetric tight-binding obstruction should not exist. In the tight-binding formalism, one removes the obstruction by  enlarging the tight-binding lattice model to include the higher-energy bands in $Q_c$ -- this justifies the presence of $Q'$ in $H_{b}$. \\

We know of two mechanisms for a symmetric tight-binding obstruction for fixed-rank, tight-binding Hamiltonians:\\

\noi{i} Both the tight-binding lattice sites and the Wannier centers of $P$ are rigidly fixed to \textit{distinct} positions, owing to non-translational symmetries of the tight-binding model. This type of obstruction is exemplified by the nontrivially-polarized filled band of the Su-Schrieffer-Heeger model,\cite{nogo_AAJH} and also by a band representation in a modified Kane-Mele honeycomb model.\cite{fragile_po} \\

\noi{ii} The second mechanism exists even if the point group of space group $G$ is trivial, implying that the Wannier centers of $P$  lie at generic Wyckoff positions. Naively, these Wannier centers would always be continuously tunable to lie on the tight-binding lattice sites. However, there exists finite-rank tight-binding Hamiltonians without any non-translational symmetry, but having a low-energy band representation that is topologically obstructed from a tight-binding limit  -- such is the case of the Hopf insulator,\cite{Hopfinsulator_Moore} as will be clarified in a later publication.\cite{AA_hopf}


\subsubsection{Proof of domain-wall stability criterion}\la{sec:proofdwcriterion}

Let us argue for the contrapositive restatement of the domain-wall stability criterion. Our strategy is to exclude domain-wall states for a model Hamiltonian $H_{dw}$ (specified below) whose low- and high-energy subspaces are both band-representable; this would rule out spectrally robust domain-wall states for any Hamiltonian that is deformable to $H_{dw}$ while preserving both bulk gap and $(G_{int},G)$-symmetry.\\

Our model Hamiltonian for a domain wall is  
\e{H_{dw}=(Q-P)B_{dw}(Q-P)+2Q',\la{defineHdom}} 
with  $B_{dw}$ a real-space bipartition operator that equals $+1$ on one side of the domain wall, and $-1$ on the other side. Moving away from the domain wall, $H_{dw}$ asymptotically approaches $ 2Q'+(Q-P)$ on one side of the wall, and $2Q'-(Q-P)$ on the other side; the difference is that $P$ and $Q$ are inverted with respect to the center of the bulk gap. \\

If both $P$ and $Q$ are BRs, then they are both deformable to tightly-bound BRs. Once again, the presence of $Q'$ nullifies any possible symmetric tight-binding obstruction for  $P$ and $Q$. Since $(Q-P)$ and $Q'$ act in orthogonal subspaces, the spectrum of $H_{dw}$ is obtained by independently diagonalizing the two terms on the right-hand side of \q{defineHdom}. Any one-site localized Wannier function of $P$ is an eigenstate of $(Q-P)B_{dw}(Q-P)$ with eigenvalue $+1$ on side of the domain wall, and eigenvalue $-1$ on the other side. The same can be said for the one-site localized Wannier functions of $Q$, except that the eigenvalues $\pm 1$ are inverted. We thus obtain that the spectrum of $H_{dw}$ is $\{-1,+1,+2\}$, with no eigenvalues in the relevant energy gap $(-1,+1)$, and hence no domain-wall states. This completes the proof.\\

The above two proofs can easily be generalized to include  energy bands ($P'$)  lying  lower  than $P$ on the energy axis, and transforming as a BR of $G$. (In the domain wall case, we will not assume $P'$ is inverted  across the domain wall.) Generalizing $H_{b,dw}\rightarrow H_{b,dw}-2P'$, the above steps in excluding boundary/domain-wall states would  essentially be unchanged.  \\

As a final remark, we have assumed throughout this work that a band has only the symmetry of a crystallographic space group; it is possible that other types of symmetries (e.g., particle-hole symmetry) may protect spectrally robust boundary states, even in the case of band representations.



\subsection{Application to fragile topological insulators and photonic crystals}\la{sec:boundaryftifphc}


Boundary states whose eigen-energies cover the bulk gap are a well-known feature of many stable obstructed representations.\cite{kane2005B,fukanemele_3DTI,moore2007,Rahul_3DTI,hsieh2008,Hsieh_SnTe,Xu_observeSnTe,Tanaka_observeSnTe,ChaoxingNonsymm,Shiozaki2015,unpinned,Nonsymm_Shiozaki,Poyao_mobiuskondo,Hourglass,Cohomological,Ma_discoverhourglass,GSHI} Less well known is that such boundary states also manifest  in  tight-binding models of some  fragile obstructed representations.\cite{fu2011,berryphaseTCI,TCIbyrepresentation} There is however a danger in naively extrapolating the predictions of an idealized tight-binding model to a real material. As mentioned in \s{sec:instabilitycriteria}, a tight-binding model always involves truncating the infinite-rank Hilbert space of a continuum crystal. Thus even if the low-energy band of a tight-binding model is fragile obstructed with accompanying boundary states, it is possible that the complete, continuum low-energy subspace is actually band-representable -- this would imply that boundary states are \textit{not} spectrally robust, according to the boundary stability criterion. \\

In the just-described hypothetical scenario, we may say that a BR in the continuum low-energy subspace `breaks' the fragile obstruction of the tight-binding model, in the sense that BR $\oplus$ FOR = BR'. Not just any BR can break a given fragile obstruction; to determine the appropriate BR, one can use the projected symmetry method [cf.\ \s{sec:demonstration}] or the projected position method [cf.\ \s{sec:zakinv}]. Now if FOR has boundary states covering the bulk gap, while BR' is deformable to a symmetric tight-binding limit without boundary states, the combined implication is that the hybridization of FOR with BR allows for the removal of all in-gap boundary states. This removal may be envisioned in the following thought experiment: suppose BR $\oplus$ FOR  were placed on a finite sample with boundaries, but with zero hybridization between BR and FOR. In addition to the boundary states (of FOR) which cover the bulk gap, it is also possible to have conventional boundary states (of BR) which do not cover the gap. Once the two types of boundary states hybridize, they must  be adiabatically removable from the bulk gap; this has been described in \s{sec:futci} as a representation-dependent stability of boundary states. To recapitulate, we have argued that if boundary states (of a fragile obstructed representation) covers the energy gap, they must have a  representation-dependent stability.      \\

In practice, we believe that our boundary stability criterion rules out spectrally robust boundary states for a great majority of electronic insulators and photonic crystals. To support our claim, a recent study of 26,938 stoichiometric electronic materials claimed that  none of them has a low-energy occupied subspace that is fragile obstructed;\cite{Maia_completecatalog} this was  rationalized in \ocite{zhida_fragileaffinemonoid} as there being `usually enough occupied elementary band representations' to break any fragile obstructed representation below the Fermi level. However, we caution that their claim   is based on identifying band/obstructed representations from their symmetry representations in $\bk$-space -- such an identification  method is not generally exhaustive [cf.\ \s{sec:futci}].\\

A fragile obstruction of a finite-rank, high-energy band  is even more likely to be breakable. This is because the continuum high-energy subspace formally has infinite rank, while there are only a finite number of elementary band representations which all have finite rank\cite{Zak_bandrepresentations,elcoro_EBRinBilbao} -- the existence of a  BR that can break any given fragile obstruction is overwhelmingly probable. Such considerations become relevant when evaluating the spectral robustness of domain-wall states [cf.\ domain-wall stability criterion].  \\

On the other hand, the continuum low-energy subspace  has finite rank, and this rank can be of order one if the bulk gap  lies close to the bottom of the energy spectrum. The closer to the bottom, the likelier to find unbreakable fragile obstructions. In this regard, photonic crystals have an advantage over electronic crystals; the latter have a fixed Fermi level, but photonic crystals  can be experimentally probed at any frequency.\\

\noindent \textit{Application to electronic insulators and gapped photonic crystals in class AI.} Let us assume that the topological classification by the method of `topological crystals' is complete,\cite{Song2018} with the implication that any $G$-symmetric band in class AI is either a band representation or a fragile obstructed representation of $G$. Given a crystal with an energy gap, the high-energy band is overwhelmingly likely to be band-representable, hence spectrally robust in-gap states are only possible if the low-energy subspace is fragile obstructed. Presently, this possibility remains an unproven principle, and is not realizable by any fragile obstructed representation that we know. (The reader may wonder if the rotation-invariant  topological crystalline insulators described in \s{sec:demonstration} satisfy the bill. Unfortunately, its unconventional boundary states  can be destabilized by  conventional boundary states emerging from \emph{above} the energy gap, as illustrated in \fig{fig:hexagonal_phc}(d).)

\section{Discussion and outlook}\la{sec:discussion}

In vector bundle theory, the splitting principle has been used to reduce difficult questions on multi-rank bundles to  simpler questions on sums of unit-rank bundles;\cite{may_splittingprinciple,adams_stablehomotopy} this reductionist approach is known to simplify the derivation of relations between Chern classes.\cite{Hatcher2009} One may ask if an analogous splitting principle exists in band theory, where crystallographic space-group symmetry must be incorporated into the band/bundle. \\

In this work we have reduced the difficult question -- of whether a given rank-$N$ band is a band representation -- to simpler questions on a splitting into $N$ unit-rank bands. For a majority of space groups (specified below), the necessary and sufficient condition for band representability is the existence of a splitting satisfying  \\

\noi{A} that each unit-rank band has an analytic projector  and a trivial first Chern class, and \\

\noi{B} the set of unit-rank bands are permuted by all crystallographic symmetries.\\

\noindent This statement is formalized by the crystallographic splitting theorem of \s{sec:state_equivthm}. As a shorthand, a splitting that satisfies condition (A) [resp.\ (B)] is said to be a Wannier splitting [resp.\ a symmetric splitting]; if both conditions are satisfied, it is a symmetric Wannier splitting. \\

To apply this theorem to prove band-representability, it is desirable to have a systematic method to symmetrically split a band, and then to verify if condition (A) is satisfied; once (A) is verified, the corresponding symmetric Wannier functions can be constructed via an algorithm detailed in \s{app:constructWFs}. We have proposed two methods for symmetric splitting: the first involves diagonalizing a projected position operator, applies to a limited set of space groups [specified in the \textit{symmetric splitting lemma} of \s{sec:resultzakphase}], but has the advantage that all unit-rank bands automatically have analytic projectors [proven in \app{sec:projposition}]. The second   method involves diagonalizing a projected symmetry operator, applies to a wider set of space groups [specified in \app{sec:diagnoseBR_projsymm}], but does not guarantee that each unit-rank band has an analytic projector [cf.\ \s{sec:obstructP}]. Of the two methods, only the projected symmetry method can be used for space groups with a three- (or four-fold) rotational symmetry, and this is how we proved in \s{sec:demonstration} that the hexagonal (or tetragonal) topological insulator in Wigner-Dyson class AI is fragile; the implications of fragility for the analogous photonic crystals are discussed in \s{sec:photonic}. A general methodology that would apply to any space group is still lacking, and in our opinion would be a significant advance.\\

It should be clarified that $P$ being band representable implies the existence of {a} symmetric Wannier splitting, but does not imply that \textit{all} Wannier splittings of $P$ are symmetric, or that \textit{all} symmetric splittings of $P$ are Wannier. To exemplify a symmetric, non-Wannier splitting, some \textit{elementary} band representations\cite{Zak_bandrepresentations,Evarestov_bandrepresentations,Bacry_bandrepresentations,elementaryenergybands} are each splittable into  unit-rank bands  which are \textit{trivially} permuted by the space group and have analytic projectors,  and it is guaranteed that at least two of the unit-rank bands must have \textit{nontrivial} first Chern class.\cite{nogo_AAJH}\\

One implication of our splitting theorem is that obstructed representations -- defined as not band representations -- cannot have a splitting that is simultaneously Wannier and symmetric. This has varied implications for topological insulators whose filled bands are  obstructed representations: if (A) is satisfied, then [not (B)] can be interpreted as a symmetry obstruction for exponentially-localized Wannier functions [cf.\ \s{sec:wannierobstruction}]. If (B) is satisfied, then [not (A)] can be interpreted  as the absence of exponential localization, or as a holonomy obstruction for Bloch functions. The holonomy interpretation is encapsulated by the Zak winding theorem in \s{sec:resultzakphase}, provides a rigorous justification for the concept of `individual Chern numbers',\cite{prodan_spinchern,alexey_smoothgauge,AA2014,z2pack} and may be  useful as a design principle for model Hamiltonians of topological insulators. In future work, it will be interesting to generalize the Zak winding theorem to three spatial dimensions, where notions such as the `nested Wilson loop'\cite{benalcazar_quantizedmultipole} might come into play.\\

Our splitting theorem  provides an equivalent description of band representations for nearly all crystallographic space groups and grey magnetic space groups. The only exceptional band representations have been referred to as non-monomial, and they occur for three-spatial-dimensional double space groups having cubic point groups. These exceptions reflect that some half-integer-spin representations of cubic point groups are fundamentally two-dimensional, that is to say, they cannot be induced from a one-dimensional representation of a subgroup of the cubic point group [cf.\ \s{sec:whichBRmonomial}]. It is interesting to speculate on a generalization of our splitting theorem that equivalently describes non-monomial band representations -- perhaps the splitting of a rank-$N$ band must allow for sums of rank-two bands as well.  \\

Condition (A) in our  theorem involves the triviality of the first Chern class, which guarantees that each unit-rank band is topologically trivial as a  \textit{complex} vector bundle. Topological notions in \textit{real} vector bundle theory have been fruitfully applied\cite{DeNittis_classifyAI,ahn_bandtopologylinkingstructure,nogo_AAJH,Ahn2018a,Ahn2019} to bands with spacetime inversion symmetry.\footnote{The symmetry must be antiunitary, square to the identity, and have a trivial action in $\bk$-space.}
Being topologically nontrivial as a unit-rank, real vector bundle is equivalent to a nontrivial first Stiefel-Whitney class; this by itself does not imply an obstruction to symmetric Wannier functions,\footnote{It is worth remarking that having a nontrivial second Stiefel-Whitney class also does not generally obstruct band-representability\cite{Po2019} -- except for rank-two bands, which are anyway better classified by the Euler class\cite{ahn_bandtopologylinkingstructure,Ahn2019}} however it does imply that the Wannier center cannot lie on a pre-specified spatial origin.\cite{nogo_AAJH,ahn_bandtopologylinkingstructure}  It may be that a real analog of our splitting theorem exists for band representations whose Wyckoff positions are fixed to the origin.



\begin{acknowledgments}
We are grateful to Barry Bradlyn and Nicholas Read, who independently encouraged us to generalize our theorem to bundles of arbitrary rank; Barry Bradlyn first pointed us to an example of a non-monomial point group. Zhida Song provided expert advice on fragile topological insulators. LL thanks Hao Lin and Thomas Christensen for comparing numerical results. Finally, we are indebted to Simon Altmann for clarifying his semidirect product decomposition of space groups, and for mailing his book titled `Induced Representations in Crystals and Molecules'.

AA was supported initially by the Yale Postdoctoral Prize Fellowship, and subsequently by the Gordon and Betty Moore Foundation EPiQS Initiative through Grant No. GBMF4305 at the University of Illinois. AA is grateful to the hospitality of Brewlab at Champaign-Urbana, where most of this paper was written.
JH obtained financial support from NSF grant No. 1724923.
LL was supported by the National key R\&D Program of China under Grant No. 2017YFA0303800, 2016YFA0302400 and by NSFC under Project No. 11721404. CW acknowledge support from Ministry of Science and Technology of China, Grant No.\ 2016YFA0301001, the National Natural Science Foundation of China, Grants No.\ 11674188 and 51788104.
\end{acknowledgments}

\appendix

\section{Review of bands, bundles and space groups }\la{sec:preliminaries}


\subsection{Bands, vector bundles, and topological triviality} \la{sec:bandbundle}

A rank-$N$ band is given by $N$ orthogonal Bloch functions at each wavevector $\bk$ in the Brillouin torus. The Bloch functions span an $N$-dimensional vector space at each $\bk$, and the union of all such vector spaces (over the base space of the Brillouin torus) defines a rank-$N$ vector bundle. We typically use $P(\bk)$ to denote the rank-$N$ projector to the vector space at $\bk$, and $P=\sum_{\bk}P(\bk)$ as the projector to the full band; $\sum_{\bk}$ is our shorthand for a direct integral splitting\cite{Reed} over the Brillouin torus, and $P(\bk)$ is periodic in reciprocal-lattice translations.\\

To minimize notation, we sometimes use $P$ to denote the band itself, and not just the projector to the band.  As a case in point, suppose $P$ is an energy band (of a tight-binding or Schr\"odinger-type Hamiltonian) that is energetically isolated, that is, having an energy gap separating $P$ from higher- and lower-energy bands, at each $\bk$. If the tight-binding Hamiltonian has matrix elements that decay exponentially in real space, then $P(\bk)$ is an analytic function of $\bk$ throughout the Brillouin zone.\cite{Cloizeaux1964} As a shorthand, we will just say that $P$ \textit{is analytic throughout the Brillouin zone}. This analyticity condition also holds given a physically reasonable condition on the Schr\"odinger-type Hamiltonian (${\bp}^2/2m+V(\br)$), namely that the potential term $V$ is square-integrable over the unit cell.\footnote{See the Kato-Rellich theorem in \ocite{Reed}.}\\

Given a rank-$N$ band/bundle, if there exist $N$ Bloch functions which span the $N$-dimensional vector space at each $\bk$ and are both continuous and periodic over the Brillouin torus, then the band is said to be \textit{topologically trivial} as a complex vector bundle. The distinct notion of topological triviality for real vector bundles is only mentioned briefly in \s{sec:discussion}.  Everywhere else, `topologically trivial' should always be understood as in the category of  complex vector bundles.\\

In spatial dimension $d\leq 3$ (which is assumed throughout), a band is topologically trivial if and only if it \textit{has trivial first Chern class};\cite{Brouder2007,Panati2007} for $d=3$, this means that the first Chern number vanishes over any two-dimensional submanifold of the Brillouin three-torus; for $d=1$ the first Chern class is always trivial.\\

In this work, all bands are assumed topologically trivial, unless otherwise specified. Applying the Oka-Grauert theorem,\cite{Grauert1958,Huckleberry2013} topological triviality implies that the $N$ Bloch functions can also be chosen analytic in $\bk$, which further implies that their Fourier transforms -- known as Wannier functions -- are exponentially localized in real space.\cite{Brouder2007} Therefore, a topologically trivial rank-$N$ band is equivalent to $N$ orthogonal Wannier functions in each unit cell, with other Wannier functions related by discrete Bravais lattice translations. The specification of such Wannier functions which equivalently span a band shall be called a \textit{Wannier basis}.

\subsection{Space groups, point groups, and Wigner-Dyson symmetry classes}\la{sec:spacegroupwignerdyson}

Let $g$ denote an isometry in $d$  spatial dimensions, possibly combined with the reversal of time.  $g=(\bt_g|\check{g})$ may be decomposed into  point-preserving and translational components, with  $\check{g}$ a $d\times d$  orthogonal matrix and $\bt_g \in \R^d$. $g$ has the following action  on spacetime: $\br \rightarrow g\circ\br=\check{g}\br+\bt_g$ and
 $t\rightarrow s_g t$; $\check{g}\br$ should be understood as a matrix multiplying a $d$-component vector $(x,y,\ldots)$, and $s_g=-1$ for $g$ that reverses time. \\

Not including time reversal, all spatial isometries that preserve a crystal form a \textit{crystallographic space group} $G$;  there are 230 such groups in  spatial dimension $d=3$, and 17 such groups in $d=2$; the latter are also known as wallpaper groups. $\calt_d$ denotes the translational subgroup of $G$, where the subscript $d$ equals the number of linearly-independent translation vectors. The  \textit{crystallographic point group} of $G$ is defined as the quotient group $\calp=G/\calt_d$. There are 32 crystallographic point groups in $d=3$, which are further categorized into 6 crystal families. For example, the cubic crystal family consists of the three tetrahedral and two octahedral point groups, which are therefore also referred to as the five \textit{cubic} point groups.\\

A \textit{grey magnetic space group}, denoted $G_T$, is a direct product of any crystallographic space group $G$ with $\Z_2^T$, the order-two group generated by time reversal $T$. (In general, $\Z_n^g$ will denote a cyclic group of order $n$ and generated by $g$, i.e., $g^n=e$ with $e$ the identity element.)  $T^2=e$ corresponds to \textit{Wigner-Dyson symmetry class AI}. The point group of $G_T$ will be referred to as a \textit{grey magnetic point group}.\\

Throughout this work we are concerned only with linear representations of groups, and not their projective representations.  In particular, linear representations of a crystallographic point group transform with integer spin. We shall also only concern ourselves with the half-integer-spin, linear representations of \textit{double crystallographic group} $\tilde{G}$ and the \textit{double grey magnetic space group} $\tilde{G}_T$; these groups are respectively the double covers of $G$ and $G_T$. In a double cover, we introduce an additional element $\tilde{e}$ that squares to the identity and  physically corresponds to a $2\pi$ rotation. For double grey magnetic groups, $T^2=\tilde{e}$
corresponds to  Wigner-Dyson class AII; in particular, the double cover of $\Z_2^T$ is $\Z_4^T$. \\ 

Except in \s{sec:localization_obstruction}, all of our results (including the splitting theorem of \s{sec:equivalencetheorems}) hold for both symmorphic and nonsymmorphic space groups. A symmorphic space group is a semidirect product of a point group with a translational subgroup, which shall be denoted $\calt_d \rtimes \calp$; a nonsymmorphic space group is a space group that is not symmorphic.\\

\noindent \textit{Example of symmorphic space group} $\calt_3\rtimes C_{4v}\times \Z_2^T$, a grey magnetic space group, is the symmetry of the tetragonal photonic crystal in  \s{sec:tetragonal}. The crystallographic point group $C_{4v}$ is generated by a four-fold rotation $C_4$ and a mirror reflection $\mir_x$ with mirror plane containing the rotational axis.  It is convenient to adopt Cartesian coordinates
with $C_4: (x,y,z)\rightarrow (-y,x,z)$, and $\mir_x: (x,y,z)\rightarrow (-x,y,z)$. $\calt_3\rtimes C_{4v}$ can only be the crystallographic space group $P4mm$ (number 99). \\

\noindent Implicit in the semidirect notation is an action of $\calp$ on $\calt_d$, and inequivalent actions may result in inequivalent space groups, as we next illustrate.\\

\noindent \textit{Example} $\calt_3\rtimes C_{3v}\times \Z_2^T$ is the symmetry of the hexagonal photonic crystal in  \s{sec:hexagonal}. $C_{3v}$ is generated by a three-fold rotation $C_3$ and a mirror plane $\mir_d$ containing the rotational axis. There are two symmorphic crystallographic space groups ($P31m$ and $P3m1$) with the point group $C_{3v}$, and we will use $\calt_3\rtimes C_{3v}$ as a synonym for $P31m$. $P31m$ is distinguished by having  $\mir_d$ relate two of three rotation-invariant Wyckoff positions, the third position being reflection-invariant.\\

If `space group' is used in a sentence without any of the above qualifiers, it is safe to assume that the sentence applies to all categories of space groups. As a case in point, to any space group $G$ we may associate a \textit{Wyckoff position} $\bvarpi\in \R^d$ and a \textit{site stabilizer} $G_{\varpi}=\{g\in G|g\circ \bvarpi=\bvarpi\}$. The site stabilizer consists of all elements in $G$ that preserve  the Wyckoff position.

\subsection{Representations of space groups}\la{sec:repspacegroup}

A spacetime isometry $g\in G$ may be represented by an operator $\hat{g}$, which acts on functions of real space as  $\hat{g}f(\br){=} \overline{f(g^{\mo}{\circ} \br)}^{s_p}$, where $\bar{a}^{1}{:=}a$  and $\bar{a}^{{-}1}{:=}\bar{a}$ (the complex conjugate).\\

The symmetry representation of $g$ on Bloch functions $\{\ket{\psi_{j,\bk}}\}_{j=1}^N$ is defined by a unitary $N\times N$ `sewing' matrix $U_g(\bk)$:
\e{ \hat g \ket{ \psi_{j,\bk} } = U_g(\bk)_{j' j} \ket{ \psi_{j',-s_g \check g \bk} }.  \la{ksymm1} }
$U_g(\bk)$ can be explicitly expressed in terms of the normalized cell-periodic component of  Bloch functions: $u_{j,\bk}(\br) = \mathrm{e}^{-i \bk \cdot \br} \psi_{j,\bk}(\br)$, as
\e{ U_g(\bk)_{j' j} = \braket{ u_{j',s_g \check g \bk}}{ \hat g(\bk) | u_{j,\bk} }_{\mathrm{cell}}, \la{ksymm} }
with $\hat g(\bk) := \mathrm{e}^{-i s_g \check g \bk \cdot \bt_g} \hat g$ and $\braket{ . }{ . }_{\mathrm{cell}}$ denoting an integral over $\br$ in one unit cell (possibly with a summation over spin).\\


We say that a band (with projector $P$) transforms as a \textit{representation of the space group} $G$,  if $[\hat{g},P]=0$ for all $g\in G$. In short,  $P$ is referred to as a representation of $G$. \\

Let $P_0$ and $P_1$ be two representations of $G$ with equal rank, and being both analytic throughout the Brillouin zone. $P_0$ and $P_1$ are said to be  \emph{equivalent} if there exists a continuous interpolation $\{P_t \}_{t\in [0,1]}$  that preserves analyticity (throughout the Brillouin zone) and the symmetry condition $[\hat{g},P_t]=0$, for all $g\in G$ and all $t\in [0,1]$. In bundle language, this means that $P_0$ and $P_1$ are isomorphic as $G$-vector bundles, as elaborated in \app{app:atomicBR}. \\


Space-group representations fall into two categories: band representations (cf\ \app{sec:zakdefinesbr}, \app{app:locallysymmetricWannierbasis}) and obstructed representations (cf\ \app{sec:toprep}).

\subsubsection{Zak's definition of  band representations}\la{sec:zakdefinesbr}

In the standard definition by Zak, a \textit{band representation of a space group} $G$, denoted BR$(G,\bvarpi,D)$, is a representation of $G$ that is induced from a representation ($D$) of a site stabilizer $G_{\varpi}$.\\ 

We briefly review this \textit{induction process}:  begin with a set of exponentially-localized Wannier functions centered on the Wyckoff position $\bvarpi$, and transforming in a representation $D$ of the site stabilizer $G_{\varpi}$. By application of $G$ on these Wannier functions, we generate an infinite set of Wannier functions which form a representation of $G$.  Such Wannier functions that are obtained by induction will be said to form a \textit{locally-symmetric Wannier basis} for the BR; we elaborate on this point of view next.

\subsubsection{Equivalent formulation of band representations by the locally-symmetric Wannier basis}\la{app:locallysymmetricWannierbasis}

In various proofs throughout this work, it is useful to have an equivalent definition of band representations that emphasizes the symmetry properties of Wannier functions:  $P$ is a BR of $G$ if and only if $P$ is a (finite) direct sum of (infinite-dimensional) subspaces, each of which is spanned by a \textit{locally-symmetric Wannier basis}. This equivalent definition of BRs has been proven in Appendix A of \ocite{nogo_AAJH}. \\

\noindent A \textit{locally-symmetric Wannier basis $\{\ket{w_{n,\bR}^{\alpha}}\}_{n, \alpha, \bR \in \mathcal{T}_d}$ with Wyckoff position $\bvarpi_1$} is an orthonormal basis of an infinite-dimensional representation of $G$, which satisfies the following properties for all $n=1,...,M$, $\alpha=1,...,A$, $(\bR|e) \in \calt_d$: 
	\begin{enumerate}
		\item $\ket{w_{n,\bR}^{\alpha}}=\widehat{(\bR|e)} \ket{w_{n,\bze}^{\alpha}}$,
		\item $\ket{w_{n,\bze}^{\alpha}}$ is exponentially-localized,
		\item $\{\ket{w_{1 \bze}^{\alpha}} \}_{\alpha=1}^A$ spans an $A$-dimensional representation of $G_{\varpi_1}$,\la{onsiterep}
		\item $\{\ket{w_{n,\bze}^{\alpha}}\}_{\alpha}$ and $\hat g_n \{\ket{w_{1,\bze}^{\alpha}}\}_{\alpha}$ span the same $A$-dimensional representation of $G_{\varpi_n} = g_n G_{\varpi_1} g_n^{-1}$,\la{equiv}
	\end{enumerate}
	where $G/(\calt\rtimes G_{\varpi_1}) = \{ [g_1=(\textbf{0}|e)], [g_2], ..., [g_M] \}$ is a coset decomposition of $G$. \\ 

\noindent It is worth clarifying that property \ref{equiv} was not stated explicitly in  Definition 4 in Appendix A of \ocite{nogo_AAJH}, however the property was implicitly assumed.

\subsubsection{Fragile vs stable obstructed representations}\la{sec:toprep}


An \textit{obstructed representation of  a space group} $G$ is a representation of $G$ that is not a band representation of $G$ (in Zak's definition).  The filled, low-energy band  of a $G$-symmetric topological insulator is an obstructed representation of $G$.\\

If $G$ has a trivial point group, then an obstructed representation of $G$ is equivalent to\cite{Brouder2007} the topological nontriviality of the corresponding complex vector bundle. This is not generally true if the point group of $G$ is nontrivial, e.g., for $G=\calt_2\times \Z_4^T$ (Wigner-Dyson class AII) all its representations necessarily constitute a topologically trivial complex vector bundle, owing to the time-reversal symmetry. In particular, the filled band of the $\Z_2$ Kane-Mele topological insulator is both topologically trivial (as a complex vector bundle) and an obstructed representation (of $G$); a proof of the latter statement is given in \s{sec:zakphasewind}.  \\


Obstructed representations of a space group $G$ may be further subdivided into fragile obstructed and stable obstructed. A \textit{fragile obstructed representation} (FOR) of $G$ is an obstructed representation of $G$, with the property that a BR of $G$ exists, such that the direct sum of this BR with the FOR is a higher-rank band representation. A \textit{stable obstructed representation} of $G$ is an obstructed representation of $G$ that is not fragile obstructed.

\section{Monomial representations of finite groups}\la{app:finitegroup}

In the main text we have amply used the notion of monomial representations of point groups. Monomial representations can equivalently be viewed as induced representations (from one-dimensional representations of subgroups) [cf.\ \app{app:induced}] or as complex permutation representations [cf.\ \app{app:complexpermrep}]. Both views are elaborated pedagogically in this appendix, and their equivalence established in \app{app:equivcomplexperminduced}. Lastly, we prove a useful lemma for monomial direct-product groups in \app{app:monomialdirectproduct}.\\

Throughout this \app{app:finitegroup}, we let $H$ denote a finite group; a representation $(U,V)$ of $H$ on an $n$-dimensional representation space $V$ is given by a map
$h \rightarrow U(h)$, with $U(h)$ generally an $n$-dimensional unitary matrix. If $H$ is a point group (consisting of discrete spatial isometries that preserve a point in space), then $h_1h_2\rightarrow U(h_1)U(h_2)$. If $H$ is a magnetic point group, 
\e{ h_1h_2 \rightarrow U(h_1)\overline{U(h_2)}^{s(h_1)},\la{corep}}
where $\bar{a}^{s(h)}$ equals the complex conjugate of $a$ if $h$ involves time reversal, and otherwise  $\bar{a}^{s(h)}=a$. \q{corep} is the multiplication rule for corepresentations.\cite{wignerbook,magnetic_groups}

\subsection{Induced representations}\la{app:induced}

Let $H$ be a finite group with subgroup $A$, and $(\Pi,V)$ be a representation of $A$, with $V$ an $n$-dimensional representation space. For every $a\in A$ and basis vector $v_{\alpha} \in V$, $a$ acts on $v_{\alpha}$ as $a\circ v_{\alpha} = \sum_{\beta=1}^n\Pi(a)_{\ab}v_{\beta},$ with $\Pi(a)$ an $n$-dimensional matrix.  We define $D$ as the index of $A$ in $H$, and $\{h_1=e,h_2\ldots,h_D\}$ as a full set of representatives of the left cosets $H/A$, such that $H$ can be decomposed as $H=\cup_{i=1}^Dh_i A$. For any $g\in H$ and representative $h_i$, $gh_i\in H$, and therefore by the coset decomposition there exists $h_{\sigma_g(i)}$ [with $\sigma_g(i)\in \{1,\ldots,D\}$] and $a_i\in A$ such that $g h_i =h_{\sigma_g(i)}a_i$.   It will be useful to show that  $\sigma_g$ is a permutation of $\{1,\ldots,D\}$.\\

\noindent \textit{Proof of permutation} Let us first prove that $\sigma_g$ is injective. Suppose it were not, i.e.,  $\sigma_g(i)=\sigma_g(i')$ for $i\neq i'$. It would follow that $h_ia_i^{-1}= h_{i'}a_{i'}^{-1}$. Since $a_i,a_{i'}\in A$,  $h_ia_i^{-1}\in h_iA$ and $h_{i'}a_{i'}^{-1}\in h_{i'}A$ must belong in distinct cosets of $H/A$, which contradicts the just-stated equality. An injective map from  $\{1,\ldots,D\}$ to itself must also be surjective, hence $\sigma_g$ is a permutation.\\

Let $Ind_A^H(\Pi,V)$ denote the \textit{induced representation} of $(\Pi,V)$. The representation space of $Ind_A^H(\Pi,V)$ is $W=\oplus_{i=1}^Dh_iV$, with basis vectors $\{h_iv_{\alpha}|i=1\ldots D, \alpha=1\ldots n\}$;  $g\in H$ is defined to act on  the basis vector as
\e{ g \circ h_iv_{\alpha} = \sum_{\beta=1}^n[\Pi(a_i)]_{\ab}h_{\sigma_g(i)} v_{\beta}. \la{inducedrep}}

\subsection{Complex permutation representations}\la{app:complexpermrep}

A \textit{complex permutation representation}  is a representation of $H$ where every element of $h\in H$ is mapped to a complex permutation matrix $U(h)$, which satisfies the multiplication rule of \q{corep}.  \\

In what follows we maintain a basis $\{ v_1, \ldots, v_n\}$ for the representation space $V$ such that each $h\in H$ is represented by a complex permutation matrix. It is useful to introduce the notion of a \textit{transitive} complex permutation representation. By `transitive', we mean that for every pair of basis vectors  ($v_i,v_j$), there exists $h\in H$ such that $\big[ U(h) \big]_{ij}$ is nonzero. \\

\noindent \textit{Claim.} A complex permutation representation is either transitive, or it is a direct sum of transitive complex permutation representations.\\

\noindent \textit{Proof.} For every complex permutation representation $U \colon h \mapsto U(h)$ of $H$, there exists a (real) permutation representation $U'$ of $H$ obtained by replacing every nonzero matrix element in $U(h)$ by $1$. 
Then $H$ has the following permutation group action, denoted $\circ$, on the set of basis indices $N:=\{1,\ldots,n\}$: $h \circ i = j$ for the unique $j$ for which  $[U'(h)]_{ij}=1$ 
-- in this case, we say that $i$ is in the orbit of $j$. (The orbit of $j$ is the equivalence class of $j$.) The set of equivalence classes (or orbits) forms a partitioning of $N$. This partitioning then implies a splitting of $V$, where each summand is spanned by all basis vectors with indices in one orbit. By construction, the restriction of $\{U(h)\}_{h\in H}$ to one summand is transitive. \hfill\(\Box\) 

\subsection{Complex permutation representations as induced representations}\la{app:equivcomplexperminduced}

A transitive complex permutation representation of $H$ is equivalently a representation of $H$ induced from a one-dimensional representation of a subgroup of $H$. The proof of the forward direction (transitive complex permutation representation $\imp$ induced representation) may be found in the proof of Theorem 2.6 in \ocite{Bray1982}. Here we provide an elementary proof of the backward direction, which we did not find in the standard literature.\\

\noindent \textit{Proof of equivalence} Consider $Ind_A^H(\Pi,V)$ with $V$ a one-dimensional vector space. For every $h\in H$, $\Pi(h)$ is a unimodular phase factor. As a particular case of \q{inducedrep}, the action of $g$ on a basis vector is $g\circ h_iV = \Pi(a_i) h_{\sigma_g(i)} V,$ with $\sigma_g$ a permutation on $\{1\ldots D\}$. The representation of $g$ in the basis of $\{h_1V,h_2V,\ldots,h_D V\}$  must therefore be a complex permutation matrix; since this is true of all $g$,  $Ind_A^H(\Pi,V)$ must be a complex permutation representation. Moreover this complex permutation representation is transitive, since for any pair of basis vectors $h_iV$ and $h_jV$, there exists $h_ih_j^{-1}\in H$ which relates the two vectors (modulo a phase), and therefore $\big[ U(h_ih_j^{-1}) \big]_{ij}\neq 0.$ \hfill\(\Box\)\\

From this equivalence it follows that (i) a complex permutation representation of $H$ (being a direct sum of transitive complex permutation representations) is equivalently  (ii) a direct sum of representations of $H$ (each induced from a 1D representation of a subgroup of $H$). (i-ii) may be taken as equivalent definitions of  a \textit{monomial representation} of $H$.

\subsection{Direct-product groups that are monomial}\la{app:monomialdirectproduct}

As a reminder, a \textit{monomial group} is a group for which all irreducible representaitons (irreps) are monomial. For example, abelian finite groups are monomial because all their irreps are 1D. \\

\noindent \textbf{Lemma for monomial direct-product groups.} 
For $H$ a group and $A$ an abelian group, $H$ is monomial if and only if $H \times A$ is monomial. \\

\noindent \textit{Proof of Lemma.} 
It is well-known that \textit{all} irreducible representations of a direct product (of two groups) are obtained by the tensor product of irreps  of the individual groups.\cite{tinkhambook} In our application, $A$ being Abelian implies it has only one-dimensional irreps which we label by $\eta$: any $a\in A$ is mapped to complex phase factor $\eta(a)\in U(1)$. We label an irrep of $H$ by $D$, which maps $h\in H$ to a unitary matrix $D(h)$. Any irrep of $H\times A$ can then be labelled by $(D,\eta)$ and maps $(h,a)\in H\times A$ to the unitary $\eta(a)D(h)$.\\

Let us first prove the forward direction of the lemma: $H$ monomial $\Rightarrow$ $H\times A$ monomial.  By assumption, for any irrep $D$ of $H$, there exists a basis for $D$ such that $D(h)$ is a complex permutation matrix for all $h\in H$. The tensor product of such a basis with $\eta$ gives a basis for $(D,\eta)$ where $\eta(a)D(h)$ is a complex permutation matrix for all $(h,a)\in H\times A$. This is because any complex permutation matrix that is multiplied by a complex number [here, $\eta(a)$] remains a complex permutation matrix. Since the above holds for all irreps of $H\times A$, we deduce that $H\times A$ is monomial. \\

Lastly, we will prove the backward direction, which is contrapositively restated as: $H$ not monomial   $\Rightarrow$ $H\times A$ not monomial. By assumption, there exists at least one irrep $D$ (of $H$) having no basis in which $D(h)$ is a complex permutation matrix for all $h\in H$. This implies, for $a$ being the identity element $e\in A$, that no basis exists for $(D,\eta)$ where $\eta(e)D(h)=D(h)$ is a complex permutation matrix for all $\{(h,e) | h\in H\}$. Consequently, no basis exists for which $\eta(a)D(h)$ is a complex permutation matrix for all $(h,a)\in H\times A$; hence $(D,\eta)$ is a non-monomial irrep of $H\times A$, which completes the proof.\hfill\(\Box\)

\section{Proof of crystallographic splitting theorem}  \la{app:permthm}

This appendix contains the proof of the crystallographic splitting theorem for monomial band representations, as stated in \s{sec:state_equivthm}. Below, steps 1-3  outline the proof of the forward arrow (existence of splitting satisfying (A-B) $\imp P$ is a monomial BR of $G$), and 4 the backward arrow.\\

\noi{1.} In \s{sec:partition}, we prove the existence of a  splitting $P=\oplus_{i}P^{(i)}$ (the sum over $i$ is finite), with each $P^{(i)}$ a single orbit under $G$. By this, we mean that (a) $P^{(i)}$  is a direct sum of a subset of $\{P_j\}_{j=1}^N$, (b) $P^{(i)}$ forms a representation of $G$, and (c) the action of $G$ on members of $P^{(i)}$ is transitive: for any $P_j,P_{j'}$ in the direct sum of $P^{(i)}$, there exists $g\in G$ such that $gP_jg^{-1}= P_{j'}$. \\

\noi{2.}  Since each unit-rank $P_j$ is analytic with trivial first Chern class, it has a Wannier representation with a corresponding Wannier center (defined up to lattice translations). By `Wannier center, we mean the expected position of a Wannier function in a Wannier basis for $P_j$. It is possible that the Wannier centers for different $P_j$ (contained in the same orbit) are identical. We will show in \s{sec:divisibility} that for each orbit, the number of distinct Wannier centers ($A$) divides the rank of $P^{(i)}$. This means that there are (in each unit cell) the same number ($M$) of Wannier functions with the same Wannier center (denoted as $\bvarpi_{\alpha}$, with $\alpha=1,\ldots,A$); we introduce $\mu=1,\ldots,M$ as an additional label to distinguish Wannier functions centered at the same position. It follows from this discussion that  we can always decompose
\e{ P^{(i)}= \oplus_{\alpha=1}^{A} \oplus_{\mu=1}^{M} P^{(i)}_{\alpha,\mu}, \la{decomposeagain}}
such that $P^{(i)}_{\alpha,\mu}$ has unit rank and projects to  Wannier functions indexed by $(\alpha,\mu)$.  $P^{(i)}_{\alpha,\mu}$ then gives us  a convenient relabelling of $P_j$.  \\


\noi{3.} In \s{sec:constructLSWB}, we construct a rank-$N$ Wannier basis by induction from a single Wannier function  arbitrarily chosen from $P^{(i)}_{\alpha,\mu}$; the choice of $P^{(i)}_{\alpha,\mu}$ among $\{P^{(i)}_{\alpha,\mu}\}_{\alpha,\mu}$ is also arbitrary. 
It will be proven that this Wannier basis spans $P^{(i)}$, and is induced from a mononimal representation of a site stabilizer under $G$. Having thus proven that $P^{(i)}$ is a monomial BR (of $G$) completes the proof of the forward direction. \\

\noi{4.} In the proof of the backward arrow, we then assume that $P$ is a representation of $G$ induced from a monomial representation $D$ of a site stabilizer $G_{\bvarpi_1}$. We then choose a basis for the representation space of $D$ such that  each $g\in G_{\bvarpi_1}$ is represented as a complex permutation matrix. We will demonstrate that this basis gives a splitting of $P$ into single-rank projectors which are permuted by any element of $G$, thus proving the backward direction.

\subsection{Partitioning of band into space-group orbits}\la{sec:partition}

We would like to decompose the band projected by $P$ into subbands which are individually invariant under $G$. For this purpose,  it is useful to define $H$ as the group of \textit{all} symmetries (contained in $G$) that has a trivial action on each of $P_j$:
\e{ H =\{ g \in G |\forall j\in \{1\ldots N\}, \sigma_g(j)=j \} < G. \la{definemaximalsubgroup}}
Let us prove that $H$, as defined in \q{definemaximalsubgroup}, is a normal subgroup of $G$.\\

\noindent \textit{Proof of normality.} By definition of $H$,
\e{ \forall h \in H,\as \forall j=1,\ldots,N, \as[h,P_j]=0.}
Since $h$ commutes with the right-hand side of 
\e{ g P_j g^{-1} = P_{\sigma_g(j)},}
it follows that
\e{   [h,g P_j g^{-1}]=0 \imp [g^{-1}hg,P_j]=0.}
Since the above is true for all $j=1,\ldots,N$, $g^{-1}hg$ must act as the trivial permutation, and therefore belongs in $H$. This holds for all $g \in G$ and $h\in H$; therefore, $gH=Hg$ as desired. \hfill\(\Box\)  \\

Since $H$ is a normal subgroup, the quotient $G/H$ is a group whose order $|G/H|$ is defined as the \textit{index} of $H$.  Let each equivalence class in $G/H$ be represented by an element $f_j\in G$, such that
\e{ G/H=\{[f_1=e],[f_2],\ldots [f_{|G/H|}]  \};}
$e$ above is the identify element in $G$, so $[e]$ consists of all elements in $H$. Because $H$ acts as the trivial permutation, $\sigma_f=\sigma_{[f]}$ depends only on the equivalence class of $G/H$.
It is useful to view $\sigma_f$ as defining a group action for $G/H$  on $\{P_j\}_j$, with $[e]$  acting trivially, and the compatibility condition given by 
\e{  &P_{\sigma_{f_2f_1}(j)} = \hat{f}_2 \hat{f}_1 P_j (\hat{f}_2\hat{f}_1)^{-1} \lin
	&=\hat{f}_2 (\hat{f}_1 P_j \hat{f}_1^{-1})\hat{f}_2^{-1} = P_{\sigma_{f_2}\sigma_{f_1}(j)}. \la{compatibility}}


The \textit{orbit} of $P_j$ is defined as the subset of $\{P_j\}_{j=1}^N$ to which $P_j$ can be moved by elements in $G/H$:
\e{ G/H\cdot P_j := \{ P_{\sigma_{[f]}(j)} | [f] \in G/H\}. }
The set of orbits of $\{P_j\}_{j=1}^N$ under $G/H$ (a group) form a partition of $P=\oplus_i P^{(i)}$ (a grouping of $\{P_j\}_{j=1}^N$ into non-empty subsets $P^{(i)}$, such that each element of $\{P_j\}_{j=1}^N$ is included in one and only one subset). Every orbit is an invariant subset on which $G/H$ acts transitively, i.e., for every pair $P_{j'},P_j$ in the orbit, there exists  $[f]\in G/H$ such that $\sigma_{f}(j')=j$). \\

Let us focus on one orbit in the partition  $P=\oplus_{i=1}P^{(i)}$
with rank $N^{(i)}$.
Since the following proof would be valid for any orbit, we may simplify notation by dropping the orbit index ($i$): 
without loss of generality, we relabel
$P=\oplus_{j=1}^NP_j$ as the projector for a \textit{single} orbit under $G$.


\subsection{Lemma on the group action on Wannier centers}\la{sec:divisibility}

Since each of $P_j$ is analytic with trivial first Chern class, it must be localizable, i.e., it has a Wannier representation -- with a corresponding  Wannier center that is uniquely determined modulo lattice translations. Since $P_j$ is invariant under $H$ [cf.\ \q{definemaximalsubgroup}], $P_j$ must be a BR of $H$ (according to the unit-rank splitting theorem), and its associated Wannier center is invariant (modulo lattice translations) under $H$.  It is possible that the Wannier centers of distinct $P_j$ are equivalent (modulo lattice translations); this  defines a  surjection  $P_j \mapsto \bvarpi_{S(j)}$, with $j=1,\ldots,N$, $S(j)= 1,\ldots, A$, and $A\leq N$. \\

\noindent \textbf{{Lemma 1.}} There exists a group action of $G/H$ on the set of single-rank projectors and the set of Wannier centers (defined modulo lattice translations), i.e., for any $[g]\in G/H$
\e{ g: (\bvarpi_{\alpha},P_j) \mapsto  (\bvarpi_{g\cdot \alpha},P_{\sigma_g(j)}), \la{groupactionprojcenter}}
with $g\cdot \alpha$ a permutation on $\alpha\in \{1\ldots A\}$, defined through
\e{  \bvarpi_{g\cdot \alpha} \equiv g\circ \bvarpi_{\alpha}. \la{definegcdot}}
The group action of \q{groupactionprojcenter} is transitive, and satisfies
\e{ g\cdot (S(j)) =S ( \sigma_g(j)), \ins{for} j=1,\ldots,N. \la{fakecompatibility}}
From this we will show that  $A$ divides $N$. \\

\begin{center}
\textit{Proof of Lemma 1}
\end{center}

Since $P_j$ is localizable, so would be $gP_jg^{-1}$ for any $g\in G$. This is because any crystallographic symmetry acts as an isometry in real space: $\br \rightarrow g \circ \br$, and therefore cannot change the exponential decay of Wannier functions. Consequently, if $P_j$ has a Wannier center $\bvarpi_{\alpha=S(j)}$,  $gP_jg^{-1}$ would have the Wannier center $g\circ \bvarpi_{\alpha}$. We write this as in \q{groupactionprojcenter}.
Since $G$ has an action on $\{P_j\}_j$, and $\{P_j\}_j$ a surjection to $\{\bvarpi\}_{\alpha}$, it must be that 
$g\circ \bvarpi_{\alpha}\equiv \bvarpi_{\alpha'}$ for the unique $\alpha'\in \{1,\ldots,A\}$ satisfying $\alpha'=S(\sigma_g(j))$. We define the permutation $g\cdot$ through \q{definegcdot},
so that \q{fakecompatibility} follows immediately.\\ 

Now, we would show that $g\cdot$ defines a group action of $G/H$ on $\{\bvarpi_{\alpha}\}_{\alpha}$. Indeed, the identity element  $[e]\in G/H$ includes all   $h\in H$, and  $h\circ \bvarpi_{\alpha}\equiv \bvarpi_{\alpha}$ because $H$ trivially permutes $\{P_j\}_j$ [cf.\ \q{definemaximalsubgroup}]. The compatibility axiom is also satisfied:
\e{ \bvarpi_{ (g_1g_2)\cdot \alpha}\equiv (g_1g_2)\circ \bvarpi_{\alpha}=g_1\circ (g_2\circ \bvarpi_{\alpha}) \equiv \bvarpi_{g_1\cdot (g_2\cdot \alpha)}. }

Let us show that $g\cdot$ is transitive as a group action, i.e., for any $\bvarpi_{\alpha}$ and  $\bvarpi_{\alpha'}$, there exists $[g]\in G/H$ such that $\alpha'=g\cdot \alpha$. This $g$ is determined (possibly non-uniquely) by the transitive group action of $G/H$ on $\{P_j\}_j$. To clarify, if $S(j)=\alpha$ and $S(j')=\alpha'$, then we determine $g$ through $\sigma_g(j)=j'$. \\

Finally we apply the transitivity property to prove that $A$ divides $N$. Indeed, suppose $S$ maps $M$  elements (denoted $\{{j_1},{j_2},\ldots,{j_{M}}\}$)  to a single element $\alpha$. By the transitivity property, for any element $\alpha'$ distinct from $\alpha$, there exists a nontrivial element $[p]\in G/H$ such that $p\cdot \alpha=\alpha'$. By the condition of \q{fakecompatibility}, 
\e{ S: \{ \sigma_p(j_1),\ldots, \sigma_p(j_{M})\} \mapsto \alpha'. }
Crucially, $\{ \sigma_p(j_1),\ldots, \sigma_p(j_{M})\}$ must not intersect $\{j_1,\ldots, j_{M}\}$, because the two sets map to distinct elements under $S$. If we repeat the logic for all other distinct elements of $\{\bvarpi_{\alpha}\}_{\alpha=1}^A$, we conclude that for any element $\alpha''$ (distinct from $\alpha$ and $\alpha'$), there corresponds $M$ elements in $\{1,\ldots,N\}$ which do not intersect with   $\{j_1,\ldots,j_{M}\}$ or $\{ \sigma_p(j_1),\ldots, \sigma_p(j_{M})\}$. It follows that if $\{\bvarpi_{\alpha}\}_{\alpha=1}^A$ has $A$ distinct elements, then $N=MA$ as desired. This completes the proof of the lemma. \hfill\(\Box\) \\

\begin{center}
\textit{Implications of Lemma 1}
\end{center}

Here we collect some useful implications of the lemma and introduce the definitions of certain stabilizer groups, as will be applied in \s{sec:constructLSWB}.\\

The lemma implies that we are able to decompose $P$ (of a single orbit) into a sum of single-rank projectors $P_{\alpha,\mu}$ [cf.\ \q{decomposeagain}], with $P_{\alpha,\mu}$ a relabelling of $P_j$; we remind the reader that the orbit index $i$ has been dropped for notational simplicity.\\

Due to the transitivity of the group action [cf.\ \qq{groupactionprojcenter}{definegcdot}], for any pair $P_{\alpha,\mu}$, $P_{\alpha',\mu'}$, there must exist  $[p]\in G/H$ such that
\e{ pP_{\alpha,\mu}p^{-1} = P_{\alpha',\mu'}, \as p \circ \bvarpi_{\alpha} = \bvarpi_{\alpha'}\la{pickp}}
holds.  In particular, for $\alpha=\alpha'=1$, 
\e{ &g_{\mu} P_{1,1}g_{\mu}^{-1}=P_{1,\mu},  \;\; g_{\mu}\circ \bvarpi_1= \bvarpi_1,\la{defgmu}}
defines $g_{\mu}$; if more than one element of $G$ satisfies \q{defgmu}, then we may arbitrarily denote one representative as $g_{\mu}$, and we may as well take $g_1$ to be the identify operation. The second equality in \q{defgmu} identifies $g_{\mu}$ as an element in the site stabilizer
\e{G_{\bvarpi_1}:= \bigg\{g\in G\;\bigg|\; g \circ \bvarpi_1 = \bvarpi_1\bigg\}.  \la{defGvarpi} }
Similarly, restricting \q{pickp} 
to $\mu=\mu'=1$,
\e{ &p_{\alpha} P_{1,1}p_{\alpha}^{-1}=P_{\alpha,1},  \;\; p_{\alpha}\circ \bvarpi_1= \bvarpi_{\alpha},\la{defgalpha}}
defines $p_{\alpha}$, with $p_{1}$ the identity operation. Due to the assumed transformation of the Wannier center $p_{\alpha}\circ \bvarpi_1 = \bvarpi_{\alpha}$, it must be that 
\e{p_{\alpha}P_{1,\mu}p_{\alpha}^{-1}= P_{\alpha,p_{\alpha}\cdot \mu},\la{palphaacts}} 
with $p_{\alpha}\cdot\mu $ a permutation on the $\mu$ index; $p_{\alpha}\cdot 1=1$ according to \q{defgalpha}.\\

It will be useful to define $G_{\alpha,\mu}$ as the stabilizer of $P_{\alpha,\mu}$ under $G$:
\e{ G_{\alpha,\mu}:=\bigg\{g\in G \;\bigg|\;  [g,P_{\alpha,\mu}]=0\bigg\},\la{defG1mu}}
and the site stabilizer of $\bvarpi$ under $G_{\alpha,\mu}$ as
\e{G_{\alpha,\mu,\bvarpi}:= \bigg\{g\in G_{\alpha,\mu}\;\bigg|\; g \circ \bvarpi = \bvarpi\bigg\}.  \la{defG1muvarpi} }
It follows from \q{palphaacts} that 
the stabilizers $G_{1,\mu}$ and $G_{\alpha,p_{\alpha}\cdot \mu}$ are conjugate:
\e{  G_{\alpha,p_{\alpha}\cdot \mu}=p_{\alpha}G_{1,\mu}p_{\alpha}^{-1}.}
Combining the above equation with $p_{\alpha}\circ \bvarpi_1 = \bvarpi_{\alpha}$, we derive a conjugacy condition on the site stabilizers:
\e{G_{\alpha,p_{\alpha}\cdot \mu,\bvarpi_{\alpha}}=p_{\alpha}G_{1,\mu,\bvarpi_1}p_{\alpha}^{-1}. \la{conjugacysite}}

\subsection{Inducing Wannier basis for single-orbit band}\la{sec:constructLSWB}

Beginning from $P_{1,1}$ that represents $G_{1,1}$ [the stabilizer of $P_{1,1}$ under $G$; cf.\ \q{defG1mu}], we will deduce the existence of a one-dimensional Wannier representation  of the site stabilizer $G_{1,1,\bvarpi_1}$  [cf.\ \q{defG1muvarpi} above and \q{defineW1mu} below]. This one-dimensional representation will be induced to an $M$-dimensional monomial representation of $G_{\bvarpi_1}$ [cf.\ \textit{Lemma 2} below], which is then induced to an infinite-dimensional representation of $G$ -- we will identify the latter as $P$ for a single orbit under $G$. This would complete the proof of the forward arrow.\\

Since $P_{1,1}$ is a unit-rank representation of $G_{1,1}$, with assumed analytic projector and trivial first Chern class, $P_{1,1}$ must be a BR of $G_{1,1}$, according to our unit-rank splitting theorem. There must therefore exist a locally-symmetric Wannier basis $\{W_{1,1,\bR}\}_{\bR\in \calt}$ for $P_{1,1}$, a BR of $G_{1,1}$. We remind the reader [cf.\ \app{app:locallysymmetricWannierbasis}] that being locally-symmetric means that $W_{1,1,\bR}$ has a Wannier center at $\bvarpi_1+\bR$ and forms a one-dimensional representation of the site stabilizer $G_{1,1,\bvarpi_1+\bR}$ [cf.\ \q{defG1muvarpi}], for all $\bR$. In particular,
\e{ \forall g \in G_{1,1,\bvarpi_1}, \as g\ket{W_{1,1,\bze}}=\rho(g)\ket{W_{1,1,\bze}} \la{onedimrepG11} }
with $\rho(g)$ a unimodular phase factor.\\

A set of $M$ Wannier functions (lying in $P$) may be defined by
\e{\ket{W_{1,\mu,\bze}}:=g_{\mu}\ket{ W_{1,1,\bze}}, \;\;\mu=1,\ldots,M, \la{defineW1mu}}
with $g_{\mu}$ defined through \q{defgmu}. \\

\noindent \textbf{{Lemma 2.}} With $\ket{W_{1,\mu,\bze}}$ given by \q{defineW1mu}, (a)  $\{W_{1,\mu,\bze}\}_{\mu=1\ldots M}$ forms an orthonormal basis for a {monomial  representation}   of $G_{\bvarpi_1}$, and (b) each $W_{1,\mu,\bze}$ is a 1D representation of the site stabilizer $G_{1,\mu,\bvarpi_1}$, as defined in \q{defG1muvarpi}.\\

\noindent Lemma 2 is proven below; let us first finish the proof of the forward direction of the crystallographic splitting principle.\\

By application of lattice translations in $\calt$ and the symmetry transformation $p_{\alpha}$ [defined in \q{defgalpha}], we generate a set of Wannier functions from $W_{1,\mu,\bze}$:
\e{ \ket{W_{\alpha,p_{\alpha}\cdot \mu,\bR}}:=(\bR|e)p_{\alpha}\ket{W_{1,\mu,\bze}} \in P_{\alpha,p_{\alpha}\cdot \mu}.\la{Wrel}}
That $\ket{W_{\alpha,p_{\alpha}\cdot \mu,\bze}}$ belongs in $P_{\alpha,p_{\alpha}\cdot \mu}$ follows from \q{palphaacts}; that $\ket{W_{\alpha,p_{\alpha}\cdot \mu,\bR\neq \bze}}$ \emph{also} belongs in $P_{\alpha,p_{\alpha}\cdot \mu}$ follows from $\calt(< G)$ being a subgroup of $G_{\alpha,p_{\alpha}\cdot \mu}$ [the stabilizer defined in \q{defG1mu}].\\

Since the band spanned by $\{W_{\alpha,p_{\alpha}\cdot\mu,\bR}\}_{(\bR|e)\in \calt}$ is of unit rank, and so is $P_{\alpha,p_{\alpha}\cdot \mu}$ by assumption, we may identify $P_{\alpha,p_{\alpha}\cdot \mu}=\sum_{\bR}\ketbra{W_{\alpha,p_{\alpha}\cdot\mu,\bR}}{W_{\alpha,p_{\alpha}\cdot\mu,\bR}}$. In combination, we have thus found a Wannier basis $\{W_{\alpha,\mu,\bR}\}_{\alpha,\mu,(\bR|e)\in \calt}$ for the entirety of $P$ (corresponding to a single orbit). \\

We now conclude that $P$ spans a monomial BR of $G$, induced from a finite-dimensional monomial representation of $G_{\varpi_1}$ spanned by $\{ W_{1,\mu,\bze} \}_{\mu}$ (the representation in terms of complex permutation matrices is explicitly given in \q{complexpermrep}). This is because the induction procedure (to derive a monomial BR) consists of defining new Wannier functions through \q{Wrel}, where $\{ p_{\alpha} \}_{\alpha}$ are representatives of the following coset decomposition:
\e{G/(\calt\rtimes G_{\bvarpi_1})=\{[p_{\alpha}]|\alpha=1,\ldots, A\}. \la{coset}} 
We briefly review how \q{coset} arises: $\calt\rtimes G_{\bvarpi_1}$ is the subgroup of $G$ that trivially maps $\bvarpi_1$ (modulo lattice translations). Since the orbit of $\bvarpi_1$ under $G$ comprises $A$ Wannier centers (modulo lattice translations),  $G/(\calt\rtimes G_{\bvarpi_1})$ must have $A$ elements, each represented by $p_{\alpha}$ that maps $\bvarpi_1\mapsto \bvarpi_{\alpha}$.\\

This finishes the proof of the forward direction of the crystallographic splitting principle. \hfill\(\Box\)


\begin{center}
\textit{Proof of Lemma 2}
\end{center}

\emph{Proof of statement (a) in Lemma 2.} The orthonormality condition  $\braket{W_{1,\nu,\bze}}{W_{1,\nu',\bze}}=\delta_{\nu\nu'} $
follows from $P_{1\nu}P_{1\nu'} =P_{1\nu} \delta_{\nu\nu'}.$ Recall that $G_{\bvarpi_1}$ has a transitive action on $\{P_{1,\mu}|\mu=1,\ldots, M\}$, which is therefore an orbit (of any of its elements) under $G_{\bvarpi_1}$:
\e{ Orb[P_{1,1}] = \{P_{1,\mu}|\mu=1,\ldots, M\}.}
Since $G_{1,1}$ is the stabilizer of $P_{1,1}$ under $G$, it follows that $G_{1,1,\bvarpi_1}$ is the stabilizer of $P_{1,1}$ under $G_{\bvarpi_1}$:
\e{Stab[P_{1,1}] = G_{1,1,\bvarpi_1}.}
By the orbit-stabilizer theorem, 
\e{ \f{|G_{\bvarpi_1}|}{|Stab[P_{1,1}]|}=|Orb[P_{1,1}]|=M.\la{apporbstab} }
There must therefore be $M$ elements in the coset
\e{  G_{\bvarpi_1}/G_{1,1,\bvarpi_1} = \{ [g_{\mu}]|\mu=1,\ldots,M \}. \la{cosetexap}}
To prove that each element can be represented by
$g_{\mu}\in G_{\bvarpi_1}$ defined in \q{defgmu}, it suffices to show that  $g_{\mu}$ and $g_{\mu'}$ lie in different equivalence classes if $\mu\neq \mu'$. (Supposing the contrary, there would exist $g_{1,1}\in G_{1,1,\bvarpi_1}$ such that 
\e{ g_{\mu}=g_{\mu'}g_{1,1}\imp  P_{1,\mu}=g_{\mu}P_{1,1}g_{\mu}^{-1}= P_{1,\mu'},}
which contradicts our assumption that $P_{1\mu}P_{1\mu'}=0$.)  It follows from \q{cosetexap} that the following coset decomposition holds:
\e{ G_{\bvarpi_1} =\cup_{\mu=1}^{M}g_{\mu} G_{1,1,\bvarpi_1}.\la{cosetdecom}}
Now we derive the desired representation: consider that for any $g\in G_{\bvarpi_1}$
\e{  g\ket{W_{1,\mu,\bze}}=gg_{\mu}\ket{W_{1,1,\bze}}. }
Since $g_{\mu}\in G_{\bvarpi_1}$, the closure property of groups ensures
$gg_{\mu}\in G_{\bvarpi_1}$. We may therefore apply the coset decomposition of \q{cosetdecom} to express $gg_{\mu}=g_{{\mu'}}g_{1,1} $, for some ${\mu}'=1,\ldots,M$ and some $g_{1,1}\in G_{1,1,\bvarpi_1}$. Consequently,
\e{g\ket{W_{1,\mu,\bze}}=g_{\mu'}g_{1,1} \ket{W_{1,0,\bze}}=\overline{\rho(g_{1,1})}^{s(g_{\mu'})}\ket{W_{1,\mu',\bze}} \la{complexpermrep} }
where $\overline{a}^{s(g)}=\overline{a}$ (the complex conjugate of $a$) if $g$ is antiunitary, and otherwise  $\overline{a}^{s(g)}=a$. \q{complexpermrep} defines a unitary representation of $G_{\bvarpi_1}$  where each $g\in  G_{\bvarpi_1}$ is mapped to a complex permutation matrix, with nonzero matrix elements given by the unimodular phase factor: $\overline{\rho}^s$. We thus conclude that $\{W_{1,\mu,\bze}\}_{\mu=1\ldots M}$ spans a complex permutation representation of $G_{\bvarpi_1}$, or equivalently a monomial representation of $G_{\bvarpi_1}$; this equivalence has been proven in \app{app:complexpermrep}.\\


\textit{Proof of statement (b) in Lemma 2.} It follows from \q{defgmu} that $W_{1,\mu,\bze}$ lies in the vector space projected by $P_{1,\mu}$. By definition of the stabilizer $G_{1,\mu}$ [cf.\ \q{defG1mu}], $P_{1,\mu}$ is a representation of $G_{1,\mu}$. This implies that for any $g_{1,\mu,\bvarpi_1}\in G_{1,\mu,\bvarpi_1}< G_{1,\mu}$ [cf.\ \q{defG1muvarpi}], $g_{1,\mu,\bvarpi_1}\ket{W_{1,\mu,\bze}}$ remains in $P_{1,\mu}$. and is therefore orthogonal to $W_{1,\mu'\neq \mu,\bze}$. Further applying that $g_{1,\mu,\bvarpi_1}\in G_{1,\mu,\bvarpi_1} < G_{\bvarpi_1}$, and that $\{W_{1, \mu,\bze}\}_{\mu}$ forms a complex permutation representation of $G_{\bvarpi_1}$ [cf.\ \q{complexpermrep}], we deduce that   
\e{g_{1,\mu,\bvarpi_1}\ket{W_{1,\mu,\bze}}=\rho(g_{1,\mu,\bvarpi_1})\ket{W_{1,\mu,\bze}},} with $\rho$ a unimodular phase factor. Since this is true for any $g_{1,\mu,\bvarpi_1}\in G_{1,\mu,\bvarpi_1}$, we arrive at the desired claim. \hfill\(\Box\) \\

\begin{center}
\textit{Proof of backward arrow of crystallographic splitting principle}
\end{center}

Suppose we have a BR of $G$ induced from a complex permutation representation $D$ of the site stabilizer $G_{\bvarpi_1}$, with the Wyckoff position $\bvarpi_1$ having multiplicity $A$. Let $\{W_{1\mu}\}_{\mu=1\ldots M}$ be basis vectors of the representation space of $D$, such that 
\e{\forall \,h\in G_{\bvarpi_1}, \;\; h\ket{W_{1\mu}} = \rho(h;\mu)\ket{W_{1,h\cdot \mu}}, }
with $\rho$ a unimodular phase factor, and $h\cdot$ a permutation on the $\mu$ index. \\

Given the coset decomposition in \q{coset}, we define a set of $A$ Wannier functions by
\e{ \ket{W_{\alpha,\mu}}:=p_{\alpha}\ket{W_{1\mu}}, \as \alpha=1\ldots A,}
and the unit-rank projection to their Bravais-lattice translates as
\e{ P_{\alpha\mu}:=\sum_{(\bR|e)\in \calt} (\bR|e) \ket{W_{\alpha,\mu}} \bra{W_{\alpha,\mu}}(\bR|e)^{-1}.}
Since each Wannier function is assumed to be exponentially-localized, each projector $P_{\alpha\mu}$ must be analytic with trivial first Chern class,\cite{Brouder2007,Panati2007} and gives a splitting for   
\e{P =\oplus_{\alpha=1}^A\oplus_{\mu=1}^M P_{\alpha,\mu}.}
Let us  prove that $\{P_{\alpha,\mu}\}_{\alpha,\mu}$ is permuted by each element of $G$, which would complete the  proof of the backward arrow. \hfill\(\Box\) \\

\noindent \emph{Proof of permutation.} The action of $g$ on $P_{\alpha,\mu}$ is 
\e{gP_{1\mu}g^{-1} =  \sum_{(\bR|e)\in \calt}g (\bR|e)\ket{W_{\alpha,\mu}}\bra{W_{\alpha,\mu}}(\bR|e)^{-1} g^{-1}.\la{actiongp1mu}}
Utilizing the coset decomposition in \q{coset}, any $g\in G$ can be expressed as $g=(\bR'|e)p_{\alpha'}h'$, for one $h'\in G_{\bvarpi_1}$, one $\alpha'\in \{1\ldots A\}$, and one $(\bR'|e)\in \calt$. It follows that
\e{ g(\bR|e)\ket{W_{\alpha\mu}}=(p_{\alpha'}h'\circ \bR +\bR'|e)p_{\alpha'}h'p_{\alpha}\ket{W_{1\mu}}. \la{applycoset}}
Since $p_{\alpha'}h'p_{\alpha}\in G$, it also has the coset decomposition
\e{p_{\alpha'}h'p_{\alpha}=(\bR''|e)p_{\alpha''}h''.}
Substituting the above equation into \q{applycoset}, we derive
\e{ g(\bR|e)\ket{W_{\alpha\mu}}\eq (p_{\alpha'}h'\circ \bR +\bR'+\bR''|e)\lin
	&\times \rho(h'';\mu)\ket{W_{\alpha'',h''\cdot\mu}}. \la{applycoset2}}
Substituting the above equation into \q{actiongp1mu}, and applying that $p_{\alpha'}h'\circ \bR +\bR'+\bR''$ is a Bravais-lattice vector, we derive the desired claim:
\e{ gP_{\alpha\mu}g^{-1}=P_{\alpha''\in\{1\ldots A\},h''\cdot \mu\in \{1\ldots M\}}.}
\hfill\(\Box\) \\

\section{Methods of symmetric splitting}\la{app:methods_symmetricdecomp}

Let $P$ project to a rank-$N$ representation of a  space group $G$; in this section  we shall not distinguish between between space groups, magnetic space groups, and double space groups.   \\

\noindent We define a \emph{symmetric splitting with respect to $G$} as a splitting $P=\oplus_{j=1}^NP_j$ into single-rank projectors satisfying the symmetry condition (B) of the splitting theorem, namely, that for all $g\in G$, $g:P_j\rightarrow P_{\sigma_g(j)}$ with $\sigma_g$ a permutation on $\{1,\ldots,N\}$. \\
 
\noindent Given a symmetric splitting, the splitting theorem states that if $P_j(\bk)$ is analytic in $\bk$ (over the Brillouin torus) and has trivial first Chern class, then $P$ is a monomial band representation (BR) of $G$. Beside offering a method to prove band-representability, a symmetric splitting automatically gives a set of Wannier functions which are permuted by the space group. \\

For a given $P$ there is no unique symmetric splitting, but we will describe two methods which involve diagonalizing various operators: (i) the projected symmetry operator in  \s{sec:diagnoseBR_projsymm}, and (ii) the projected position operator  in \s{sec:projposition}.

\subsection{The projected symmetry method}\la{sec:diagnoseBR_projsymm}

We have exemplified the projected symmetry method for fragile obstructed insulators in \s{sec:demonstration}. Here we  describe our  method in greater generality: suppose we are given a tight-binding Hamiltonian $h(\bk)$ defined with respect to a 
L\"owdin-orthonormalized\cite{lowdin1950,slater1954} basis of Wannier functions. $h(\bk)$ is assumed to have the symmetry of a space group $G$:
\e{ g \in G, \as \hat{g}h(\bk)\hat{g}^{-1}=h(g\circ\bk),}
with $\hat{g}$ the matrix representation of $g$ in the Wannier basis. We define the Wannier-center operator  $\hbr$ as a diagonal matrix with each diagonal element equal to the central  position of each Wannier function, such that
\e{h(\bk+\bG)=e^{-i\bG\cdot \hbr} h(\bk)e^{i\bG\cdot \hbr} \la{translatehk}}
for any reciprocal vector $\bG$. Finally we assume that the real-space matrix elements of the tight-binding Hamiltonian decay exponentially; this guarantees that $h(\bk)$ is analytic in $\bk$ throughout the Brillouin torus.\cite{Cloizeaux1964} Moreover, if a rank-$N$ energy band of $h(\bk)$ is spectrally isolated  (i.e., separated by all other energy bands by a nonzero spectral gap at each $\bk$), it is guaranteed that the rank-$N$ projector $p(\bk)$ is also analytic in $\bk$ throughout.\cite{Panati2007,Read2017} Our goal is to symmetrically decompose  $p(\bk)$. \\


We would like to identify  a Hermitian operator $\tilde{s}$ (in the tight-binding basis of Wannier functions), such that the  eigenbands of the $\tilde{s}_{\bk}:=p(\bk)\tilde{s}p(\bk)$ give a symmetric splitting with respect to $G$.  $\tilde{s}_{\bk}$ is the \textit{projected symmetry operator}; by construction it is Hermitian and analytic throughout the Brillouin torus. It is necessary that $\tilde{s}_{\bk}$ has the same translational property [cf.\ \q{translatehk}] as $h(\bk)$, so we impose $[\tilde{s},e^{i\bG\cdot\hbr}]=0$ for every reciprocal vector $\bG$. We further impose that for any $g\in G$, $\hat{g}$ commutes or anticommutes with $\tilde{s}$; in the former case $\hat{g}$ trivially permutes the eigenspaces of $\tilde{s}$, and
\e{ \hat{g}\tilde{s}_{\bk}\hat{g}= \tilde{s}_{g\circ \bk};} 
in the latter case, 
\e{ \hat{g}\tilde{s}_{\bk}\hat{g}= -\tilde{s}_{g\circ \bk},}
and $\hat{g}$ interchanges the eigenspaces of $\tilde{s}$ that have nonzero eigenvalues.  \\

In some models, $\tilde{s}$ is simply obtained by deforming the eigenvalues of one of the unitary symmetry operators $\hat{g}$. 
An example of this kind, with $g$ the four-fold rotation, was provided in our case study of fragile obstructed insulators in Wigner-Dyson class AI [cf.\  \s{sec:fu_diagnosefragility} and \s{sec:fu_diagnosefragility}]
Generally, $\tilde{s}$ need \textit{not} correspond to a symmetry of $h(\bk)$, as we exemplify below.  \\

We define the eigenvalue problem
\e{ [\tilde{s}_{\bk}-\lambda_j(\bk)]\ket{u_j(\bk)} = 0,\;\text{with}\; \lambda_1(\bk)\geq \lambda_2(\bk)\geq\ldots  \la{eigenproblemgen}}
for all $\bk$. Generically, the band dispersion $\lambda_j$ should be nondegenerate except on a zero-measure set of $\bk$, e.g., at a conical (Dirac-Weyl) band touching.  If each band is  spectrally isolated, then we have obtained a symmetric splitting $P=\oplus_{i=1}^NP_j$ with respect to $G$, with each $P_j(\bk)$ being analytic throughout the Brillouin torus; note $P_j(\bk)$ is the projector to the Bloch state $e^{i\bk\cdot(\hbr+\bR)}\ket{u_j(\bk)}$. Furthermore if each $P_j$ has trivial first Chern class, then $P$ must be a monomial band representation (BR), according to our splitting theorem.\\

On the other hand if $P=\oplus_{j}P_j$ is not a monomial BR, then either (a) there exists $P_j(\bk)$ that is non-analytic at a set of $\bk$ where  $\lambda_j$ is degenerate, or (b) each $P_j$ is analytic throughout the Brillouin torus, and at least one of $\{P_j\}$ must have a nontrivial Chern class. Colloquially speaking,  $\tilde{s}_{\bk}$ is the Hamiltonian of either  a topological semimetal or a Chern insulator. Note if each $\lambda_j$ is  nondegenerate for all $\bk$, then case (b) is implied; however, the converse statement -- namely that (b) implies nondegeneracy -- is not generally valid. \\

Except for certain double space groups with cubic point groups [cf. \s{sec:equivalencetheorems}], not being a monomial BR means that $P$ is an obstructed representation. \\

\noindent \textit{Example: $\Z_2$ topological insulator in Wigner-Dyson class AII.} Let $P$ project to the filled, rank-two band of the Kane-Mele $\Z_2$ topological insulator,\cite{kane2005A,kane2005B} with space group. $\calt_2\times \Z_4^T$. Following Prodan,\cite{prodan_spinchern} one may pick $\tilde{s}=\vec{n}\cdot \bS$, with $\bS$ the spin operator and $\vec{n}$ an arbitrary directional vector. Since time reversal inverts $\bS$, the bands of $\tilde{s}(\bk)$ would be nontrivially permuted by $T$. Thus $\tilde{s}=\vec{n}\cdot \bS$ gives a symmetric splitting with respect to $\calt_2\times \Z_4^T$, despite not generally being a symmetry  of the Hamiltonian. If $\tilde{s}_{\bk}$ were spectrally gapped at each $\bk$, then $P$ being an obstructed representation guarantees that the two bands of $\tilde{s}_{\bk}$ have opposite and nonzero first Chern numbers -- this is nothing more than the spin Chern number formulated for infinite samples by Prodan.\cite{prodan_spinchern} Our projected symmetry method may be viewed as the  generalization of Prodan's projected spin method  to include crystallographic space-group symmetry within class AII, and also to go beyond symmetry class AII. An example of the latter -- a symmetric splitting in class AI -- has been given in \s{sec:outlineprooffragile}. To exemplify the former, we consider  the Kane-Mele \textit{honeycomb} model, whose symmetry is the double-group extension  of $p6mm \times \Z_2^T$, which we will denote by $\tilde{G}_6$.  While $\tilde{s}=\vec{n}\cdot \bS$ no longer gives a symmetric splitting of $\tilde{G}_6$ for arbitrary $\vec{n}$, $\vec{n}=\vec{z}$ (the out-of-plane direction) would give the desired splitting. The reason is that $S_z$ commutes with all rotations in $\tilde{G}_6$, and anticommutes with all reflections.


\subsection{Symmetric splitting by the projected position operator}\la{sec:projposition}

We have proven in \s{sec:resultzakphase} that the splitting $P=\oplus_{j=1}^N P^x_j$ into bands of the projected position operator [cf.\  \qq{projectedpositioneigen}{pjpxp})] is symmetric with respect to certain two-space-dimensional space groups  [satisfying conditions (i-ii) in \textit{Lemma 1} of  \s{sec:resultzakphase}]. 
In this section we will prove a statement in \textit{Lemma 1} that is needed to derive the Zak winding theorem  [cf.\ \s{sec:resultzakphase}], namely that each $P^x_j$ is analytic throughout the Brillouin zone. \\ 

By `analyticity throughout the Brillouin zone', we mean that the restriction of $P^x_j$ to $\bk$: 
\e{P^x_j:=\int \f{d^2 k}{(2\pi)^2} P^x_j(\bk), \;\; P^x_j(\bk):=\ketbra{\psi^x_{j\bk}}{\psi^x_{j\bk}}.\la{decomposewilson}} 
is both analytic in $\bk$ (for all $\bk$ in the Brillouin zone), and periodic in reciprocal lattice translations $\bk\rightarrow \bk+\bG$. 
The Bloch function $\psi^x_{j\bk}$ is obtained by 1D Fourier transform of the eigenfunctions of the projected position operator [cf.\ \q{projectedpositioneigen}]: 
\e{ \psi^x_{j\bk} = \sum_{R\in \Z} e^{ik_xR} h_{j,k_y,R}. \la{definecheckpsi}}

\subsubsection{Proof of analyticity}\la{app:proveanalyticity}

Given that $P$ has trivial first Chern class, a basis for the Bloch functions $\{\psi_{n\bk}\}_{n=1}^N$ exist that is analytic in $\bk$ throughout the Brillouin torus, and is periodic under translation by any reciprocal vector: $\psi_{n\bk}=\psi_{n\bk+\bG}$; a review of this well-known fact may be found in \app{sec:bandbundle}. \\

Since both $\{\psi_{n\bk}\}_{n=1}^N$ and $\{\psi^x_{j\bk}\}_{j=1}^N$ [cf.\  \q{definecheckpsi}] span the same rank-$N$ band $P$, there exists a $U(N)$ transformation $Q(\bk)$ that relates the two sets of Bloch functions:
\e{ \psi^x_{j\bk} =  e^{-ik_xx_j(k_y)}\sum_{n=1}^N \big[ Q(\bk) \big]_{nj}\psi_{n\bk}. \la{relatecheckpsitopsi}}
As shown in App.\ D of \ocite{AA2014}, the columns of $Q(\bk)$ are given by the eigenvectors of the Wilson loop $\W(\bk)$,  which is defined as a path-ordered exponential of the Berry connection [cf.\ \q{Bconn} with $u_{n\bk}=e^{-i\bk\cdot \br}\psi_{n\bk}$]:
\e{ \W(\bk) = \mathcal{P} \mathrm{exp}\bigg[i\int_{k_x}^{k_x+2\pi}A_x(s,k_y)ds \bigg]. \la{defwilsonkxky}}
The above integral is over a $\bk$-loop with base point $(k_x,k_y)$ and end point $(k_x,k_y+2\pi).$ [$\W(\bk)$ slightly differs in definition from $\W(\calc)$ in \q{definewilsonloop}.] $Q$ in \q{relatecheckpsitopsi} is the unitary transformation that diagonalizes the Wilson loop:
\e{ \W(\bk) = Q(\bk)D(k_y)Q(\bk)^{-1},\la{diagwilsonkxky}}
with $D$ a diagonal matrix equal to
\e{ D(k_y) = \text{diag}[e^{i2\pi x_1(k_y)},\ldots, e^{i2\pi x_N(k_y)}];}
$2\pi x_j$ is referred to as a Zak phase [cf.\ \q{definezakphase}], and depends only on $k_y$; this dependence is because 
for a given $\bk$-loop, the spectrum of $\W$ is independent of the base point.\cite{AA2014}\\

While we have flippantly claimed the columns of $Q(\bk)$ are the eigenvectors of $\W(\bk)$, beware that an eigenvector, if nondegenerate in eigenvalue, is only defined up to a phase. (If $x_j=x_{j'}$ is degenerate at isolated $k_y$, then the the two eigenvectors associated to $x_j$ and $x_j'$ can still be defined up to a phase by continuity in $k_y$.) This phase ambiguity is reduced by the following procedure: 
since there is no topological obstruction to analyticity of an eigenvector over the base space $S^1$, each column of $Q(0,k_y)$ can be made analytic and periodic in $k_y$. Moreover, from \qq{defwilsonkxky}{diagwilsonkxky}, one deduces that   
 $Q(k_x,k_y)$ and $Q(0,k_y)$ can always be related by a Wilson line:\cite{AA2014}
\e{ Q(k_x,k_y)=\mathcal{P} \mathrm{exp} \bigg[i\int^{k_x}_{0}A_x(s,k_y)ds \bigg]Q(0,k_y);\la{qkgauge}}
these conditions on $Q$ are henceforth adopted. \\

\noindent \textit{Analytic properties of the $Q$ matrix.} The analyticity of $\psi_{n\bk}$ implies that the Berry connection $A_x(\bk)$  is also analytic in $\bk$. Since  $Q(0,k_y)$ is analytic in $k_y$, and  \q{qkgauge} holds as well, we deduce that $Q(\bk)$ is analytic in $\bk$. The periodicity of $\psi_{n\bk}=\psi_{n\bk+\bG}$ implies $A_x(\bk)=A_x(\bk+\bG)$. The periodicity of $A_x$, combined with the periodicity of $Q(0,k_y)$ in $k_y$, implies that $Q(\bk)$ in \q{qkgauge} is periodic in $k_y$. However, $Q$ is generically aperiodic in $k_x$, i.e., $Q(k_x+2\pi,k_y)=\W(\bk)Q(\bk).$ \\

\noindent \textit{Analytic properties of the Zak phase.} The analyticity and periodicity of $A(\bk)$ imply that $\W(\bk)$ is likewise analytic and periodic. This implies that  the spectrum $\{e^{i2\pi x_j(k_y)}\}_{j=1}^N$ of $\W(\bk)$ is analytic and periodic. If we further assume the Zak permutation order $Z_{2\pi \be_x}=1$, then  each eigenvalue $e^{i2\pi x_j(k_y)}$ can be made analytic and periodic, too. Beware, however, that $2\pi x_j$ can wind with respect to $k_y$; the associated Zak winding number $W_{j,2\pi\be_x}$ has been defined in \q{windingnumber}.\\

Given the above-stated analytic properties of $Q$, $x_j$ and $\psi_{n\bk}$, we are then able to deduce the analytic properties of $\psi_{j\bk}^x$, as defined in \q{relatecheckpsitopsi}. Namely $\psi_{j\bk}^x$ is analytic in $\bk$ throughout the Brillouin zone, periodic in $k_x$,\cite{AA2014} but aperiodic in $k_y$ if the Zak winding number $W_{j,2\pi\be_x}$ [cf.\ \q{windingnumber}] is nonzero:
\e{ &\psi^x_{j,k_x+2\pi,k_y} = \psi^x_{j,k_x,k_y}, \lin
&\psi^x_{j,k_x,k_y+2\pi} =e^{-ik_xW_{j,2\pi \be_x} }\psi^x_{j,k_x,k_y}. \la{psijx}}
An alternative (and numerically-motivated) construction of such a basis is described in \ocite{alexey_smoothgauge}, where it is referred to as a `cylindrical gauge'. 
While $\psi_{j\bk}^x$ is possibly aperiodic under $k_y\rightarrow k_y+2\pi$, one deduces from \q{psijx} that the projector $P^x_j(\bk)=\ketbra{\psi_{j\bk}^x}{\psi_{j\bk}^x}$ is always periodic. Combining this with the analyticity of $\psi^x_{j\bk}$, we derive that $P^x_j(\bk)$ is both periodic and analytic throughout the Brillouin zone.
\hfill\(\Box\) \\


\section{Proving the fragility of rotation-invariant topological crystalline insulators \label{app:numerics}}

\subsection{Fragility of tetragonal  TCI}\la{app:provefragileC4v}

Liang Fu's tight-binding model\cite{fu2011} for the $\calt_3\rtimes C_{4v}\times \Z_2^T$-symmetric TCI is spanned by two pairs of $p_x,p_y$ orbitals in each unit cell. The reduced real-space coordinates  of the two pairs of orbitals are $(0, 0, 0)$, in an orthogonal basis of Bravais lattice vectors. To remove the symmetry obstruction of the filled rank-two band, a unit-rank BR [induced by an $s$ orbital at $(0.5, 0.5, 0.0)$] is introduced to the model. The original parameters in \ocite{fu2011} are adopted. Additionally, the on-site energy of the $s$ orbital is set to $-4.0$, in units where the nearest-neighbor hopping between $p$ orbitals (in the $x-y$ plane) equals $1$; this ensures that the $s$-type BR lies below the energy gap. The hopping between the $s$ orbital and the two $p_x$ orbitals in the home unit cell is continuously increased to $0.375$, with all other hoppings determined by translational symmetry and $C_{4v}$. During this interpolation, the bulk gap never closes.\\

\begin{figure}[h]
	\centering
	\includegraphics[width=0.9\columnwidth]{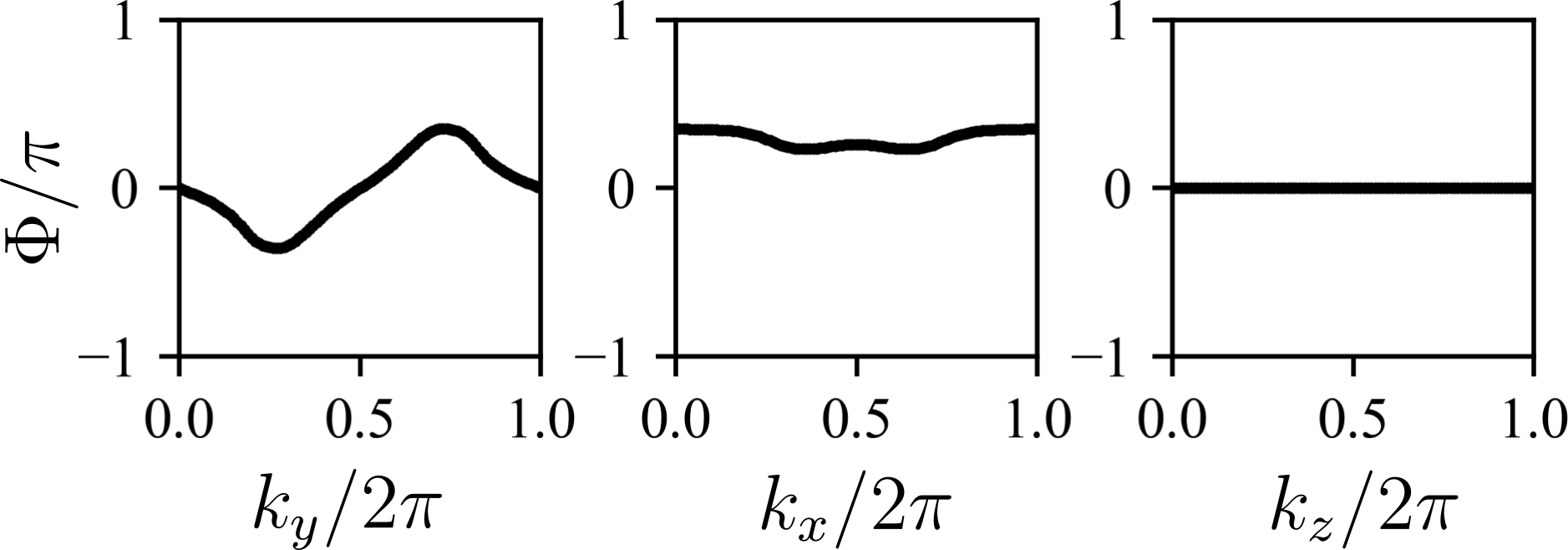}
	\caption{Zak phase (divided by $\pi$) of the lowest band of the projected symmetry operator in three independent planes ($k_z=\pi$, $k_y=0$ and $k_x=0$ from left to right) in the Brillouin zone.\label{fig:Fu_wilson}}
\end{figure}

Before the introduction of the $s$-like BR, the spectrum of the projected symmetry operator is gapless along a nodal line, as illustrated in Fig. \ref{fig:Fu_phase}(a). Upon the introduction of the $s$-like BR, the projected symmetry operator consists of three bands whose dispersion are nondegenerate throughout the Brillouin zone [cf.\ \fig{fig:Fu_phase}(b)]. Each unit-rank band has trivial first Chern class, as verified by computing the winding of the Zak phase in three independent $\bk$-directions. For illustration, the Zak phase of the lowest band (of the three) is presented in Fig. \ref{fig:Fu_wilson}. 

\subsection{Fragility of hexagonal TCI}\la{app:provefragileC3v}

In \ocite{AAchen}, a $\calt_3\rtimes C_{3v}\times \Z_2^T$-symmetric topological insulator was proposed on a triangular Bravais lattice with primitive vectors: $\boldsymbol{a}_1=(1, 0, 0)$, $\boldsymbol{a}_2=(-1/2,\sqrt{3}/2,0)$ and $\ba_3=(0,0,1)$, and with the following tight-binding model Hamiltonian 
\begin{widetext}
	\begin{align}
	H(\bk) = & [\frac{5}{2}-\cos(k_1+2\pi/3)-\cos(k_2+2\pi/3)-\cos(k_1+k_2-2\pi/3)-\cos(k_3)]\Gamma_{30}\nonumber\\
	& + \left\{ 0.3[e^{i\pi/3}\cos(k_1) +e^{-i\pi/3 }\cos(k_2)-\cos(k_1+k_2)]\Gamma_{1+}+h.c.\right\} + \sin(k_3)\Gamma_{20}.
	\end{align}
\end{widetext}
Here, $k_j=\bk\cdot\ba_j$ for $j=1,2,3$,  $\Gamma_{ij}=\sigma_i\otimes \tau_j$ (with $\sigma_i$ and $\tau_j$ being two sets of Pauli matrices), and $\Gamma_{1+}=\sigma_1\otimes(\tau_1+i\tau_2)$. The tight-binding basis for the above Hamiltonian is given by two sets of $p_x\pm ip_y$ orbitals, both located at $(0, 0, 0)$.\\

\begin{figure}
	\includegraphics[width=0.9\columnwidth]{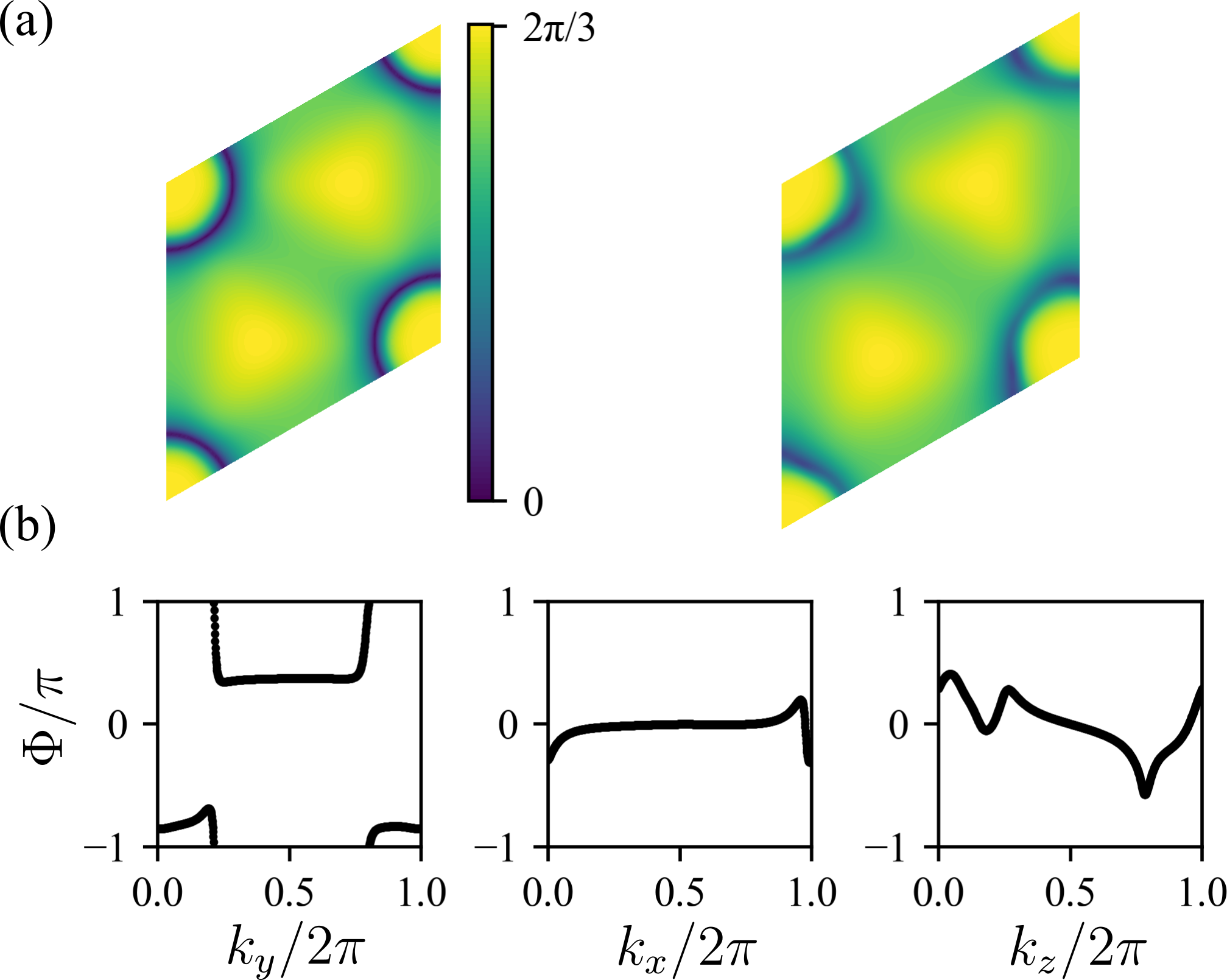}
	\centering
	\caption{(a) Left panel: half the spectral gap between the two bands of the projected $C_3$-rotation operator; right panel: the spectral gap between the lowest band and middle band of the projected $C_3$-rotation operator. (b) Zak phase  (divided by $\pi$) for the lowest band of the projected $C_3$ operator, for three independent $\bk$-planes ($k_x=0$, $k_y=0$ and $k_z=0$ from left to right) in the Brillouin zone. \label{fig:AA_phase}}
\end{figure}

The low-energy band of the above model is an obstructed representation of  $\calt_3\rtimes C_{3v}\times \Z_2^T$, as deducible from the nontrivial Zak phase described in \ocite{berryphaseTCI}. Associated to this obstruction is an integer-valued topological invariant $\chi$  -- the  halved-mirror chirality --  which equals $1$ in this model. The obstruction also manifests in the projected $C_3$-rotation operator  as a nodal line in the spectrum, as illustrated in the left panel of \fig{fig:AA_phase}(a).\\

To break the symmetry obstruction on Wannier functions, we add to the low-energy subspace a unit-rank BR induced by an $s$ orbital at the Wyckoff position $(0, 0, 0)$, with the tight-binding hoppings tabulated in Tab. \ref{tab:C3v-hopping}. The projected $C_3$ operator now consists of three unit-rank bands whose dispersions are nondegenerate throughout the Brillouin zone [cf.\ right panel of \fig{fig:AA_phase}(a)]; each band has trivial first Chern class [cf.\ \fig{fig:AA_phase}(b)].

\begin{table}[]
	\begin{tabular}{|c|c|c|}
		\hline
		$\boldsymbol{R}$        & $m$   & hopping             \\
		\hline
		$[1,0,0]$  & $p_+^{(2)}$ & $i/6$                \\
		& $p_-^{(2)}$ & $i/6$                \\
		$[1,1,1]$  & $p_+^{(1)}$ & $-0.116667-0.202073i$ \\
		& $p_-^{(1)}$ & $0.233333$            \\
		$[1,-1,1]$ & $p_-^{(2)}$ & $0.4$                 \\
		$[1,2,0] $  & $p_+^{(1)}$ & $-i/6$ \\
		\hline        
	\end{tabular}
	\caption{Hoppings between the $s$ orbital and other orbitals in the format of $\langle\boldsymbol{0}m|H|\boldsymbol{R}s\rangle $, where $\boldsymbol{R}$ is a three-element vector denoting the unit cells of the $s$ orbital and $m\in \{1,2\}$ is an index for the $p$ orbital. $p_{\pm}^{(2)}$, for example, denotes the $p_x\pm ip_y$ for the second set of $p$ orbitals.\label{tab:C3v-hopping}}
\end{table}

\section{Proof that certain point groups are monomial}\la{app:pointgroupsmonomial}

Here we show that the following point groups are monomial: 
\begin{enumerate}
    \item[(1)] $32$ crystallographic point groups,\cite{tinkhambook}
    \item[(2)] $27$ noncubic double point groups, 
    \item[(3)] grey magnetic point group generalizations of (1) and (2) (which correspond to the Wigner-Dyson symmetry classes AI and AII, respectively). 
\end{enumerate}
 Noncubic point groups are crystallographic point groups that are neither tetrahedral nor octahedral, i.e., they are not any of $T$, $T_h$, $T_d$, $O$, and $O_h$. Grey crystallographic point groups are of the form $\mathcal{P} \times \Z_2^T$ where $\mathcal{P}$ is a point group in (1), and $\Z_2^T$ is a cyclic group of order two generated by time-reversal symmetry; grey double point groups are the double covers of $\mathcal{P} \times \Z_2^T$.\\ 

Bacry\cite{Bacry1988} has claimed that all $32$ crystallographic point groups, i.e., the point groups in (1), are monomial, but he did not provide a reference or proof. 
We have not seen the claim (or proof) of monomiality for point groups in (2)-(3) anywhere in the literature. \\

The rest of \app{app:pointgroupsmonomial} is organized as follows. \\

\noi{i} We begin in \app{app:huppert} by reviewing   Huppert's theorem,\footnote{The original work is found in \ocite{Huppert1953}. An English translation and proof can be found in Theorem 3.10, chapter 2 of \ocite{Bray1982}.} 
which gives sufficient conditions for a finite group to be monomial; we will prove two corollaries of Huppert's theorem that are useful for point groups. \\


\noi{ii} Basic properties of the crystallographic point groups are reviewed in \s{classes}. Each class of point groups labelled (1-3) will further be divided into sub-classes: 
\begin{enumerate}
    \item[A.] proper rotation groups, 
    \item[B.] improper rotation  groups with inversion,
    \item[C.] improper rotation groups  without inversion.
\end{enumerate}
Proper rotation groups consist purely of rotations, while improper rotation groups include at least one reflection  or inversion element.\\

\noi{iii} This sub-classification was used by Altmann\cite{Altmann1961} to show that every crystallographic point group in (1) can be written as a triple semi-direct product of a normal, abelian subgroup and two cyclic subgroups ({any one of} which might be trivial). Altmann's result, together with the two previously-mentioned corollaries and Wigner's theorem\cite{Wigner_ontheoperationoftimereversal}, allow us to prove the monomiality of (1-3) in \app{semidirect}, \app{noncubic} and  \app{typeI}, respectively.\\

\noi{iv} Finally in \app{app:nonmononial}, we show that the five double cubic point groups are not monomial, and exemplify  a non-monomial representation for the double tetrahedral group.


\subsection{Huppert's theorem for monomial groups, and two corollaries}\la{app:huppert}

To prepare the reader for Huppert's theorem, we briefly review the standard definitions of solvability, supersolvability and Sylow subgroups. \\

A finite group $G$ is \textit{solvable} if there exists a series of normal groups, i.e.,
\e{ C_1 =G_0 \triangleleft G_1 \triangleleft G_2 \ldots \triangleleft G_k=G \la{normalseries} }
for a $k \ge 1$, such that $G_{j+1}/G_j$ is abelian for all $j=1,\ldots, k-1$. Here, $G_j \triangleleft G_{j+1}$ means that $G_j$ is normal in $G_{j+1}$. \\

$G'$ is \textit{supersolvable} if there exists a series of normal groups, i.e., 
\e{ C_1=G'_0 \triangleleft G'_1 \triangleleft G'_2 \ldots \triangleleft G'_n=G' \la{definesupersolvable}}
for a $n \ge 1$, such that $G'_j\triangleleft G'$ and $G'_{j+1}/G'_j$ is cyclic. Supersolvability is a stronger condition than solvability. \\

A $p$-group is a group for which every element has order equal to an integer power of a prime $p$. A \textit{maximal} subgroup of a group $G''$ is a subgroup that is not contained in any larger subgroup (that is not $G''$ itself). Lastly, a \textit{Sylow subgroup} of $G''$ is a maximal $p$-group. \\

\noindent \textbf{Huppert's theorem} Let $H'$ be a finite group with normal, solvable subgroup $N'$, and with supersolvable quotient group $H'/N'$. If all subgroups of $N'$ that are Sylow are also abelian, then $H'$ is monomial. \\

\noindent \textbf{Corollary 1.} If a finite group $H$ has a decomposition $H = N \rtimes C^{(1)} \rtimes C^{(2)} \rtimes \ldots \rtimes C^{(n)}$ for a $n\ge 1$, where $C^{(j)}$ are cyclic subgroups, $N$ is an abelian normal subgroup of $H$, and $\rtimes$ is associative, then $H$ is monomial.\\

\noindent $\rtimes$ being associative  means that
\e{& A\rtimes B \rtimes C \rtimes D = A\rtimes (B \rtimes C \rtimes D)\lin 
\eq (A\rtimes B) \rtimes (C \rtimes D)= (A\rtimes B \rtimes C) \rtimes D\la{normalseries2}}
which implies $A$,  $A\rtimes B$,  and $A\rtimes B \rtimes C $ are all normal subgroups of $A\rtimes B \rtimes C \rtimes D$.\\

\noindent\textit{Proof of Corollary 1.}  Any abelian group $N$ is also solvable, because there exists a normal series $C_1 \triangleleft N$ with $N/C_1=N$ that is abelian. Furthermore, $H/N=C^{(1)} \rtimes C^{(2)} \rtimes \ldots \rtimes C^{(n)}$ has the normal series  given in \q{definesupersolvable}, with the identifications $G'_j= C^{(1)}\rtimes \ldots \rtimes C^{(j)}$ and $G'=G_n'=H/N$. The associativity of $\rtimes$ implies that $G'_j\triangleleft H/N$ for any $j=1,\ldots,n$. Furthermore, $G_{j+1}/G_j=C^{(j+1)}$ is cyclic. Finally, since $N$ is abelian, all its subgroups are abelian; therefore, if $N$ has Sylow subgroups, such subgroups must also be abelian. In combination of the above facts, we find that $H$ and $N$ satisfy all conditions of $H'$ and $N'$ in Huppert's theorem, respectively; hence $H$ is monomial.  \hfill\(\Box\)  \\ 

\noindent \textbf{Corollary 2.} 
If a finite group $H$ has an abelian, normal subgroup $N$, such that $H/N$ is a cyclic subgroup of $H$, then $H$ is monomial.\\

\noindent \textit{Proof of Corollary 2.} Since $N$ is abelian, $N$ is solvable (see beginning of Proof of Corollary 1). Furthermore, $H/N$ is supersolvable, i.e., it has a normal series $C_1 \triangleleft H/N$ such that $(H/N)/C_1=H/N$ is cyclic. Because $H/N$ is abelian, all its Sylow subgroups are also abelian; therefore, Huppert's theorem applies.  \hfill\(\Box\) 


\subsection{Review of  proper vs improper point groups}\la{classes}


Here, we elaborate on the sub-classification  of point groups given in point (ii) of the outline of \app{app:pointgroupsmonomial};\\

A review of crystallographic point groups [class (1)] is given here, with emphasis on its  sub-classification into proper rotation groups [A], improper rotation groups with inversion symmetry [B], and improper rotation groups without inversion symmetry [C]. Class and subclass labels will be combined as (1)A, (1)B, (1)C. The sub-classification cubic double crystallographic point groups will be described subsequently in \app{noncubic}, after we clarify the meaning of a double group.\\

There are 11 crystallographic point groups which consist only of rotations [class (1)A]:  the trivial point group $C_1$,  the cyclic groups $\{ C_n \}_{n=2,3,4,6}$, the dihedral groups $\{ D_n \}_{n=2,3,4,6}$, the tetrahedral group $T$, and the octahedral group $O$. \\

Class (1)B consists of improper rotation groups with inversion which can be constructed by direct products of the above $11$ proper rotation groups $\mathcal{P}$ with $\Z_2^{\inv}$ -- the order-two group generated by inversion $\inv$; we denote these by $\mathcal{P}_{\inv} = \mathcal{P} \times \Z_2^{\inv}$. The direct-product structure reflects that inversion squares to identity and commutes with every point-group operation.
The $11$ point groups constructed in this way are $S_2$, $C_{2h}$, $C_{3i}$, $C_{4h}$, $C_{6h}$, $D_{2h}$, $D_{3d}$, $D_{4h}$, $D_{6h}$, $T_h$ and $O_h$. Here, and throughout this work, we employ the standard notation for point groups that is reviewed in \ocite{tinkhambook}.\\

The remaining crystallographic point groups  in class (1)C are improper rotation groups without inversion, and will be denoted by  $\mathcal{P}_{\neg \inv}$. Such groups may be constructed from $8$ out of the $11$ proper rotation groups [(1)A] which have at least one index-2 subgroup. The three (1)A groups without index-two subgroups are the trivial group $C_1$, $C_3$ and $T$. $D_4$ and $D_6$ each has two index-two subgroups, as given in the third column of \tab{tab:groups}. All other (1)A groups have exactly one index-two subgroup [cf.\ \tab{tab:groups}]. Denoting a (1)A group by $\calp$ and its index-two subgroup by $H$, a (1)C group is constructed as $\mathcal{P}_{\neg \inv}=H + \inv \, \mathcal{P}\backslash H$, where $\mathcal{P}\backslash H$ denotes all elements of $\mathcal{P}$ that are not in $H$. Two 1(C) groups can be constructed from $\calp=D_4$ (and also $D_6$), which has two index-two subgroups. Altogether there are ten (1)C groups which we tabulate in \tab{tab:groups}.\\


We will eventually use that $\mathcal{P}_{\neg \inv}$ is isomorphic to $\mathcal{P}$. The proof is as follows. First note that $H$ is both a subgroup of $\mathcal{P}$ and $\mathcal{P}_{\neg \inv}$. We define a map $\varphi \colon \mathcal{P} \to \mathcal{P}_{\neg \inv}$ that is the identity map on $H < \mathcal{P}$, and maps an element $p \in \mathcal{P}\backslash H$ bijectively onto $\inv p \in \mathcal{P}_{\neg \inv}$. Since $\inv$ squares to the identity and commutes with all point-preserving spatial isometries (including all elements in $H$ and $\calp$), the bijection $\varphi$ preserves the multiplication rule, and hence constitutes a group isomorphism. \\


\begin{table}
\centering 
$\begin{array}{|c|c|c|}
\hline
  \mathcal{P}_{\neg \inv} & \mathcal{P} & H \\
  \hline \hline
  C_s & C_2 & e \\
  \hline
  S_4 & C_4 & C_2 \\
  \hline
  C_{3h} & C_6 & C_3 \\ 
  \hline
  C_{2v} & D_2 & C_2 \\ 
  \hline
  C_{3v} & D_3 & C_3 \\
  \hline
  C_{4v}, D_{2d} & D_4 & C_4, D_2 \\
  \hline
  C_{6v}, D_{3h} & D_6 & C_6, D_3\\
  \hline
  T_d & O & T \\ 
  \hline
\end{array}$
    \caption{Improper point groups without inversion ($\mathcal{P}_{\neg \inv}$; first column) can be constructed from proper rotation groups ($\mathcal{P}$; second column) and an index-2 subgroup ($H< \calp$; third column). \label{tab:groups} }
\end{table}

\subsection{Crystallographic point groups are  monomial}\la{semidirect}

To show that crystallographic point groups [class (1)] are  monomial, we will apply Altmann's {semidirect-product decomposition}\footnote{See page 220-222 of \ocite{Altmann1961}.} of the crystallographic point groups.\\

\noindent \textit{Review of the semi-direct product.} $N \rtimes C$ is a group that is constructed from two groups $N$ and $C$ for which $C$ acts on $N$ by conjugation, i.e., $n \rightarrow c n c^{-1}$, for all $n \in N$ and $c \in C$. As a set, $N \rtimes C = N \times C$; as a group, elements in $N \rtimes C$ are multiplied as $(n,c) \cdot (n',c') = (n \, c n' c^{-1}, c c')$. The subgroup $N \times C_1$ -- henceforth referred to as $N$ -- is a \textit{normal} subgroup of $N \rtimes C$; the subgroup $C_1 \times C$ -- simply denoted by $C$ -- is generally just a subgroup of $N \rtimes C$.\\

Altmann showed that all crystallographic point groups can be expressed as $\mathcal{P}=N \rtimes C$, where $N$ is a maximal normal subgroup of $\mathcal{P}$ and $C$ a cyclic subgroup. Both $N$ and $C$ are subgroups of the group $O(3)$ of isometries in  3D real space, so the action of $C$ on $N$ is uniquely defined within $O(3)$.  \\ 

For $28$ of the $32$ crystallographic point groups that are not $\{T_h,T_d,O_h,O\}$, $N$ can further be shown to be abelian. Let us give an example for each sub-class of (1): in class (1)A, $\mathcal{P}=D_n$, $N=C_n$, $C=C_2'$ ($C_2'$ is generated by a two-fold rotation with rotational axis perpendicular to the rotational axis defined for $C_n$); in class (1)B, $\mathcal{P}=C_{4h}$, $N=C_4$, $C=\Z_2^{\inv}$; in class (1)C, $\mathcal{P}=C_{4v}$, $N=C_4$, $C=C_s'$ ($C_s'$ is generated by a mirror plane which is parallel to the rotational axis of $N$, and thus acts trivially on $N$). To recapitulate, each of these 28 crystallographic point groups is an extension of an abelian group ($C$) by an abelian group ($N$); such groups are called \textit{metabelian}, and it is known that all metabelian groups are monomial.\footnote{See Theorem 52.2 in \ocite{Curtis2006}.} \\

For the remaining $4$ crystallographic point groups -- $T_h$, $T_d$, $O_h$ and $O$ -- all maximal normal subgroups $N$ are non-abelian.\footnote{See page 222 of \ocite{Altmann1961}.}  But since every such $N$ is also a crystallographic point group, it has itself a decomposition $N = N' \rtimes C'$ with $C'$ cyclic and $N'$ a maximal normal subgroup of $N$. Altmann\cite{Altmann1961} showed that every crystallographic point group $\mathcal{P}=N\rtimes C$ for which $N=N'\rtimes C'$ is non-abelian, has a decomposition with $N'$ abelian and $\rtimes$ associative [cf.\ \q{normalseries2}]. The latter also implies that $N'$ is a normal subgroup of $\mathcal{P}$. \\

\noindent \textit{Example of triple semi-direct product:} $O=T \rtimes C_{2}''= D_2 \rtimes (C_3' \rtimes C_2'')$.\cite{Altmann1961} In review, the octahedral group $O$ consists of the orientation-preserving symmetries of a cube. Let the $x$-, $y$- and $z$-axes go through the center of the three faces of the cube; these three axes also define the axes of the two-fold rotational symmetries, which generate the subgroup $D_2$. A cube also has a three-fold rotational axis going through the corner $(1,1,1)$ of the cube; this three-fold rotational symmetry generates the group $C_3'$. (Incidentally, $D_2 \rtimes C_3' = T$ are  the orientation-preserving symmetries of a tetrahedron.) Finally, the cube has another two-fold rotational symmetry with axis going through the center of the vertex $(1,1,0)$. This two-fold rotational symmetry generates $C_2''$. Altogether, the mirror, three-fold and two-fold rotational symmetries generate the group $O$.\\


To recapitulate, for all crystallographic point groups $\mathcal{P}$, there exists an abelian normal subgroup $N'$ of $\mathcal{P}$ such that $\mathcal{P} = N' \rtimes (C' \rtimes C)$ where $C, C'$ are cyclic subgroups of $\mathcal{P}$; $C$ or $C'$ may be the trivial subgroup.\footnote{See page 228 of \ocite{Altmann1961}.} Therefore, Corollary 1 of \app{app:huppert} implies that $\mathcal{P}$ is monomial.


\subsection{Noncubic double point groups are monomial}\la{noncubic}

In this section, we will apply Corollary 2  of \app{app:huppert} to prove that all $27$ noncubic double point groups [class (2)]  are monomial. Given also that the five cubic double point groups are non-monomial [as proven in \app{app:nonmononial}], we conclude that a double crystallographic point group is monomial if and only if it is noncubic. \\

After giving a brief review of double crystallographic point groups in \app{app:reviewdoublegroup}, we tackle the proof of monomiality for class (2)A, (2)B and (2)C separately, in \app{app:class2A}, \app{app:class2B} and \app{noninv}.

\subsubsection{Review of double point groups}\la{app:reviewdoublegroup}

It is well-known from the study of angular momentum that the
double cover of $SO(3)$ is $SU(2)$. $SU(2)$ may be viewed as the unsplit central extension of $SO(3)$ by $\Z_2$. The  $\Z_2$ group is generated by $\tilde e$, which commutes with every element in $SU(2)$, and  has the physical interpretation of a $2\pi$ rotation. 
$SU(2)$ being a double cover means  there exists a 2-1 surjection $\phi \colon SU(2) \to SO(3)$. \\

Analogously, for each of the 11 proper rotation point groups (denoted $\calp$) that are subgroups of $SO(3)$, the double cover $\tilde{\mathcal{P}} = \phi^{-1}(\mathcal{P})$ is a subgroup of $SU(2)$.
The identity $e \in \mathcal{P}$ lifts (via $\phi^{-1}$) to two elements in $\tilde{\mathcal{P}}$ -- $e$ and $\tilde{e} \neq e$ -- which both commute with all elements in $\tilde{\mathcal{P}}$, and satisfy $\tilde{e}^2 = e$. For any $g\in \mathcal{P}$, there exist two corresponding elements, $g$ and $\tilde g = g \tilde{e}$, in $\tilde{\mathcal{P}}$. The multiplication rule of any two elements in $\tilde \calp$ is determined by the multiplication rule of the same elements in $SU(2)$. \\


More generally, the double covers of the
crystallographic point groups $(\mathcal{P},\calp_{\inv},\calp_{\neg \inv})$ are referred to as the \textit{double crystallographic point groups}, and denoted by $(\tilde{\mathcal{P}}, \tilde \calp_{\inv}, \tilde \calp_{\neg \inv})$.  We shall only concern ourselves with half-integer-spin representations of the double crystallographic point groups, in which $\tilde{e}$ is represented by $-1$ times the identity matrix.

\subsubsection{Proof for proper double point groups}\la{app:class2A}


Of the eleven proper double crystallographic point groups, only two of them ($\tilde{T}$ and $\tilde O$) are cubic. The remaining nine groups form class (2)A, and are proven here to be monomial.\\

The double covers of $C_n$ {for $n=1,2,3,4,6$} are still cyclic but with twice as many elements, i.e., $\tilde C_n \cong \Z_{2n}$. This reflects that a $2\pi$ rotation is not the identity element, but a $4\pi$ rotation is [cf.\ \app{app:reviewdoublegroup}].  Abelian groups such as  $\Z_{2n}$ are monomial because all their irreps are 1D (and 1D irreps are trivially induced from 1D irreps of the group itself). \\

The only remaining double proper crystallographic point groups  are the double covers ($\tilde D_n$) of $D_n = C_n \rtimes C_2'$, where { $n=2,3,4,6$} and the $C_2'$   axis is perpendicular to the $C_n$ axis. The generators of $C_2'$ and $C_n$ are denoted $c_2'$ and $c_n$ respectively.  As an element of $\tilde D_n$, $c_n$ generates a subgroup isomorphic to  $\Z_{2n}$.  This subgroup is normal in $\tilde D_n$ because $c_2' c_n c_2'^{-1} = c_n^{-1}$ (recall here that $c_2'$ inverts the $C_n$ axis), and the quotient group $\tilde D_n/\Z_{2n}=\{[e],[c_2'] \}$ is cyclic and isomorphic to $\Z_2$;  note that  $c_2'^2 = \tilde e \in \Z_{2n}$ lies in $[e]$. 
Therefore, $\Z_{2n}$ is an abelian and normal subgroup of $\tilde D_n$ with cyclic quotient group $\Z_2$. Corollary 2 then implies that $\tilde D_n$ is monomial.

\subsubsection{Proof for improper double point groups with inversion}\la{app:class2B}

The nine groups (denoted $\tilde{\mathcal{P}}_{\inv}$) in class (2)B are obtained by including inversion  symmetry ($\inv$) for each of the nine groups (denoted $\tilde \calp$ in class (2)A. $\inv$ squares to the identity and commutes with all double point-group operations,\cite{heinebook,kosterbook} we have the direct-product form 
$\tilde{\mathcal{P}}_{\inv}=\tilde{\mathcal{P}} \times \Z_2^{\inv}$. We have already proven in \app{app:class2A} that all $\mathcal{P}$ in class (2)A are monomial; then, according to the \textit{Lemma for monomial direct-product groups} in \app{app:monomialdirectproduct}, $\tilde{\mathcal{P}}_{\inv}=\tilde{\mathcal{P}} \times \Z_2^{\inv}$ must also be monomial.

\subsubsection{Proof for improper double point groups without inversion}\la{noninv}



Of the ten improper double crystallographic point groups without inversion, only one of them ($\tilde{T}_d$) is cubic, and the rest have the denotation
$\tilde{\mathcal{P}}_{\neg \inv}$ and form class (2)C.\\

To prove the monomiality of $\tilde{\mathcal{P}}_{\neg \inv}$, we first prove its isomorphism with   $\tilde{\calp}$ in class (2)A [cf.\ \app{app:class2A}].  $\tilde{\mathcal{P}}_{\neg \inv}\cong \tilde{\calp}$ will be derived from the isomorphism ${\mathcal{P}}_{\neg \inv}\cong {\calp}$, for $\calp$ a proper, noncubic crystallographic point group having an index-two subgroup $H$.  (To clarify, of the nine noncubic proper crystallographic point groups, two of them [$C_1$ and $C_3$] have no index-two subgroups, for two of them [$D_4$ and $D_6$] each has two index-two subgroups, while the rest each has one index-two subgroup. This means that the nine groups in class (2)C will be shown to be isomorphic to seven groups in class (2)A.)\\

We remind the reader of the set decompositions $\calp = H+\calp\backslash H$ and  ${\mathcal{P}}_{\neg \inv}=H+\inv \calp \backslash H$, as reviewed in \app{classes}. Under  the 2-1 surjection $\phi:SU(2) \rightarrow SO(3)$, the preimage of $H$ is a subgroup of both $\tilde \calp$ and $\tilde{\mathcal{P}}_{\neg \inv}$.  On the other hand, $\phi^{-1}(\calp \backslash H)$ is a subset of $\calp$, while $\inv \phi^{-1}(\calp \backslash H)$ is a subset of $\tilde{\mathcal{P}}_{\neg \inv}$. There is therefore a bijection of group elements between $\tilde \calp$ and $\tilde{\mathcal{P}}_{\neg \inv}$, where in the  direction $\tilde{\mathcal{P}}_{\neg \inv}\rightarrow \tilde \calp$ one merely drops the $\inv$ label. Moreover, this bijection preserves the multiplication rule, because $\inv$ commutes with every point-group operation.\cite{heinebook,kosterbook} \\ 

Now we combine the just-stated isomorphism with a result established in \app{app:class2A}, namely that
all \textit{noncubic}     double proper rotation groups [$\tilde \calp$ in class (2)A] are monomial. Since each noncubic double improper rotation group without  inversion [$\tilde \calp_{\neg \inv}$ in class (2)A] is isomorphic to one of  $\tilde \calp$ in class (2)A,  we deduce that $\tilde \calp_{\neg \inv}$ must also be monomial. This follows because if two groups $A\cong B$ are isomorphic, then $A$ is monomial if and only if $B$ is monomial. Indeed, every representation of $A$ gives a representation of $B$   via the group isomorphism, and vice versa. So if all representations of $A$ are induced from 1D irreps of subgroups of $A$, so must all representations of $B$  be induced from 1D irreps of subgroups of $B$, and vice versa.\\

\subsection{Grey magnetic point groups and grey magnetic double point groups are monomial}\la{typeI}

Here we prove that all $32$ grey magnetic point groups (denoted $\calp_T$), and all $27$ grey magnetic noncubic double point groups ($\tilde \calp_T$) are monomial.\\

Our proof relies on Wigner's seminal result,\cite{Wigner_ontheoperationoftimereversal} namely that \textit{all} irreps of $\calp_T=\calp\times \Z_2^T$ are induced from irreps of the crystallographic point group $\calp$. Similarly, \textit{all} half-integer-spin irreps of    $\tilde \calp_T$ (the double-group extension of $\calp\times \Z_2^T$) are induced from irreps of the double point group $\tilde \calp$.  (In fact, Wigner goes further to show that a representation $D$ of $\calp$ is either (a) compatible with time-reversal $T$ symmetry, or (b) incompatible with $T$ symmetry, but $D\oplus D^*$ [$D^*$ being the complex conjugate of $D$] is compatible. Which case holds  depends on whether $D$ is an integer-spin or half-integer-spin representation, and whether $D$ is real, complex, or quaternionic. Such considerations, however, lie outside the scope of our proof.) \\

Since {any} irrep (denoted $D_T$) of $\calp_T$ is induced from an irrep ($D$) of $\calp$,  the question of whether $D_T$ is monomial reduces to the question of whether $D$ is monomial. In other words, if $D$ is induced from a one-dimensional representation of a subgroup $H< \calp$, then it follows that $D_T$ is also induced from the same one-dimensional representation of  $H< \calp < \calp_T$. Such a one-dimensional representation always exists for any representation $D$ of $\calp$, because  of our previously-established result that all 32 crystallographic point groups are monomial; cf.\ \app{semidirect}. Thus we conclude that all 32 grey magnetic point groups are also monomial. \\

By similar reasoning, one concludes that all 27 grey magnetic noncubic double point groups are monomial, based on our result that all 27  double noncubic point groups are monomial; cf.\ \app{noncubic}.\\

\subsection{Double cubic point groups are non-monomial} \la{app:nonmononial}

We will show that the double-group extensions of the cubic crystallographic point groups $T,T_d,T_h,O,O_h$ are non-monomial.\\

$\tilde T$ and $\tilde O$ 
are standard examples of \textit{non}-monomial groups.\footnote{See Example 5 on page 57 of \ocite{Bray1982}, where we used that $T$ is isomorphic to the alternating group and $O$ to the symmetric group of four elements.} \\

\noindent \textit{Example of non-monomial irrep of double cubic point group $\tilde{T}$.} Let us consider the non-monomial\footnote{See Example 2 on page 54/55 of \ocite{Bray1982}.} 2D irrep $\bar E$ of the double group $\tilde T$. In a representation basis with spin quantization axis that is parallel to the $C_2$ axis, the three generators of $\tilde T$ are represented as 
\es{ C_2 = \mathrm{e}^{-\pi i \sigma_z/2}, \, C_2' = \mathrm{e}^{-\pi i \sigma_x/2} , \, C_3' =  \mathrm{e}^{-\pi i (\sigma_x+\sigma_y+\sigma_z)/(3 \sqrt{3})}. }
$C_2$ and $C_2'$ are complex permutation matrices but $C_3'$ is not; an analogous statement holds in an eigenbasis of $C_2'$. In a basis where $C_3' =  \mathrm{e}^{-\pi i \sigma_z/3}$ is diagonal, we find instead that $C_2 = \mathrm{e}^{-\pi i (-\sigma_x+\sigma_y-\sigma_z)/(2 \sqrt{3})}$ is not a complex permutation matrix. \\



Of the three remaining double cubic point groups, two have the direct-product form: $\tilde{T}_h = \tilde T \times \Z_2^{\inv}$ and $\tilde O_h = \tilde O \times \Z_2^{\inv}$. (The direct-product structure was explained in \app{noncubic}.) Since $\tilde O$ and $\tilde T$ are non-monomial, it follows that  $\tilde O_h$ and $\tilde T_h$ must also be non-monomial,  according to the \textit{Lemma for monomial direct-product groups} in \app{app:monomialdirectproduct}. \\

Finally, to show that $\tilde T_d$ is non-monomial, we will use that double-group extensions of isomorphic crystallographic point groups are also isomorphic, and monomiality (as well as non-monomiality) is preserved by group isomorphisms -- both of these claims have been proven in \app{noninv}.
Thus $T_d\cong O$ implies their double-group extensions  are also isomorphic: $\tilde T_d \cong \tilde O$; moreover, since $\tilde O$ is known to be non-monomial, so must $\tilde T_d$ be non-monomial.


\section{Tightly-bound BRs and the existence of the symmetric tight-binding limit }\la{app:atomicBR}

We have used in \s{sec:zakphasewind} and \s{sec:norobustsurface} that every BR has a symmetric tight-binding limit to a tightly-bound BR. 
We remind the reader that a tightly-bound BR is a BR for which all Wannier functions are one-site localized. The goal of this appendix is to describe tightly-bound BRs in the language of $G$-vector bundles (in \app{VBBR}), so as to rigorously prove the existence of a symmetric tight-binding limit for any BR (in \app{ALproof}). In \app{TBlatt} we provide definitions for $G$-vector bundles and discuss their applications in tight-binding (TB) lattice models.

\subsection{$G$-vector bundles and tight-binding lattice models}\la{TBlatt}

We have heuristically introduced (complex) vector bundles in \app{sec:bandbundle} from the perspective of band theory. Here, we review some basic bundle notions from the mathematical perspective, and describe their application to tight-binding lattice models. \\ 

\noindent A \textit{complex vector bundle} is a continuous surjection $p\colon E \to B$ from a (topological) space $E$, called \textit{total space}, to a (topological) space $B$, called \textit{base space}. Furthermore, it has a \textit{local trivialization}: For every $b \in B$ there exists a neighbourhood $U_b \subset B$  and a continuous bijection $h_b \colon p^{-1}(U_b) \to U_b \times \C^N$ with continuous inverse such that $\left. h_b \right|_{p^{-1}(b)}$ is a linear isomorphism of vector spaces. $E_b=p^{-1}(b)$ is called the \textit{fiber} over $b$, and $N$ the \textit{rank} of the vector bundle. \\ 

The total space $E$ can also be viewed as a disjoint union of all fibers, i.e., $E\equiv \sqcup_{b \in B} E_b$. The local trivialization implies that every fiber of a complex rank-$N$ vector bundle is isomorphic to $\C^N$.  \\

The notion of isomorphism for vector spaces is well-known, e.g., any complex $N$-dimensional vector space is isomorphic to $\C^N$. There is an analogous notion of isomorphism for vector bundles: \\ 

\noindent For two vector bundles $E$, $E'$ over the same base space $B$, a \textit{vector bundle isomorphism} is a continuous bijection $f \colon E \to E'$ with continuous inverse such that $\left.f \right|_{E_{b}}$ is a linear isomorphism from the fiber $E_b$ to the fiber $E'_b$ for all $b\in B$. \\ 

For example, a vector bundle that has nontrivial first Chern class is not isomorphic a \textit{product bundle}; the latter has total space $BZ \times V$,  with $BZ$ the Brillouin zone and  $V$ an $N$-dimensional complex vector space. \\

Let us apply these bundle notions to \textit{tight-binding lattice models}. A tight-binding lattice model corresponds to a finite-dimensional vector space (in each unit cell indexed by Bravais-lattice vector $\bR$)  spanned by $N_{tot}$  tight-binding basis functions $\vec{\phi}_{\alpha,\bR}$, indexed by $\alpha=1,\ldots,N_{tot}$.
The Fourier transforms of tight-binding  basis functions span the fibers of a vector bundle with total space $E_{TB}$ over $B=BZ$ (the Brillouin zone). Since tight-binding basis functions are -- by definition -- one-site localized, {their Fourier transforms are $\bk$-independent}; hence, the fibers are $\bk$-independent $N_{tot}$-dimensional complex vector spaces, denoted $V_{TB}$, and $E_{TB}=BZ \times V_{TB}$. \\

Any rank-$N$ vector bundles $E$ over the BZ -- with $1 \le N \le N_{tot}$ -- can be embedded in this tight-binding product bundle $E_{TB}$ with $N_{tot}$ large, as exemplified by a rank-$N$ energy band of a tight-binding Hamiltonian. 
Then each fiber $E_{\bk}$ is spanned by $N$ vectors which we denote as $\vec{\mathfrak{V}}_{n}(\bk) = ( \mathfrak{V}_{n}(\bk)^{\alpha} )_{\alpha=1}^{N_{tot}}$ with $n=1,\ldots,N$. It is assumed that $\{\vec{\mathfrak{V}}_{n}(\bk)\}_{n=1}^N$ are periodic over the Brillouin torus; these vectors span the fiber $E_{\bk}$ at each $\bk$. If the vector bundle is topologically trivial (in spatial dimension $d\leq 3$, topological triviality is equivalent to having trivial first Chern class), then each $\vec{\mathfrak{V}}_{n}(\bk)$ can be chosen to be a periodic\textit{ and analytic} function of $\bk$. The Fourier transform of each such $\vec{\mathfrak{V}}_{n}(\bk)$ defines a set of Wannier functions related by Bravais-lattice translations. Especially, if $E$ is spanned by one-site localized Wannier functions then $\vec{\mathfrak{V}}_{n}(\bk)$ can be chosen $\bk$-independent, and $E$ is a product bundle. \\

Space group symmetries $g \in G$ provide additional structure to vector bundles. Especially, the total space $E$ and the base space $B$ are $G$-spaces. \\ 

\noindent For a topological group $G$ and a topological space $X$, called a \textit{$G$-space}, a continuous \textit{action} of $G$ on $X$ is given by a continuous map $\circ \colon G \times X \to X$ such that $e \circ x=x$ and $(gh) \circ x = g \circ ( h \circ x)$ for all $x\in X$ and all $g,h\in G$. \\ 

In band theory, $G$ acts on the BZ by $g \colon \bk \to \check g \bk$ (modulo reciprocal lattice vectors), and acts fiberwise on $E$ by a unitary matrix $U_g(\bk)$ that is sometimes called the `sewing matrix':
\e{ \hat g \colon \vec{\mathfrak{V}}_{n}(\bk) \in E_{\bk} \mapsto \sum_{n'=1}^N \big[ U_g(\bk) \big]_{n' n} \vec{\mathfrak{V}}_{n'}(\check g \bk) \in E_{\check g \bk}. \la{action} } 
(A space group $G$ is a topological group using the discrete topology.) In fact, $E$ is a $G$-bundle, defined as follows.
\begin{definition} \normalfont
For $G$-spaces $E$ and $B$, a \textit{$G$-vector bundle} is a vector bundle for which $p\colon E \to B$ is a $G$-map, and the fiber-wise action $g \colon E_{b} \to E_{g b}$ is a linear isomorphism for all $b \in B$. \\
For $p$ to be a \textit{$G$-map} means that $g p(e) = p(g e)$ for all $e \in E$ and $g \in G$. \\
Two $G$-vector bundles $E$, $E'$ over the same base space $B$ are \textit{$G$-isomorphic} if there exists a complex vector bundle isomorphism $f \colon E \to E'$ that is also a $G$-map. 
\end{definition}
Note the notational difference between an isomorphism (as a complex vector bundle) and a $G$-isomorphism.

\subsection{BRs and tightly-bound BRs as G-vector bundles}\la{VBBR}

We now discuss how BRs and tightly-bound BRs can be expressed as $G$-vector bundles. \\

For simplicity, let us consider a rank-$N$ BR$(G,\bvarpi,D)$. Then there always exists a basis $\{ \vec{\mathfrak{V}}_{n}(\bk) \}_{n=1}^N$ of each fiber $E_{\bk}$ that is analytic in $\bk \in BZ$ (e.g. \ocite{nogo_AAJH}, section I. B). Their continuity and linear independence at each $\bk \in BZ$ implies that there exists an isomorphism (as complex vector bundles) from $E$ to the rank-$N$ product bundle (cf.\ \ocite{Hatcher2009}, p. 8).\\

For a BR$(G,\bvarpi,D)$,  the action of $G$ on $E$ is referred to as a \textit{$(G,\bvarpi,D)$-action}, which we define by \q{action} with $U_g(\bk)$ having the following canonical form (cf.\ \ocite{Zak1979}):
\e{ \big[ U_g(\bk) \big]_{n' n} = \mathrm{e}^{-i s_g \check g \bk \cdot \Delta_{g,n',n}} \big[ U_g(\textbf{0}) \big]_{n' n}. \la{symmrep} }
Here, $n'$ is uniquely defined by the Wannier center $\bvarpi_{n'}$ and by a Bravais lattice vector $\Delta_{g,n',n}$ such that $g\circ \bvarpi_n = \bvarpi_{n'} + \Delta_{g,n',n}$.
Furthermore, $U_g(\textbf{0})$ is determined by $D(\check g_{n'}^{-1} \check g \check g_n)$, as explained in detail in \ocite{nogo_AAJH}, Appendix A. To recapitulate, a BR($G,\bvarpi,D$) forms a $G$-VB $E$ with a $(G,\bvarpi,D)$-action. (Implicit in this definition is that $E$ is vector bundle isomorphic to the product bundle.) \\

A tightly-bound BR($G,\bvarpi,D$) is a BR($G,\bvarpi,D$) with the additional property that (i) it is sub-bundle of a tight-binding lattice model (given by $E_{TB}$ over $BZ$), and (ii) it is a $G$-product bundle with $(G,\bvarpi,D)$-action. 
(ii) implies there exists a $\bk$-independent basis for the $\bk$-independent fibers; the Fourier transform of this basis gives  one-site localized Wannier functions.


\subsection{Existence of symmetric tight-binding limit}\la{ALproof}

A BR and a tightly-bound BR with the same $(G,\bvarpi,D)$-action are $G$-isomorphic. \\

\noindent \textit{Proof.} 
Here we prove the more general claim that any two $G$-vector bundles $E$ and $E'$ of the same rank and with the same $(G,\bvarpi,D)$-action are $G$-isomorphic. In particular, this holds for $E$ a rank-$N$ BR with $(G,\bvarpi,D)$-action and for $E'$ a rank-$N$ tightly-bound BR with the same $(G,\bvarpi,D)$-action. \\


Let $E$ and $E'$ be two $G$-vector bundles of the same rank and with the same $(G,\bvarpi,D)$-action. Their fibres at $\bk \in BZ$ are spanned by $\{ \vec{\mathfrak{V}}_{n}(\bk) \}_{n=1}^N$ and $\{ \vec{\mathfrak{V}}'_{n}(\bk) \}_{n=1}^N$, respectively. By definition,  $E$ and $E'$ are isomorphic as complex vector bundles, which means there exists a linear isomorphism $\cali$ from the fibres of $E$ and to those of $E'$ (cf.\ \ocite{Hatcher2009}, p. 8):
\e{ \cali\big( \bk,\vec{\mathfrak{V}}_{n}(\bk) \big) = \big(\bk,\vec{\mathfrak{V}}_{n}'(\bk)\big). \la{linearisomorphism2}} 
To show that $\cali$ is also a $G$-isomorphism, it suffices to prove that $\cali$ is a $G$-map:
\m{ \hat g \, \cali \big( \bk, \vec{\mathfrak{V}}_{n}(\bk) \big) 
= \hat g \, \big( \bk, \vec{\mathfrak{V}}'_{n} \big) = \big( \check{g} \bk, \sum_{n'=1}^N \big[ U_g(\bk) \big]_{n' n} \vec{\mathfrak{V}}'_{n'}(\bk) ) \\
= \sum_{n'=1}^N \big[ U_g(\bk) \big]_{n' n} \big( \check{g} \bk, \vec{\mathfrak{V}}'_{n'}(\bk) ) \\
= \sum_{n'=1}^N \big[ U_g(\bk) \big]_{n' n} \cali \big( \check{g} \bk, \vec{\mathfrak{V}}_{n'}(\check g \bk) ) \big) \\
= \cali \big( \check{g} \bk,  \sum_{n'=1}^N \big[ U_g(\bk) \big]_{n' n} \vec{\mathfrak{V}}_{n'}(\check g \bk) ) \big) =  \cali \big( \hat g (\bk, \vec{\mathfrak{V}}_{n}(\bk) ) \big). }
For the equality in the second row we used the linearity of the fibers of $E'$, whereas for the first equality in the fourth row we used the linearity of $\cali$ and the linearity of the fibers of $E$. \hfill\(\Box\) \\ 

Let us discuss a physical interpretation of the above $G$-isomorphism, in the case that $E$ and $E'$ are sub-bundles of a larger rank-$N_{tot}$ $G$-vector bundle $E_{TB}$, with $E$ corresponding to a non-tightly-bound BR, $E'$ to a tightly-bound BR, and  $E_{TB}$ the vector bundle  of a tight-binding lattice model (as introduced in \app{TBlatt}).  The universal $G$-vector bundle theorem\cite{Atiyah1989} states that the proven $G$-isomorphism between $E$ and $E'$ corresponds {bijectively} to a $G$-homotopy, i.e., a continuous, symmetric deformation from $E$ to $E'$. \\

For a fixed tight-binding lattice model $E_{TB}$ and a subbundle $E$ that transforms as a non-tightly-bound BR$(G,\bvarpi,D$), the G-vector bundle $E'$ of a tightly-bound BR$(G,\bvarpi,D$) may not be a subbundle of $E_{TB}$. In this case, the BR would have a \textit{symmetric tight-binding obstruction}, as defined in \s{sec:norobustsurface}. To construct a $G$-symmetric homotopy between $E$ and $E'$, it is sufficient to  enlarge the tight-binding lattice model as $E_{TB}\rightarrow E_{TB}\oplus E'$, as exemplified numerically in  \ocite{fragile_po}. \\

\section{Lemma for Zak phases of tightly-bound band representations} \la{app:zakwannier}

We will prove a lemma stated in \s{sec:prelimzakphase}, namely that for the rank-$N$ projector $P$ to a tightly-bound band representation (BR), the Zak phase $2\pi x_j(k_y)$ for a set of loops $\calc(k_y)$ (given by varying $k_x$ at fixed $k_y$) is independent of $k_y$, for all $j=1\ldots N$. (As in \s{sec:prelimzakphase}, we will simplify notation by assuming a rectangular real-space lattice with lattice periods set to one.) \\

It is sufficient to prove the lemma for the tightly-bound BR$(G,\bvarpi_1,D)$, with the understanding that a general tightly-bound BR is a direct sum of tightly-bound BR's with  Wyckoff positions that are possibly symmetry-inequivalent.\\

The Wannier centers of the tightly-bound BR$(G,\bvarpi_1,D)$ are given by $\{\bvarpi_n+\bR\}_{n=1\ldots M,\bR\in BL}$, with $M$ the number of distinct Wannier centers in one unit cell, and $BL$ a shorthand for the Bravais lattice. The projector to this tightly-bound BR can be decomposed as a sum of projectors to a finite number $A$ of Wannier functions  on each site: $P=\sum_{n=1}^M\sum_{\bR\in BL}P_{n\bR}$.
We assume that the real-space support of Wannier functions on different sites do not intersect. (This is certainly true of tight-binding Wannier functions which are one-site localized.) Then the projected position operator simplifies to a sum of commuting operators
\e{ P\hat{x}P= \sum_{n\bR} P_{n\bR}\hat{x}P_{n\bR}.}
The eigenproblem for each commuting operator should then be independently solved:
\e{ (P_{n\bR}\hat{x}P_{n\bR}-\bar{x}^{\alpha}_{n\bR})\ket{W^{\alpha}_{n\bR}}=0, \la{evaluecenterproj}}
giving the complete spectrum of the projected position operator:
\e{ \text{spec}P\hat{x}P = \{\bar{x}^{\alpha}_{n\bR}\}_{\alpha=1\dots A,n=1\dots M,\bR\in BL}. }

Observe that $\bar{x}^{\alpha}_{n\bR}:=\bar{x}^{\alpha}_{n R_x}$ is independent of $R_y$, owing to the $y$-translational symmetry of $PxP$ and the spatial localization of the Wannier functions. Indeed, supposing  $\ket{W^{\alpha}_{n\bR}}$ is an eigenstate of $P_{n\bR}$ with eigenvalue $\bar{x}^{\alpha}_{n\bR}$ [cf.\ \q{evaluecenterproj}], $\widehat{(\be_y|e)}\ket{W^{\alpha}_{n\bR}}$ must be an eigenstate of $P_{n\bR+\be_y}$ with the same eigenvalue.\\

It follows that any linear combination of \{$W^{\alpha}_{n\bR}\}$ with the same $\{n,R_x,\alpha\}$ label remains an eigenstate of $P\hat{x}P$:
\e{ (P\hat{x}P - \bar{x}^{\alpha}_{n R_x})\sum_{R_y}f(R_y)\ket{W^{\alpha}_{n\bR}}=0,  }
with $f$ an arbitrary function. In particular, if we choose $f$ to be the plane-wave phase factor $e^{ik_yR_y}$, then the sum can be identified as the hybrid function $\ket{h_{j,k_y,R_x}}$ in \q{projectedpositioneigen} with $j:=(\alpha,n)$, and $\bar{x}^{\alpha}_{n R_x}$ can be identified as the eigenvalue $(x_j(k_y)+R_x)$  in \q{projectedpositioneigen}. We thus derive the desired result that $x_j(k_y)$ is independent of $k_y$ for all $j$. \hfill\(\Box\) \\

\section{Proof of localization obstruction lemma}\la{app:locobslemma}

In this appendix we would prove the localization obstruction lemma of \s{sec:localization_obstruction}.\\

Let $P$ be a representation of a space group $G$ with translational subgroup $\calt_d$. If the Wannier functions spanning $P$ are all one-site localized, then the set of all Wannier functions $\{W_{1,\bze}^{\alpha}\}_{\alpha=1}^A$  lying on a Wyckoff position $\bvarpi_1$ must form a representation of the site stabilizer $G_{\bvarpi_1}.$ Indeed, since any $g\in G_{\bvarpi_1}$ acts in real space as an isometry, $\hat{g}W_{1,\bze}^{\alpha}$ must also be one-site localized to $\bvarpi_1$, and therefore has zero overlap with any Wannier function that is not one of $\{W_{1,\bze}^{\alpha}\}_{\alpha=1}^A$.
On the other hand, $\hat{g} W_{1,\bze}^{\alpha}$ must belong in $P$ which represents $G$. Thus for any $g\in G_{\varpi}$,  $\braket{W_{1,\bze}^{\alpha}}{\hat{g}W_{1,\bze}^{\beta}}$ is a {$A$-dimensional} unitary matrix in the indices $\alpha$ and $\beta$. \\

To finish the proof, if $\bvarpi_1$ is the {Wannier center} of {precisely} $A$ linearly-independent Wannier functions $\{\ket{W_{1,\bze}^{\alpha}}\}_{\alpha=1}^A$ in $P$, then for any representatives of the coset: $G/(\calt_d\rtimes G_{\varpi_1}) = \{ [g_1=e], [g_2], ..., [g_M] \}$ (with $M=|G/(\calt_d\rtimes G_{\varpi_1})|$), the real-space position $\bvarpi_n=g_n\circ \bvarpi_1$ must likewise be the {Wannier center} for the $A$ linearly-independent Wannier functions: $\{\hat{g}_n\ket{W_{1,\bze}^{\alpha}}\}_{\alpha=1}^A$ in $P$. This is because  $P$ is assumed to be invariant under all elements of $G$. Using once again that $g$ acts as a real-space isometry, and that all Wannier functions are one-site localized, it follows that any Wannier function in $P$ with {Wannier center} $\bvarpi_n$ belongs to the span of $\{\hat{g}_n\ket{W_{1,\bze}^{\alpha}}\}_{\alpha=1}^A$, which forms an $A$-dimensional representation of $G_{\varpi_n} = g_n G_{\varpi_1} g_n^{-1}$. 
Finally, for any $(\bR|e)\in \calt_d$, $\{W_{n,\bR}^{\alpha}:=\widehat{(\bR|e)}W_{n,\bze}^{\alpha}\}_{\alpha}$ must also form an $A$-dimensional representation of the site stabilizer of $\bvarpi_n+\bR$. With this, all conditions are satisfied for $\{W_{n,\bR}^{\alpha}\}_{\alpha,n,\bR}$ to be a locally-symmetric Wannier basis [cf.\ \app{app:locallysymmetricWannierbasis}] for a BR of $G$ with Wyckoff position $\bvarpi_1$. (A stronger statement can be made if there exists a basis of Wannier functions where $\braket{W_{1,\bze}^{\alpha}}{\hat{g}W_{1,\bze}^{\beta}}$ is a complex permutation matrix (for any $g$ in the site stabilizer), namely that  $\{W_{n,\bR}^{\alpha}\}_{\alpha,n,\bR}$ would span a monomial BR of $G$. However, our proof more generally applies to non-monomial BRs as well.) \\

If $\{W_{n,\bR}^{\alpha}\}_{\alpha,n,\bR}$ spans $P$ then the proof is complete, otherwise there must exist other Wannier functions that lie at $G$-inequivalent Wyckoff positions. By iterating the above argument for the remaining Wannier functions, one would generally conclude that $P$ is a direct sum of BRs of $G$, possibly with  Wyckoff positions that are {not related by $G$ symmetry}. Finally, $P$ being a BR contradicts our initial assumption that $P$ is an obstructed representation.\hfill\(\Box\) \\

\bibliography{bib_Apr2018}

\end{document}